\documentclass{aa}
\usepackage{natbib}
\usepackage{arydshln}
\usepackage{xcolor}
\usepackage{upquote}
\usepackage{subfig}
\usepackage{amsmath}
\usepackage{nccmath}
\usepackage{float}

\bibpunct{(}{)}{;}{a}{}{,}

\begin{document}

\title{Gaia Data Release 2: Short-timescale variability processing and analysis}
\author{M. Roelens\inst{\ref{inst1}, \ref{inst2}}, L. Eyer\inst{\ref{inst1}}, N. Mowlavi\inst{\ref{inst1}, \ref{inst2}}, L. Rimoldini\inst{\ref{inst2}}, I. Lecoeur-Ta\"ibi\inst{\ref{inst2}}, K. Nienartowicz\inst{\ref{inst3}}, G. Jevardat de Fombelle\inst{\ref{inst3}}, O. Marchal\inst{\ref{inst2}, \ref{inst4}}, M. Audard\inst{\ref{inst1}, \ref{inst2}}, L. Guy\inst{\ref{inst2}}, B. Holl\inst{\ref{inst1}, \ref{inst2}}, D. W. Evans\inst{\ref{inst5}}, M. Riello\inst{\ref{inst5}}, F. De Angeli\inst{\ref{inst5}}, S. Blanco-Cuaresma\inst{\ref{inst6}}, T. Wevers\inst{\ref{inst7}, \ref{inst5}}.}
\authorrunning{Roelens et al.}
%\email{maroussia.roelens@unige.ch}
\institute{D\'{e}partement d'Astronomie, Universit\'{e} de Gen\`{e}ve, Chemin des Maillettes 51, CH-1290 Versoix, Switzerland\label{inst1} \and D\'{e}partement d'Astronomie, Universit\'{e} de Gen\`{e}ve, Chemin d'Ecogia 16, CH-1290 Versoix, Switzerland\label{inst2} \and SixSq, Rue du Bois-du-Lan 8, CH-1217 Meyrin, Switzerland\label{inst3} \and G\'EPI, Observatoire de Paris, Universit\'e PSL, CNRS, Place Jules Janssen 5, F-92190 Meudon, France\label{inst4} \and Institute of Astronomy, University of Cambridge, Madingley Road, Cambridge CB3 0HA, UK\label{inst5} \and Harvard-Smithsonian Center for Astrophysics, 60 Garden Street, Cambridge, MA 02138, USA\label{inst6}
\and Department of Astrophysics / IMAPP, Radboud University, P.O. Box 9010, NL-6500GL, Nijmegen, The Netherlands\label{inst7}}
%\and Large Synoptic Survey Telescope, 950 N. Cherry Avenue, Tucson, AZ 85719, USA\label{inst5}
\date{}

\abstract
{}
{We describe the methods used and the analysis performed in the frame of the \textit{Gaia} data processing activities to produce the \textit{Gaia} Data Release 2 (DR2) sample candidates with short-timescale variability together with associated parameters.}
{The \textit{Gaia} DR2 sample of candidates with short-timescale variability results from the investigation of the first 22 months of \textit{Gaia} $G$ per-CCD, $G_{\mathrm{BP}}$ , and $G_{\mathrm{RP}}$ photometry for a subsample of sources at the \textit{Gaia} faint end ($G \sim 16.5 - 20\,$mag). For this first short-timescale variability search exploiting \textit{Gaia} data, we limited ourselves to the case of suspected rapid periodic variability. Our study combines fast-variability detection through variogram analysis, a high-frequency search by means of least-squares periodograms, and an empirical selection based on the investigation of specific sources seen through the \textit{Gaia} eyes (e.g. known variables or visually identified objects with peculiar features in their light curves). The progressive definition, improvement, and validation of this selection criterion also benefited from supplementary ground-based photometric monitoring of a few tens of preliminary candidates  with short-timescale variability, performed at the Flemish Mercator telescope in La Palma (Canary Islands, Spain) between August and November 2017.}
{As part of \textit{Gaia} DR2, we publish a list of 3,018 candidates with short-timescale variability, spread throughout the sky, with a false-positive rate of up to 10-20\% in the Magellanic Clouds, and a more significant but justifiable contamination from longer-period variables between 19\%\ and 50\%, depending on the area of the sky. Although its completeness is limited to about 0.05\%, this first sample of \textit{Gaia} short-timescale variables recovers some very interesting known short-period variables, such as post-common envelope binaries or cataclysmic variables, and brings to light some fascinating, newly discovered variable sources. In the perspective of future \textit{Gaia} data releases, several improvements of the short-timescale variability processing are considered, by enhancing the existing variogram and period-search algorithms or by classifying the identified variability candidates. Nonetheless, the encouraging outcome of our \textit{Gaia} DR2 analysis demonstrates the power of this mission for such fast-variability studies, and opens great perspectives for this domain of astrophysics.}
{}


\maketitle

\section{Introduction}
\label{intro}

\textit{Gaia} is a cornerstone mission in the science program of the European Space Agency (ESA). Launched in December 2013 for a five-year nominal mission duration, the main aim of this astrometric successor of the \textit{Hipparcos} ESA mission is to determine highly accurate positions, parallaxes, and proper motions for more than one billion stars in the Milky Way.
\textit{Gaia} is expected to observe objects in our Galaxy and beyond, spread throughout the sky, providing precise astrometry at the $10 - 20\,$microarcsecond level, photometry and spectrophotometry down to $G \approx 20.7\,$mag (where $G$ is the \textit{Gaia} broad-band white-light magnitude) with standard errors down to a few millimagnitudes (mmag) for bright sources, and medium-resolution spectroscopy down to $G \approx 17\,$mag \citep{Brown2016GDR1, Brown2016GDR1b}.
%\footnote{For more information on \textit{Gaia} performances, please see the \textit{Gaia} webpage \url{http://www.cosmos.esa.int/web/Gaia/science-performance.}}.

The second \textit{Gaia} Data Release (\textit{Gaia} Data Release 2, hereafter \textit{Gaia} DR2)\footnote{For more details on the \textit{Gaia} DR2 contents, see \url{https://www.cosmos.esa.int/web/Gaia/dr2}}, published on April 25,$^{\mathrm{}}$ 2018, is based on the data collected during the first 22 months of the \textit{Gaia} nominal mission \citep{GDR2Summary2018}. Among all the analyses performed for this data release, the \textit{Gaia} DR2 variability processing, described in \cite{GDR2CU7Summary2018}, resulted in the publication of more than 500,000 variable sources, with associated light curves and additional variability parameters when appropriate, belonging to diverse variability classes from BY Draconis candidates to long-period variables. The publication also includes more than 3,000 candidates with short-timescale variability, that is to say, sources showing photometric variability with characteristic timescales from a few tens of seconds to a dozen hours. Various astronomical sources are known to exhibit such fast variability, including periodic and non-periodic phenomena, and they are spread throughout the Hertzsprung-Russell \citep[HR; ][]{Russell1914} diagram. The amplitudes for types with variabiltiy on such short timescales can rank from a few millimagnitudes to a few magnitudes. The diverse phenomena at the origin of the variability reach from stellar pulsations to binarity and eruptions. Hence, an improved knowledge and understanding of short-timescale variables can bring invaluable clues into several fields of astrophysics.
Until now, the discovered number of such short-timescale variables remains quite modest compared to other types of longer-term variability (such as Mira or Cepheid stars). This is directly linked to the observational constraints when dealing with fast variability, both in terms of time sampling and photometric precision. Nonetheless, since the early 2000s, high-cadence photometric monitoring surveys in space \citep[e.g. \textit{Kepler}; ][]{Borucki2010Kepler} and from ground \citep[e.g. the Optical Gravitational Lensing Experiment; ][]{Udalski1992OGLE} allowed a significant advance in the rather unexplored domain of short-timescale variability. Moreover, during the past decade, some projects specifically dedicated to the detection and characterisation of short-timescale variables arose, like the Rapid Temporal Survey \citep[RATS; ][]{Barclay2011RATS} or the OmegaWhite survey \citep{Macfarlane2015OmegaWhite,Macfarlane2016OmegaWhite}.

In this context, \textit{Gaia} offers a unique opportunity for comprehensive, fast-variability studies over the whole sky. Its peculiar time sampling in $G$ band, involving fast cadences from a few seconds to a few hours \citep{deBruijne2012}, and its high photometric precision \citep{GDR2Evans2018} enable us to probe the short-timescale variability domain, down to timescales of a few tens of seconds, including low-amplitude phenomena.

In this paper, we present the \textit{Gaia} DR2 catalogue of 3,018 candidates with short-timescale variability. This is published as part of the \textit{Gaia} DR2 archive\footnote{\url{https://gea.esac.esa.int/archive/}}.\\
In Section \ref{stsDetectionCharac} we detail the algorithms and specific metrics we used to detect short-timescale variability and select bona fide candidates. Section \ref{stsPeriodicSelection} summarises the various filtering steps and selection criteria we applied to retrieve the final 3,018 candidates with short-timescale variability that is suspected to be periodic, published in \textit{Gaia} DR2. In Section \ref{results} we present some statistical and astrophysical properties of our set of candidates with short timescale variability, together with a few specific interesting examples. Finally, Section \ref{conclu} sums up the \textit{Gaia} DR2 short-timescale variability processing and results.

%Our \textbf{fast variability} analysis is mostly based on the \textit{Gaia} per-CCD and per-Field-of-View \textbf{(FoV)} photometry in $G$ band, but also makes use of spectrophotometry in $G_{\mathrm{BP}}$ and $G_{\mathrm{RP}}$ bands, as well as \textbf{astrometric parameters}. \textbf{It is important to mention} that the use of $G$ per-CCD photometry is meant as a preliminary exercise, \textbf{and that the analysed $G$ per-CCD light-curves are not made available to the scientific community as part of \textit{Gaia} DR2, since} the publication of this data is planned for the final release of the nominal mission\footnote{For more information on the  Data Release Scenario, please see \url{https://www.cosmos.esa.int/web/Gaia/release}}.

%We emphasise on the fact that, even though extensive work has been done for calibrating the $G$ band photometry, it is not yet finalised. Hence, per-CCD $G$ photometry is not part of the \textit{Gaia} DR2, and will be published only for the end of nominal mission release. All light-curves and results derived from preliminary internal per-CCD photometry have been approved for publication by DPAC \textcolor{blue}{(maybe more official phrasing needed?)}, and the work described here is strictly limited to the CU7 processing activities.

\section{Detection and characterisation of candidates with short-timescale variability}
\label{stsDetectionCharac}

In this section, we present the \textit{Gaia} DR2 short-timescale variability processing and analysis as part of the \textit{Gaia} Data Processing and Analysis Consortium (DPAC) activities, which resulted in the identification of 3,018 candidates with short-timescale variability. For this first publication of \textit{Gaia} short-timescale variables, we limited ourselves to a selection of sources with suspected periodic variability as a test case for probing the efficiency of the foreseen \textit{Gaia} short-timescale variability processing \citep{Eyer2017}. We restricted the study to variability phenomena that can be confirmed with a sufficient level of confidence. An analysis of transient variability will be included in the \textit{Gaia} short-timescale variability treatment for future data releases. Nevertheless, parallel to this work, \cite{Wevers2018} presented specific methods and preliminary results for an investigation of transients with \textit{Gaia}.

\subsection{Input data}
\label{input}

The \textit{Gaia} short-timescale variability analysis is mostly based on \textit{Gaia} per-CCD photometric time series in $G$ band after cleaning spurious values and outlier points \citep{Eyer2017, GDR2CU7Summary2018}. We recall that on-board $G$ measurements occur in groups of (most often) nine CCD observations (one every $4.85\,$s), one such group being referred to as a field-of-view (FoV) transit.
Often, two or more FoV transits are repeated, with time intervals of $1\mathrm{h}46\mathrm{min}$ or $4\mathrm{h}14\mathrm{min}$ between successive FoVs, following the \textit{Gaia} scanning law \citep{deBruijne2012}.

It is important to mention that the use of $G$ per-CCD photometry is meant as a preliminary exercise, and that the analysed $G$ per-CCD light curves are not made available to the scientific community as part of \textit{Gaia} DR2 because the publication of these data is planned for the final release of the nominal mission\footnote{For more information on the  data release scenario, see \url{https://www.cosmos.esa.int/web/Gaia/release}}.

At this stage of the \textit{Gaia} processing, only a set of selected $G$ per-CCD light curves is investigated for short-timescale variability. For each source observed by \textit{Gaia}, a statistical test is performed to determine whether the scatter in the $G$ per-CCD points of each FoV transit is greater than a specific significance level. If more than half of the FoV transits for that source show `variability' according to this criterion, it is further analysed.
%\textbf{At this stage of the \textit{Gaia} processing, only a set of selected $G$ per-CCD time series is transmitted to the \textit{Gaia} Geneva variability team for short timescale variability analysis. Their selection relies on a p-value based criterion, indicating that the considered sources are likely to exhibit short timescale variability in their light-curve, and is done as follows}. For each single FoV transit \textbf{within the $G$ per-CCD light-curve of the considered source}, a p-value \textbf{from $\chi^{2}$ statistical test is calculated using the (at most)} nine corresponding CCD measurements. If this p-value is smaller than $0.01$, then the FoV transit is considered as `not constant' (as the low p-value \textbf{rejects the null-hypothesis, i.e. that photometric fluctuations are only resulting from random photometric noise}). If \textbf{the considered source has more than half of its FoV transits identified as  `non-constant' from this criterion, then it is selected and transmitted for short timescale variability investigation}.

Finally, in addition to these $G$ per-CCD time series, the \textit{Gaia} short-timescale analysis also made use of the filtered magnitude time series in $G$ FoV (i.e. averaging the $G$ CCD measurements within one FoV transit), $G_{\mathrm{BP}}$ , and $G_{\mathrm{RP}}$ as described in \cite{Eyer2017} and in \cite{GDR2CU7Summary2018}, benefiting from all the time-series cleaning operators developed for the global variability processing.

%In the following subsections, we describe the methods and criteria applied to detect, characterise and select bona fide, suspected periodic short timescale candidates.

We emphasise that for \textit{Gaia} DR2, we did not aim for completeness with the published sample of sources with short timescale variability, first because of the selection of analysed per-CCD time series, but also because of the intermediate status of this photometry both in terms of time span and calibration. Our initial goal was rather to provide a significant sample of at least a few thousand bona fide candidates. We are aware that we probably missed a significant fraction of the true short-timescale variables observed by \textit{Gaia}, and we expect to reach completeness at the end of nominal mission with the whole five-year time span per-CCD photometry for all the scanned sources.

\subsection{Variogram analysis}
\label{variogram}

\subsubsection{ Principle}
\label{variogramPrinciple}

\cite{Roelens2017} predicted the potential of the variogram approach in the \textit{Gaia} context for short-timescale variability analysis by means of end-of-mission light-curve simulation, and by adopting a specifically tailored variogram formulation and detection threshold definition.
This preliminary work was the cornerstone of the \textit{Gaia} DR2 search for short-timescale candidates, where the variogram method applied to \textit{Gaia} $G$ per-CCD time series was used to preselect candidates with fast variability prior to further characterisation.

\begin{figure}
\centering
\includegraphics[width=0.9\linewidth]{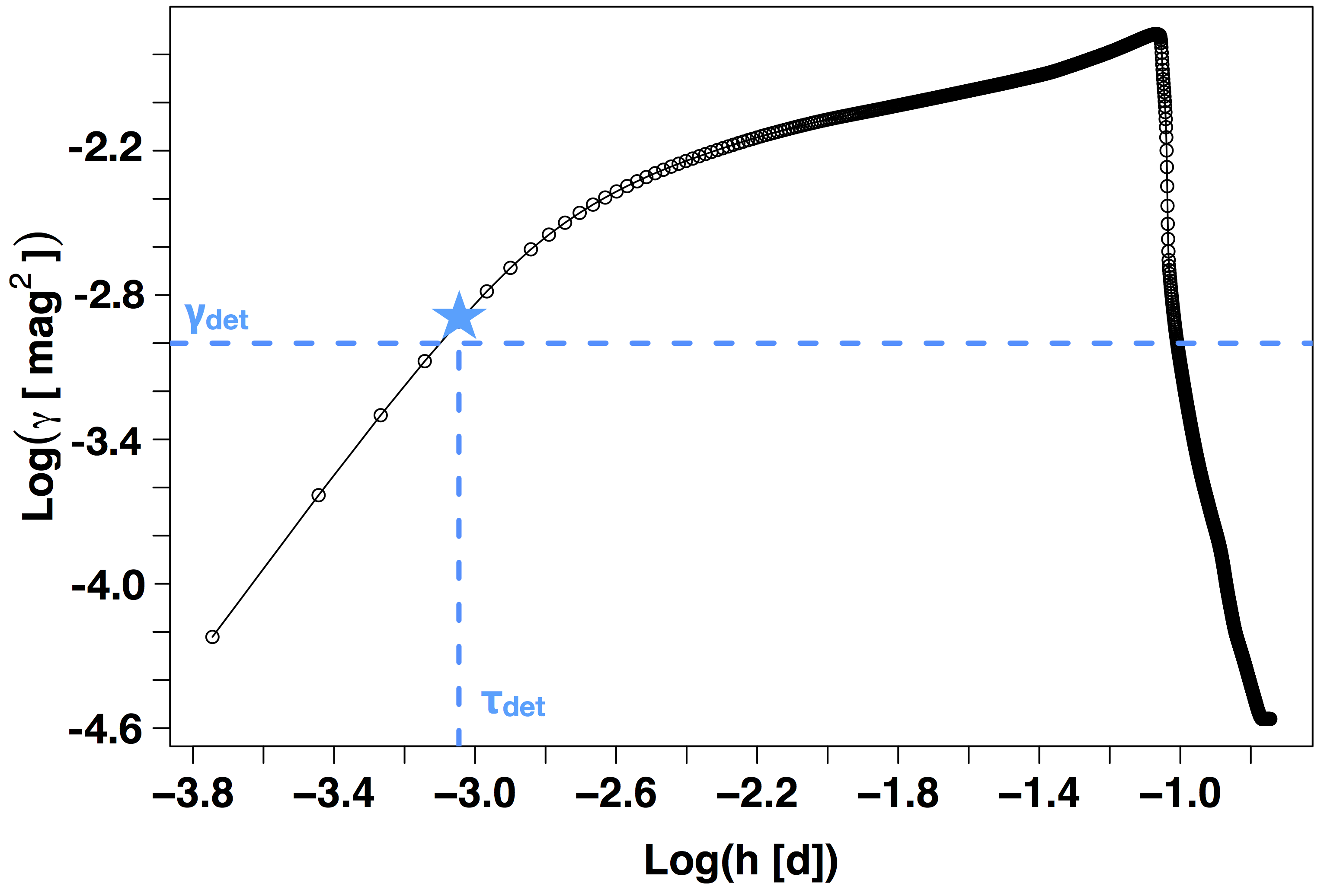}
\caption{Principle of identifying candidates with short-timescale variability  from a variogram analysis, illustrated with the variogram of a simulated transient event with an amplitude of $0.9\,$mag and a duration of $2\,$h \citep[derived from][]{Roelens2017}.}
\label{fig:sts_variogram_detection_principle}
\end{figure}

We recall that the idea of the variogram method is to quantify the magnitude variations between photometric measurements as a function of the time lag between them. The variogram value for a time lag $h$ is noted $\gamma(h)$. For a light curve with magnitudes $(m_{i})$ observed at times $(t_{i})$ for $i=1...n$, $\gamma(h)$ is calculated as a function of the magnitude differences $m_{j} - m_{i}$ on all the pairs $(i, j)$ such that $\mid t_{j} - t_{i} \mid = h~\pm~\epsilon_{h}$, with $\epsilon_{h}$ the tolerance accepted for grouping the pairs by time lag.
By exploring different time-lag values, defined by the time sampling, it is then possible to build a variogram plot (hereafter referred to as a variogram, see e.g. Figure \ref{fig:sts_variogram_detection_principle} ), which provides information on how variable the considered source is, and it informs on the variability characteristics if appropriate.

After variograms have been obtained for all the investigated sources, candidates with short-timescale variability are identified as those with a variability that is significant enough (i.e. whose variogram values are high enough), at time lags shorter than $12\,$h, in other words, verifying
\begin{ceqn}
\begin{align}
\mathrm{max}(\gamma) \geq \gamma_{det}~\&~\tau_{det} \leq 0.5\,\mathrm{d,}
\label{eqVariogDetCrit}
\end{align}
\end{ceqn}
where the detection threshold $\gamma_{det}$ defines the lowest variability level required to consider that variations are not only due to noise. $\tau_{det}$ is the shortest lag explored for which $\gamma \geq \gamma_{det}$ (if any), and it helps focusing on identified events with the fastest variability. This principle of the variogram detection is illustrated in Figure \ref{fig:sts_variogram_detection_principle}.

In the \textit{Gaia} context, we explored time lags involved in $G$ per-CCD time series, that is, the time intervals between the different CCDs within a single FoV ($4.85\,$s, $9.7\,$s, etc., up to $38.8\,$s), and the time intervals between the different FoVs ($1\,$h$\,46\,$min, $4\,$h$\,14\,$min, $6\,$h, $7\,$h$\,46\,$min, etc.), up to $h \approx 1.5\,$d.

Various variogram definitions can be found in the literature, the one adopted in \cite{Roelens2017} being an empirical weighted formulation using uncertainties on $G$ per-CCD measurements. However, for the \textit{Gaia} DR2 exercise, the estimation of the uncertainty for per-CCD photometry was not accurate enough to use this variogram definition. Consequently, we implemented an alternative formulation based on the inter-quartile range (IQR) estimation \citep[see e.g.][]{Macleod2012},
\begin{ceqn}
\begin{align}
\label{eqVariogDef}
\gamma(h) = [0.74\,\mathrm{IQR}(m_{j} - m_{i})]^{2,}
\end{align}
\end{ceqn}
 which is more robust than the classical unweighted variogram \citep[defined e.g. in][]{Hughes1992} to possibly remaining outliers, but is less time-consuming than the robust variogram described in \cite{Eyer1999}.
  
Finally, because the data we analysed to produce the \textit{Gaia} DR2 sample of candidates with short-timescale variability covered a time span of only 22 months instead of five years, we limited our analysis to sources with more than 20 available $G$ FoV transits, which ensured that we had enough data points for the variogram analysis to be reliable.

\subsubsection{Defining the detection threshold}
\label{variogramGammaDet}

As in \cite{Roelens2017}, we used a magnitude-dependent detection threshold to take into account the magnitude dependency of photometric errors in \textit{Gaia} $G$ band \citep{Evans2017}: $\gamma_{det} = \gamma_{det}(\bar{G}_{CCD}),$ where $\bar{G}_{CCD}$ is the mean of $G$ per-CCD magnitudes of the considered source.

To adopt the appropriate $\gamma_{det}$ definition, we performed the IQR-based variogram analysis on a subsample of known OGLE sources, crossmatched with objects observed by \textit{Gaia}, for which per-CCD photometry involving more than 20 FoV transits was available. This working sample contains 7419 and 380 periodic variables from the OGLE III \citep{OGLEIII2008} and OGLE IV \citep{OGLEIV2015} catalogues, respectively, as well as 459 `constant' stars from OGLE IV, that is to say, sources with the smallest variations in both $V$ and $I$ bands \citep{Eyer2017}. The OGLE periods ($P_{\rm OGLE}$) of the OGLE\ periodic variables in this data set rank from $28\,$min to $10,000\,$d.

\begin{figure*}
\centering
\subfloat[ ]{\includegraphics[width=0.49\linewidth]{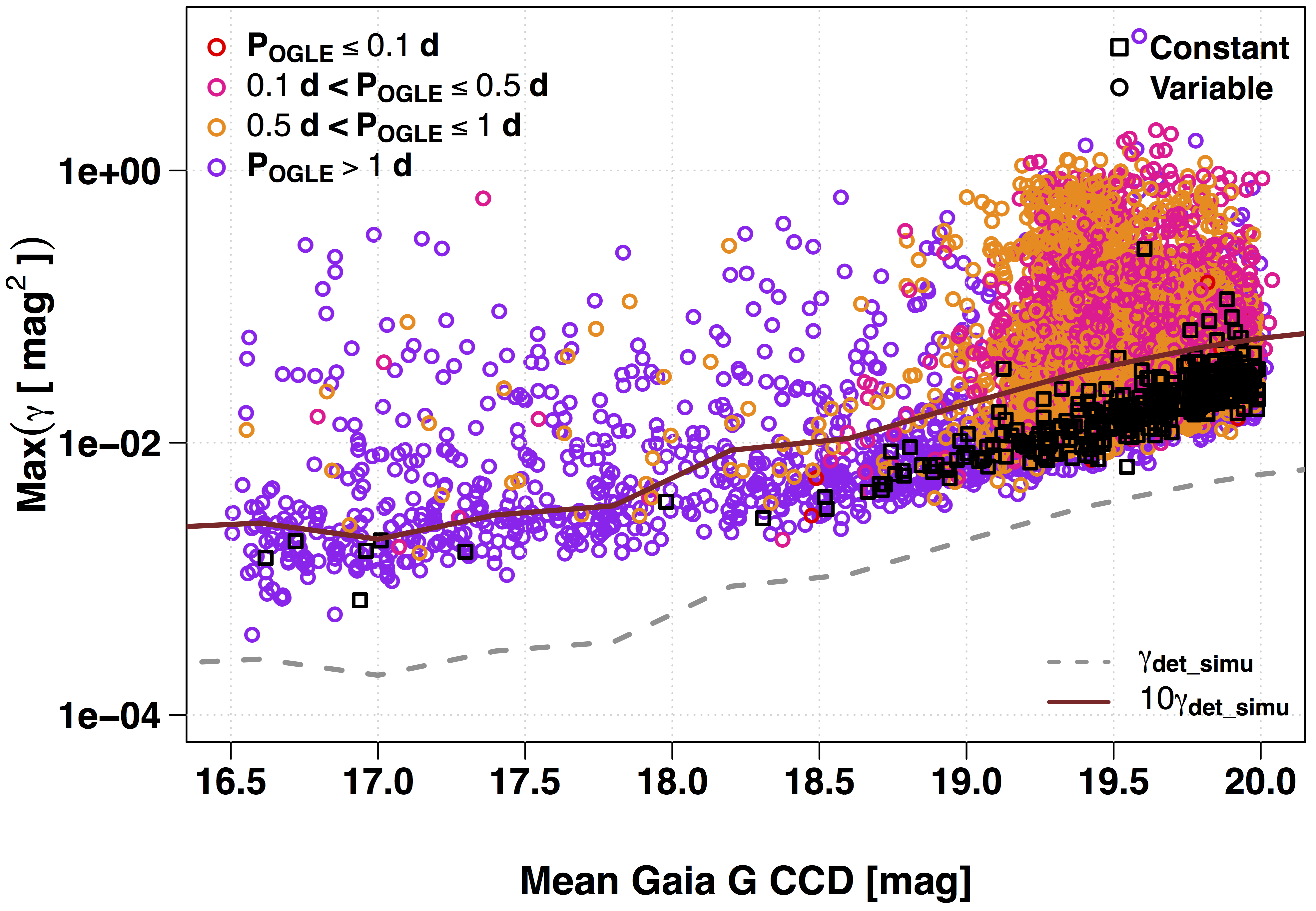}}
\subfloat[ ]{\includegraphics[width=0.49\linewidth]{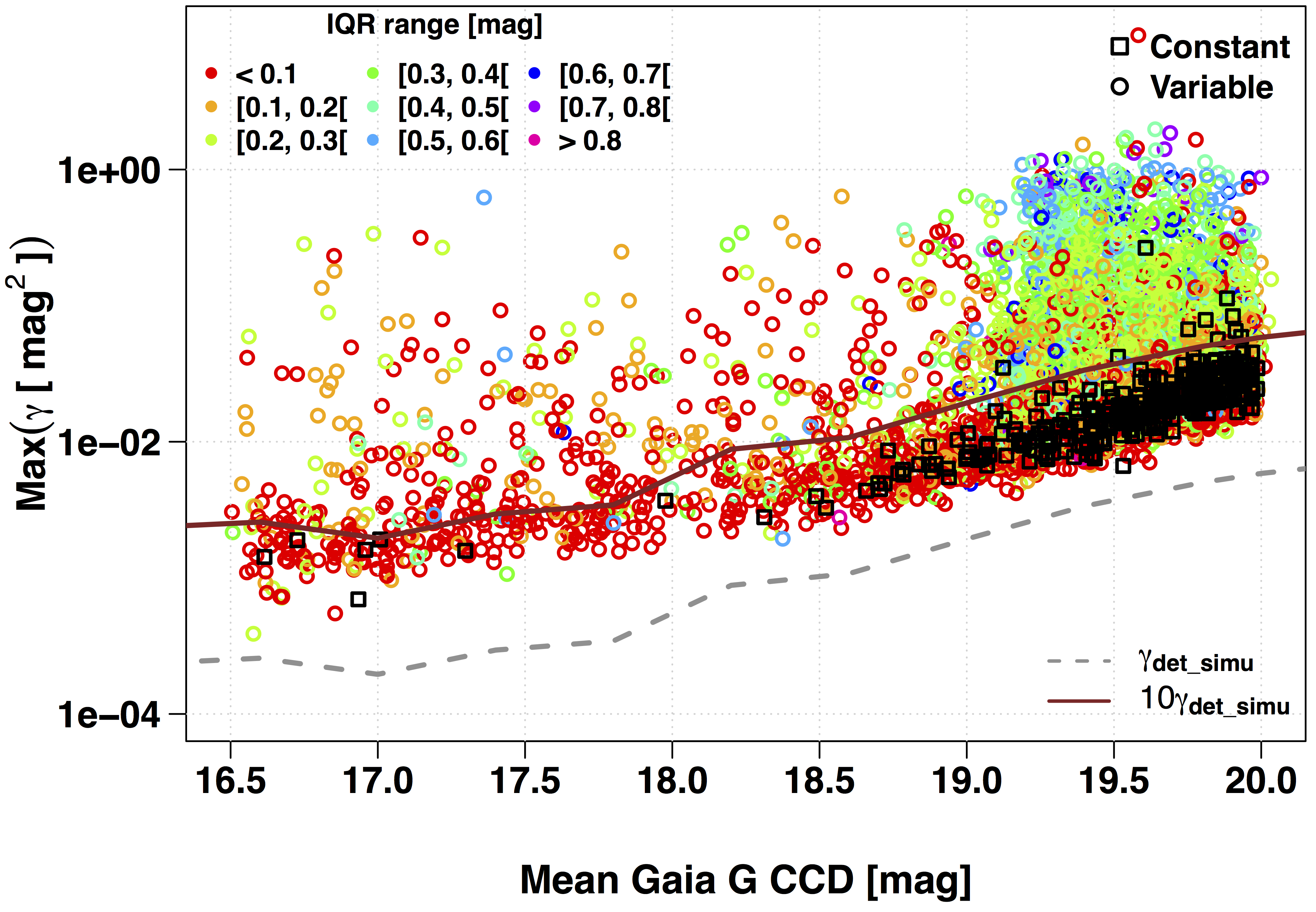}}
\caption{Maximum variogram value as a function of the mean G CCD magnitude for known constant and variable sources from the OGLE survey as observed by \textit{Gaia}. The grey line shows the detection threshold derived from simulated \textit{Gaia}-like light curves \citep[see][]{Roelens2017}. The brown line corresponds to the same threshold definition scaled by a factor of 10. (a) Colour-coded by period $P_{\rm OGLE}$ from the OGLE catalogue, and (b) colour-coded by IQR from \textit{Gaia} per-CCD $G$ photometry. It appears that with this detection threshold $\gamma_{det} = 10\gamma_{det,simu}$, (a) majority of constant stars and a significant fraction of longer-period variables are not detected as candidates with short-timescale variability when the variogram is used, and (b) the variable sources that are not detected with the variogram approach are low-amplitude sources.}
\label{fig:sts_detection_threshold}
\end{figure*}

Figure \ref{fig:sts_detection_threshold} shows $\mathrm{max}(\gamma)$ as a function of $\bar{G}_{CCD}$ for each of these crossmatched OGLE sources. The grey line in Figure \ref{fig:sts_detection_threshold} corresponds to $\gamma_{det,simu}$ , the magnitude-dependent detection threshold definition for the IQR-based variogram formulation, derived from the simulated \textit{Gaia}-like five-year time-span light curves of \cite{Roelens2017}. Clearly, $\gamma_{det,simu}$ is not adapted to real \textit{Gaia} data, first because the time span is 22 months instead of the five years in the simulations (i.e. it has fewer data points per source), and then because of the intermediate photometric calibration. However, by simply scaling $\gamma_{det,simu}$ by a factor of 10 (the brown line in Figure \ref{fig:sts_detection_threshold}), it is possible to efficiently separate constant OGLE sources from sources that are periodically variable. Additionally, with $\gamma_{det} = 10\gamma_{det,simu}$, the periodic variables that are not detected are either long-period variables (Figure \ref{fig:sts_detection_threshold}, panel a) or low-amplitude sources (Figure \ref{fig:sts_detection_threshold}, panel b).

When we applied the short-timescale detection criterion of Equation \ref{eqVariogDetCrit} with $\gamma_{det} = 10\gamma_{det,simu}$  to our OGLE working sample, we recovered about 48\% of the short-period sources (i.e. with $P_{\rm OGLE} \leq 0.5\,$d). The contamination level from false positives (namely constant sources flagged as short-timescale variables) was 2\%,  and contamination was at about 20\% from variable sources with periods longer than $1\,$d. Sources with periods between $0.5$ and $1\,$d are at the limit of our definition of short-timescale variability, but are still of interest and can be accepted as `extended' short-timescale variables.

The relatively low recovery rate of short-period variables results from the fact that most of such variables in the OGLE test sample have amplitudes below $0.1 - 0.2\,$mag, which means that they are at the limit of what can be detected according to Figure \ref{fig:sts_detection_threshold} (panel b). Moreover, because fewer data points are available in each per-CCD light curve than what is expected at the end-of-nominal mission, it is not always possible to form pairs of measurements and calculate variogram values for all the lags that correspond to the inter-FoV time intervals, where detection should be triggered for the short periods involved in our test sample (typically $P_{\rm OGLE}$ between half an hour and $1\,$d).
Although contamination from longer-period sources is quite high, it is not a great problem. The vast majority of longer-period variables that are flagged as short-timescale candidates with our variogram criterion have periods no longer than a few days, amplitudes greater than a few tenths of magnitudes, and can exhibit steep variations. Consequently, their global variation rate is sufficient to justify their detection at short timescales, even though they are not short-period variables per se.

Since the OGLE crossmatched sample we used to build the relevant variogram detection criterion contains only objects with a $G$ magnitude between about $16.5$ and $20\,$mag, we limited our analysis to this magnitude range, where the $\gamma_{det}$ definition has been validated and tested.

\subsubsection{Application to real \textit{Gaia} data}
\label{variogramApplication}

For our global search for short-timescale variability, we applied the short-timescale detection criterion detailed above to all the \textit{Gaia} sources for which per-CCD photometry was available, with more than 20 FoV transits, and with a mean $G$ magnitude of between $16.5$ and $20\,$mag. This represents a working sample of about 5.6 million sources that are to be investigated for short-timescale variability.
%Figure \ref{fig:mag_histo_with_vs_without_ccd} shows the mean $G$ magnitude distribution for all the sources \textbf{scanned by \textit{Gaia} more than 20 times over the first 22 months of science operations}, within the magnitude range $16.5 - 20\,$mag \textbf{(about 532 million sources)}, \textbf{together with the same distribution} for the sources in this sample with per-CCD data satisfying the p-value criterion described earlier in Section \ref{stsDetectionCharac} \textbf{(about 5.6 million sources)}. As can be seen, the distribution in magnitude for the sample with per-CCD data is relatively flat, whereas the total magnitude distribution is clearly skewed towards fainter sources. Consequently, the fraction of possible short timescale variables according to the p-value criterion compared to the total number of sources at the considered magnitude is higher at the bright end of our sample than at the faint end.\\
From these 5.6 million sources, a more complete crossmatch with various variable star catalogues from the literature, not only OGLE, but also Catalina \citep{Drake2014Catalina, Drake2014CRTS}, LINEAR \citep{Palaversa2013}, Kepler \citep{Debosscher2011}, EROS2 \citep{Kim2014}, or the Half-Million quasar catalogue \citep[HMQ, ][]{Flesch2015}, resulted in a set of 4747 known variables, including 439 with $P_{\rm Lit} \leq 0.5\,$d (with $P_{\rm Lit}$ the literature period), 382 with $0.5 < P_{\rm Lit} \leq 1\,$d, 1574 with $P_{\rm Lit} > 1\,$d, and 1658 non-periodic variable sources. The remaining sources are variables of periodic type, but no information on their period is available from the considered catalogue.

Of the 5.6 million sources processed with the variogram analysis, 3.9 million sources were flagged as short-timescale candidates, which is a huge fraction of the analysed sample and may question the reliability of our variogram approach in this context. However, this unexpectedly high fraction of candidates with short-timescale variability is a direct consequence of the pre-selection of the objects for which per-CCD data have been analysed. Since these sources are considered as likely to show fast variability according to the criterion described in Section \ref{input}, having about 70\% of the investigated sources flagged as candidates with short-timescale variability simply means that the variogram detection criterion is coherent with that selection.
From this list of 3.9 million candidates,  we recovered 2892 of the crossmatched sources from the catalogues mentioned above, including 356 with $P_{\rm Lit} \leq 0.5\,$d, 280 with $0.5 < P_{\rm Lit} \leq 1\,$d, 738 with $P_{\rm Lit} > 1\,$d, and 1051 non-periodic variable sources. We used these 2892 sources as a reference set to assess and improve the efficiency of our approach for finding short-timescale variables with suspected periods (see Sect. \ref{highFreqSearch} and \ref{stsSelection}). From now on, this is referred to as the reference crossmatched sample.

\subsection{High-frequency search}
\label{highFreqSearch}

For each candidate whose short-timescale variability was identified by the variogram analysis (Sect. \ref{variogram}), we additionally performed a high-frequency search to further characterise the candidates with suspected variability and to help distinguishing periodic variations from transient events. We decided to explore frequencies between $1$ and $144\,\mathrm{d}^{-1}$ (i.e. periods between $10\,$min and $1\,$d), which roughly corresponds to the period range represented in our reference crossmatched sample. It thus enables us to assess the quality of the retrieved period search results from \textit{Gaia}. Several periodogram-based methods have been considered for this purpose: the Deeming periodogram \citep{Deeming1975}, the Lomb-Scargle periodogram \citep{Scargle1982}, the least-squares periodogram described by \cite{Zechmeister2009}, the string length method \citep{Lafler1965}, and the phase-dispersion minimization (PDM) approach \citep{Czerny1989}. They were all tested on a set of 112 well-characterised short-period sources (i.e. with $P_{\rm Lit} \leq 1\,$d) from the reference crossmatched sample. We applied the considered period search algorithms on both $G$ per-CCD and $G$ FoV light curves to determine which gave the best period-recovery rate. The \textit{Gaia} period from the short-timescale analysis, hereafter $P_{\rm Gaia}$, for a given source, is defined as the inverse of the frequency $f_{\rm Gaia}$ of the highest peak in the corresponding frequencygram. For each method, we checked the fraction of sources for which we had $P_{\rm Gaia} \approx P_{\rm Lit} \pm$ 10\% or $P_{\rm Gaia} \approx P_{\rm Lit}/2 \pm$ 10\% (as it is often the case for eclipsing binary systems). The resulting numbers are summarised in Table \ref{tab:stsHighFreqSearchTest}. The least-squares periodogram appeared to be the best alternative, whether applied to $G$ per-CCD or to $G$ FoV light curves. Since the final aim of the \textit{Gaia} short-timescale variability module is to search for variability at timescales down to a few tens of seconds, which can be probed only using per-CCD photometry, and even if such very short periods were not really explored in that exercise, we decided to use $G$ per-CCD time series.

%\citep{Zechmeister2009}, similarly to what is done by the \textit{Gaia} variability characterisation module \citep{Eyer2017}, but this time applied to per-CCD $G$ time series. \textbf{We explore} frequencies between $1$ and $144\,\mathrm{d}^{-1}$ (i.e. periods between $10\,$min and $1\,$d) with a frequency step of $10^{-4}\,\mathrm{d}^{-1}$. The \textit{Gaia} period from short timescale analysis, hereafter $P_{\rm Gaia}$, for a given source, is defined as the inverse of the frequency $f_{Gaia}$ of the highest peak in the corresponding Least-Square frequencygram.

\begin{figure}
\centering
\includegraphics[width=0.9\linewidth, trim = {0 0 0 1.5cm}, clip=true]{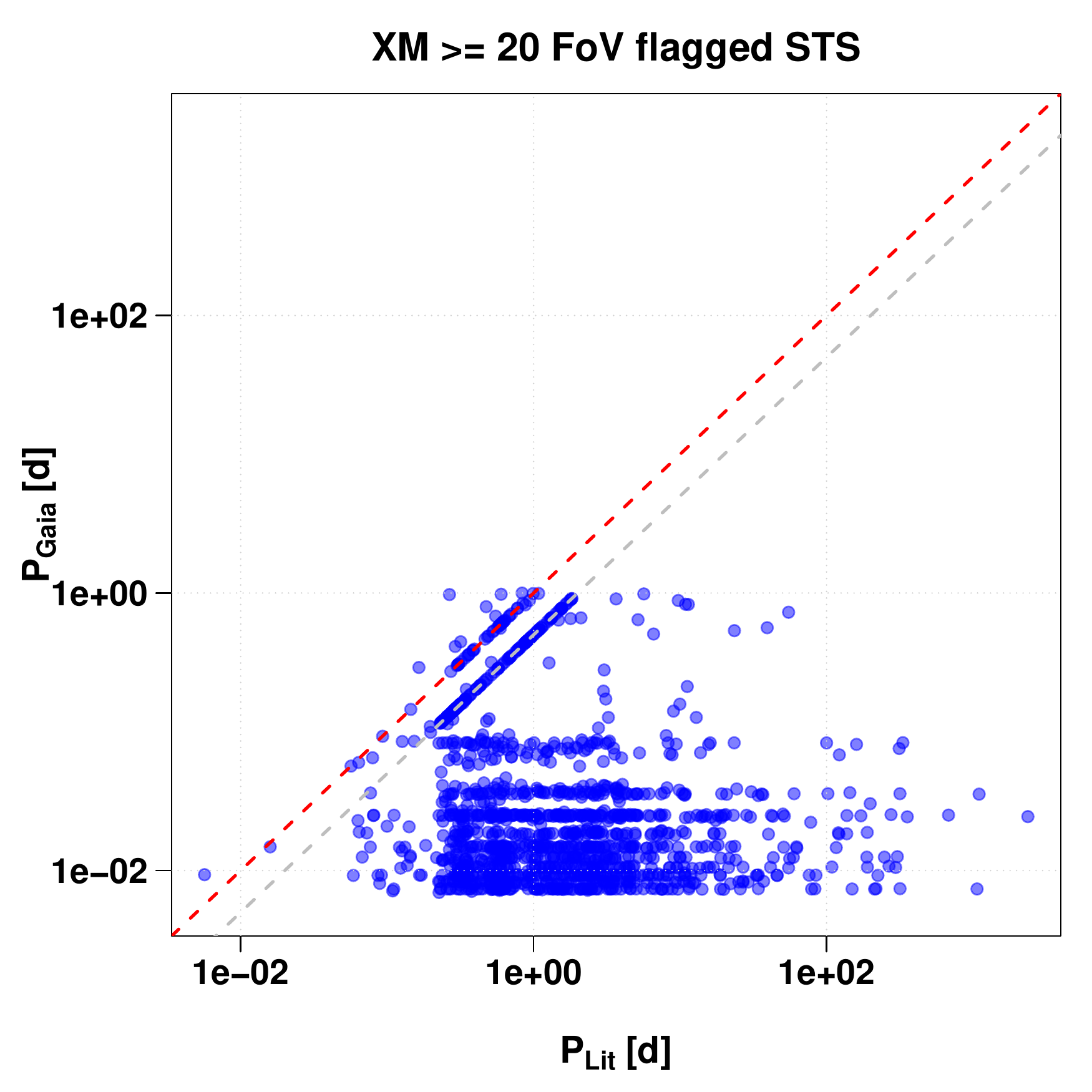}
\caption{\textit{Gaia} short-timescale period as a function of the literature period for the crossmatched periodic sources flagged as short-timescale candidates from the variogram analysis. The red and grey dashed lines correspond to $P_{\rm Gaia} = P_{\rm Lit}$ and $P_{\rm Gaia} = P_{\rm Lit}/2,$ respectively. In total, the literature period or half of the literature period is recovered only for 23\% of the crossmatched sources with $P_{\rm Lit} \leq 1\,$d.}
\label{fig:sts_period_recovery}
\end{figure}

Figure \ref{fig:sts_period_recovery} shows $P_{\rm Gaia}$ for all the periodic sources in the reference crossmatch sample described at the end of Sect. \ref{variogram} as function of their literature period $P_{\rm Lit}$. The period recovery with the least-squares method applied to \textit{Gaia} $G$ CCD photometry is not exceptional for this set of objects: of 1374 crossmatched variables that are flagged as short-timescale variables with external period information, 636 have $P_{\rm Lit} \leq 1\,$d, and for only 45 of them does the \textit{Gaia} short-timescale period recover $P_{\rm Lit}$ by 10\%. For 104 of them, mostly eclipsing binaries, the \textit{Gaia} short-timescale period recovers $P_{\rm Lit}/2$ by 10\%. 

The horizontal trends in $P_{\rm Gaia}$, visible at the shortest periods retrieved from \textit{Gaia} data, correspond to aliases due to the $6\,$h rotation period of the \textit{Gaia} satellite, that is, to frequencies of $4\,\mathrm{d}^{-1}$ and its multiples, a phenomenon that was expected, as described in \cite{Eyer2017}.

\begin{figure}
\centering
\includegraphics[width=0.95\linewidth, trim = {0 0 0 1.5cm}, clip=true, page=5]{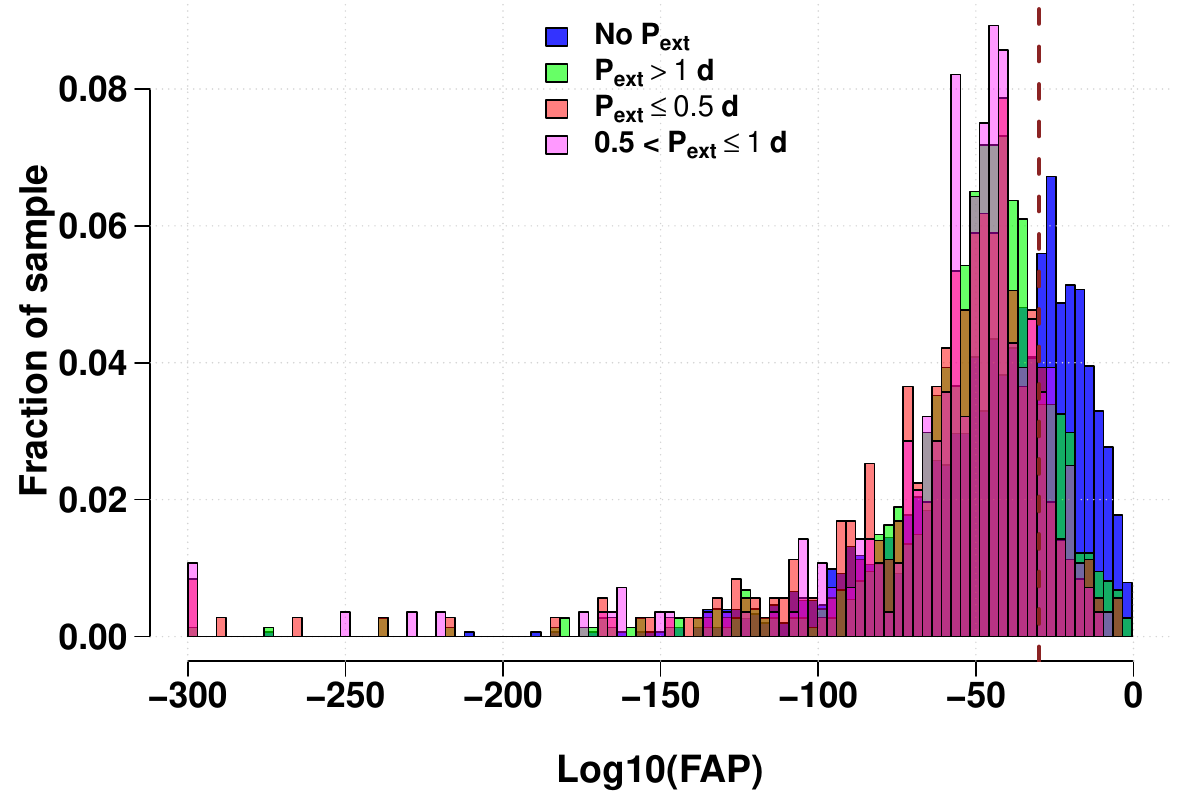}
\caption{False-alarm probability distribution from the least-squares period search for the 2892 crossmatched known variable sources flagged as short-timescale candidates from the variogram analysis.}
\label{fig:sts_fap_distrib}
\end{figure}

For each source flagged as a candidate for short-timescale variability and analysed for high-frequency search, we retrieved the false-alarm probability (FAP) of the most pro-eminent peak in the least-squares periodogram \citep{Zechmeister2009}, that is, the FAP associated to $f_{\rm Gaia}$. Figure \ref{fig:sts_fap_distrib} shows the distribution of this FAP for the reference crossmatched sample, distinguishing non-periodic variables, short-period variables, and longer-period variables. Although sources in the first category cannot be strictly separated from the others on the basis of the FAP alone, non-periodic variables tend to have higher FAP values, by definition of the FAP. Hence, rejecting candidates with FAP values greater than $10^{-30}$ , for instance, should help eliminating a significant fraction of the known non-periodic variables of the sample without loosing too many short-period sources. The overall approach adopted to focus on the suspected periodic variability is described in more detail in Sect. \ref{stsPeriodicSelection}.

\begin{table}
\centering
\caption{Summary of the period recovery results for various period-search methods, tested on a set of selected well-characterised short-period variables from the literature.}
\label{tab:stsHighFreqSearchTest}
\begin{tabular}{ccc}
\hline
\bf{Period-search} & \bf{Time-series} & \bf{Period/half-period}\\
 \bf{method} & \bf{type used} & \bf{recovery rate}\\
\hline
Deeming & $G$ CCD & 52.7\%\\
Lomb-Scargle & $G$ CCD & 75\%\\
Least squares & $G$ CCD & 78.6\%\\
String length & $G$ CCD & 70\%\\
PDM & $G$ CCD & 13.4\%\\
Deeming & $G$ FoV & 55.4\%\\
Lomb-Scargle & $G$ FoV & 77.7\%\\
Least squares & $G$ FoV & 78.6\%\\
String length & $G$ FoV & 76.8\%\\
PDM & $G$ FoV & 56.3\%\\
\hline
\end{tabular}
\end{table}

From this analysis, it was clear that the published $P_{\rm Gaia}$ values should be taken with caution, and that they are more indicative than really accurate. As described previously, at this point, period-search results are strongly affected by aliasing issues, which represents a major axis of improvement for the future \textit{Gaia} data releases. Moreover, we targeted many different variable types (e.g. ZZ~~Ceti stars, AM~~ CVn stars, $\delta$~~Scuti stars etc.) that showed various light-curve shapes, and we kept in mind the possibility of observing unknown short-timescale variable types yet to be discovered. In contrast to other variability processing modules, such as those dedicated to analyses of RR Lyrae and Cepheids \citep{GDR2Clementini2018}, we consequently did not perform any further light-curve modelling to improve the accuracy of the period we found, which partly explains the moderate short-period recovery rate.

\subsection{Other statistics and parameters}
\label{otherParams}

In addition to the variogram (Sect. \ref{variogram}) and frequency-search (Sect. \ref{highFreqSearch}) analysis, we calculated a series of statistics to characterise and identify short-timescale candidates with a suspected periodic variability (see Sect. \ref{stsSelection}).

For each short-timescale candidate, the amplitude estimate $A_{\rm G-CCD}$ was defined as the difference between the $95^{\mathrm{th}}$ and the $5^{\mathrm{th}}$ quantiles of $G$ per-CCD time series.

We also calculates the mean Abbe value per-transit, which is defined as follows:
\begin{center}
$\bar{\mathcal{A}}_{\rm per-transit} = \frac{\sum{\mathcal{A}_{\rm transit}}}{N_{\rm transit}}$ with $\mathcal{A}_{\rm transit} = \frac{\sum{(m_{i+1} - m_{i})^{2}}}{2(n-1)\sigma^{2}}$.
\end{center}
$(m_{i})_{i=1..n}$ are the per-CCD magnitudes for one transit, $n$ is the number of per-CCD measurements in the transit, $\sigma^{2}$ is the unbiased variance on the magnitudes of the transit, and $N_{\rm transits}$ is the number of transits for the considered source.
The idea of the mean Abbe value per transit is to spot sources in which several transits exhibit a smooth and significant variability at the level of one transit, that is, over a timescale of about $40\,$s. Such transits are expected to have lower $\mathcal{A}_{\rm transit}$ values than purely noisy transits, and thus have lower $\bar{\mathcal{A}}_{\rm per-transit}$. However, this statistics was not really exploited in \textit{Gaia} DR2.

For each analysed source, we defined the median variogram ratio as the median of its variogram values for lags shorter than $40\,$s divided by the median of its variogram values for lags longer than $40\,$s. This parameter quantifies how flat or how `step-shaped' the variogram is when its values at shorter lags are compared to values at longer lags. As detailed in \cite{Roelens2017}, the typical variogram plot for periodic sources is expected to first show a plateau at the shortest lags, then an increase, and oscillations for lags increasing towards and beyond the variation period. For periodic variables with periods between $10\,$min and $1\,$d, the variogram plot is therefore expected to be relatively flat for lags at CCD level (i.e. from $4.85\,$s to $40\,$s), and to show oscillations at higher values for lags at the FoV level, typically between $1\,\mathrm{h}\,46\,\mathrm{min}$ and $1.5\,$d, because depending on the variability period, the oscillations starts at different lags. For these sources, the median variogram ratio is therefore expeted to be relatively low. Conversely, variograms derived for fast transient events (e.g. flares) should be quite flat except a local increase resulting from the flare. Variograms for longer-period variables (e.g. periods longer than 5 or 10 days) are likely to be flat as well, and possibly show an increase that starts at the longest explored lags. For both transient and longer-period variables, the median variogram ratio is then expected to be higher than for short-period variables. This means that the median variogram ratio is a useful metric to distinguish short-period variability from other variability features (see Sect. \ref{stsPeriodicSelection}).
%Note that this statistics is computed as part of the short timescale processing, and used for selecting the published short timescale sample in DR2, but is not published as part of the \textit{Gaia} DR2 archive.

Finally, our short-timescale analysis process made use of some of the classical statistical parameters produced by the statistics module of \textit{Gaia} variability processing \citep{Eyer2017}, typically the IQR and the Abbe value $\mathcal{A}$ on the times series in the three photometric bands ($G$, $G_{\mathrm{BP}}$ , and $G_{\mathrm{RP}}$), as well as the Spearman correlation between the $G$ and $G_{\mathrm{BP}}$ bands and between the $G$ and $G_{\mathrm{RP}}$ bands,  noted $r_{\rm G,BP}$ and $r_{\rm G,RP}$ , respectively (Sect. \ref{stsPeriodicSelection} and Sect. \ref{otherSpurious}).

\section{Selection of bona fide candidates with short-timescale variability}
\label{stsSelection}

For this first release of short-timescale variables in \textit{Gaia}, we focused on candidates with short-timescale variability, whose variations are suspected to be periodic as a test case.
In the following sections, we describe the process for selecting the sources that had strong indications in favour of periodicity of the 3.9 million candidates flagged from the variogram analysis (see Sect. \ref{variogram}), and we also report on the validation of the criteria we used.
These two aspects involved investigating the photometry for known variable sources in the candidate sample (presented in Sect. \ref{variogram} and used in Sect. \ref{highFreqSearch}), as well as visual inspection of some candidates' light curves. In parallel, photometric monitoring of a few preliminary candidates crucially helped building the final criterion we used to select the published candidates of \textit{Gaia} DR2.
The whole selection process is summarised in Table \ref{tab:stsSelectSummary}.

\begin{table*}
\centering
\caption{Summary of the step-by-step selection criteria for candidates with a suspected periodic short-timescale variability.}
\label{tab:stsSelectSummary}
\begin{tabular}{lc}
\hline
\bf{Selection criterion} & \bf{Number of sources}\\
\hline
More than 20 transits, with per-CCD photometry, in the $G \sim 16.5 - 20\,$mag range & 5.6 million\\
\hspace{2mm} + Flagged as a candidate with a short-timescale variability from the variogram & 3.9 million\\
\hspace{4mm} + Preliminary selection of candidates with suspected fast periodic variability & 16,703\\
\hspace{6mm} + After environment filtering and after removing spurious variability and eclipsing binaries  & 3,018\\
\hline
\end{tabular}
\end{table*}

\subsection{Selection of candidates with suspected periods}
\label{stsPeriodicSelection}

The main idea for distinguishing suspected periodic variables of all the short-timescale candidates identified with the variogram was to define different cuts in the various statistics described in Sect. \ref{otherParams}, so as to exclude phenomena that we were not interested in for this first exercise. The choice of such validity intervals for short-period variability relied essentially on the analysis of the reference crossmatch sample of 2892 known variables, the idea being to limit contamination from non-periodic and longer-period variables as much as possible within the final list of candidates.
%We remind that the inclusion of some specific longer period variables, with periods of a few days and amplitudes of a few tenths of magnitudes, can be justified as periodic variables detected at short timescale even though they are not short period variables per se.
The definition of the cuts to adopt was based on simple histograms of the different statistical parameters available, similar to Figure \ref{fig:sts_fap_distrib}, comparing the distributions for non-periodic variables, short-period variables, and longer-period variables.
One very important point of our selection strategy is that we required the variability seen in $G$ band to be confirmed by consistent features in $G_{\mathrm{BP}}$ and/or $G_{\mathrm{RP}}$ bands for considering it as relevant. 
%Plots: successive histograms to illustrate the different steps and cuts. -> to do: compile them in one Keynote page, then printed as PDF for tex inclusion.

After several iterations and tests, we decided to keep in our list of candidates with suspected periodic and short-timescale variability only those sources that met the following statements:
\begin{itemize}
\item More than 18 observations in $G_{\mathrm{BP}}$ and $G_{\mathrm{RP}}$ . Even though we required  the $G$ time series to contain at least 20 FoV, the $G_{\mathrm{BP}}$ or $G_{\mathrm{RP}}$ time series can sometimes contain fewer points after cleaning. We therefore imposed a lower limit on the number of observations in these two bands to ensure that we had enough information to confirm the variability.
\item An Abbe value on the $G$ FoV time series $\mathcal{A}_{\rm G-FoV} \geq 0.7$. As explained in \cite{Mowlavi2014}, low Abbe values are expected to correspond to transient variability, not to periodic variations.
\item FAP $\leq 10^{-30}$. By definition, and as shown in Sect. \ref{highFreqSearch}, the FAP should be higher for non-periodic variables.
\item Median variogram ratio $\leq 0.26$. As detailed in Sect. \ref{otherParams}, this value is expected to be lower for short-period sources than for longer-period or transient variables.
\item $0.32 \leq \mathrm{IQR}_{\rm XP} / \mathrm{IQR}_{\rm G} \leq 3.2,$ where $XP$ stands for $BP$ or $RP$.
\item Spearman correlation coefficients $r_{\rm G,BP} \geq 0$ and $r_{\rm G,RP} \geq 0$. 
\end{itemize}

\begin{figure*}
\centering
\subfloat[ ]{\includegraphics[width=0.49\linewidth]{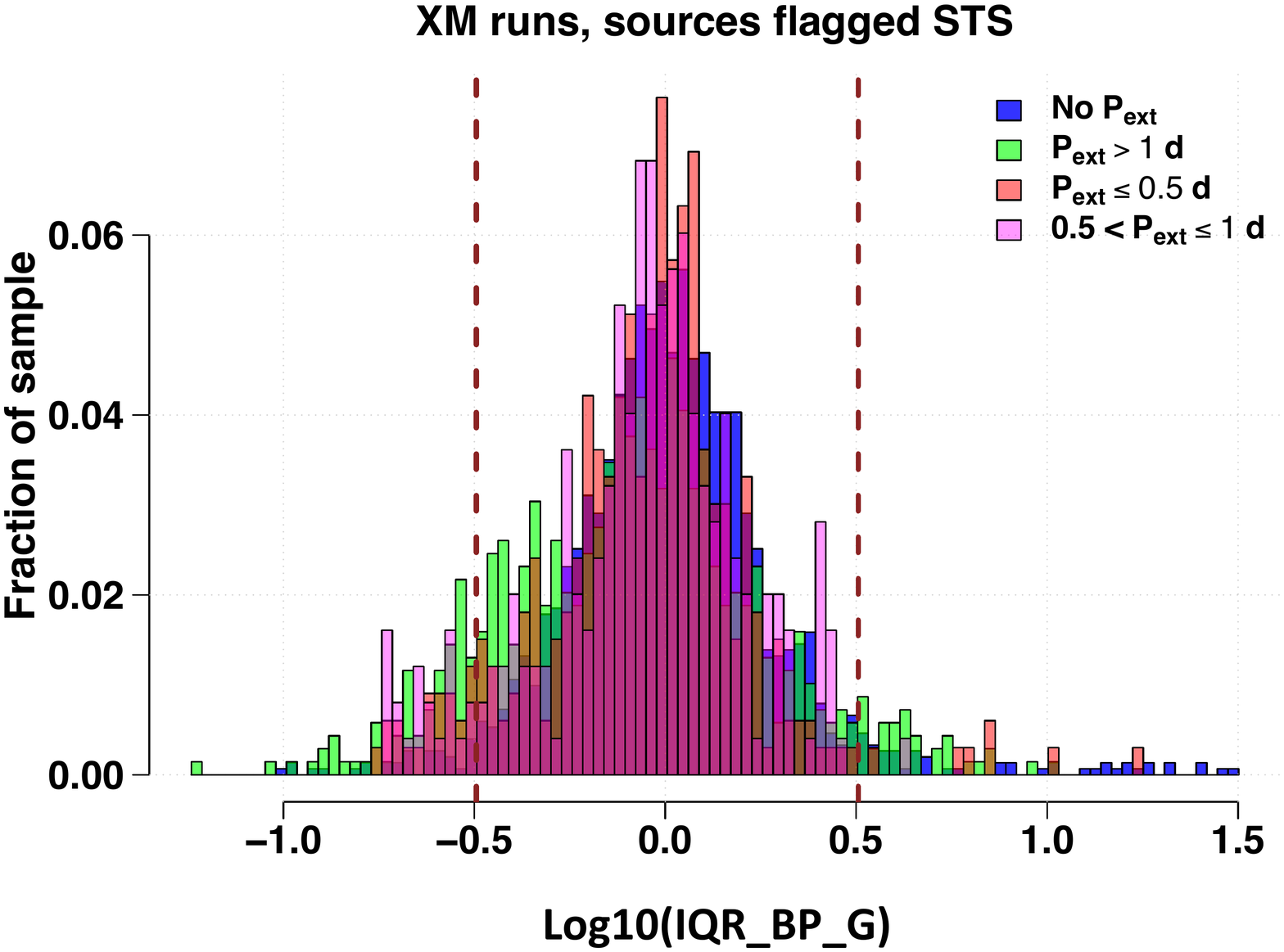}}
\subfloat[ ]{\includegraphics[width=0.49\linewidth]{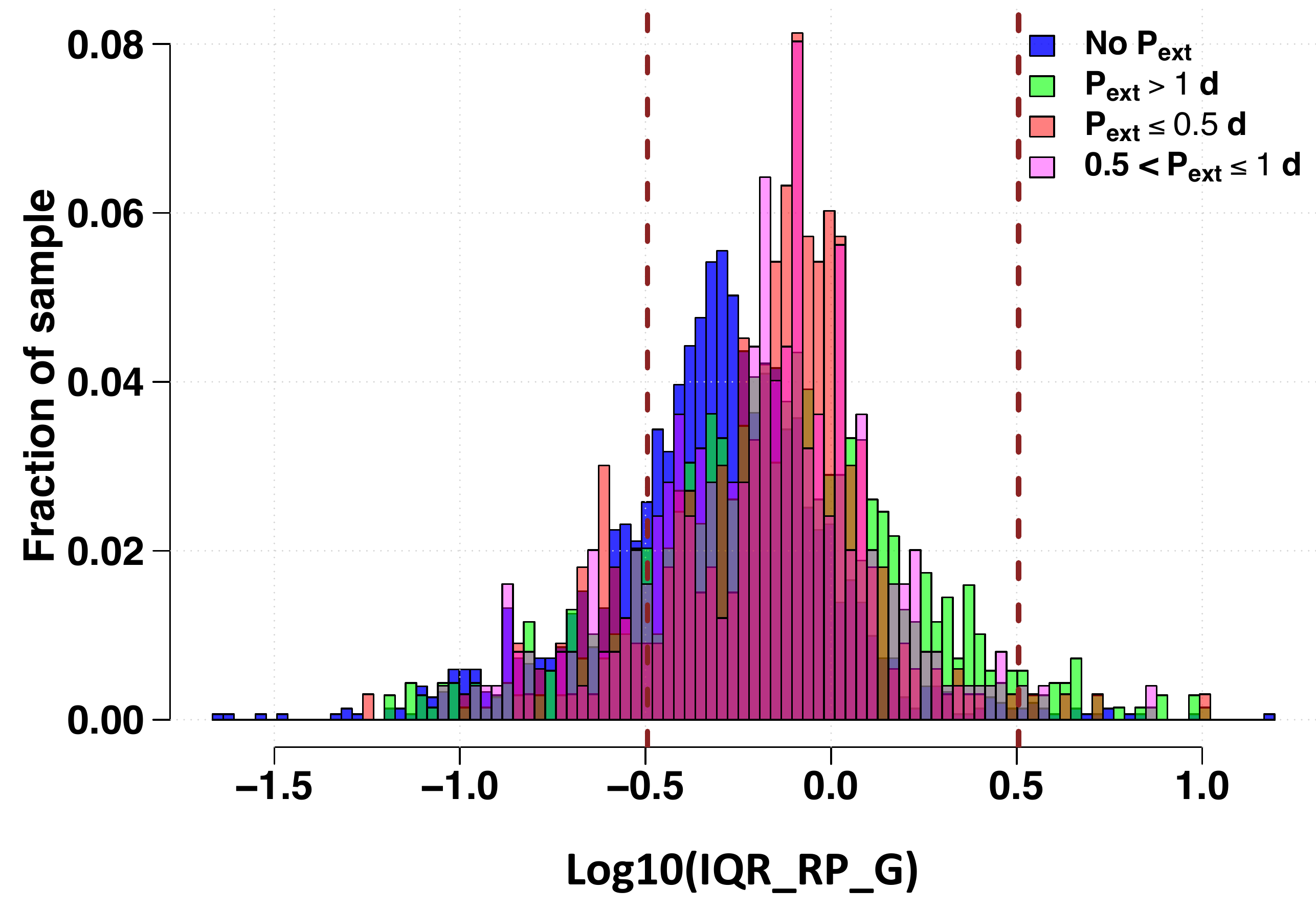}}
\caption{Distribution of the IQR ratios between $G$ and $G_{\mathrm{BP}}$ band (a) and between $G$ and $G_{\mathrm{RP}}$ bands (b) for the 2892 crossmatched known variable sources flagged as short-timescale candidates from the variogram analysis.}
\label{fig:sts_iqr_ratios_distrib}
\end{figure*}

The IQR ratio criterion is a consistency check, ensuring the coherence between behaviours in $G$, $G_{\mathrm{BP}}$ , and $G_{\mathrm{RP}}$, which was confirmed to be necessary in the study of the IQR ratio distributions for the reference crossmatch sample (Figure \ref{fig:sts_iqr_ratios_distrib}). Even though the investigation of known variables showed that the amplitudes in the three bands can be different, we chose to focus on the cases without a strong discrepancy between the $G$, $G_{\mathrm{BP}}$ , and $G_{\mathrm{RP}}$ IQRs.\\
It appeared that for this data set, the statistical behaviour of the $G_{\mathrm{BP}}$ and $G_{\mathrm{RP}}$ bands was different. However, we did not wish to bias our analysis towards a certain type of known variables (e.g. RR Lyrae stars or eclipsing binaries, which represent the majority of the sources in the crossmatch sample). In particular, we aimed to avoid missing any less well-known or unknown short-period variable types that could be observed by \textit{Gaia}. Hence, we kept a symmetric criterion for both $G_{\mathrm{BP}}$ and $G_{\mathrm{RP}}$ photometry.

Similarly, the last cut on the correlation values between the $G$ and $G_{\mathrm{BP}}$ / $G_{\mathrm{RP}}$ bands is a first (and not very strict) constraint to ensure that the variability phenomena observed in $G$ are compatible with the $G_{\mathrm{BP}}$ and / or $G_{\mathrm{RP}}$ time series. This criterion has been tightened afterwards (see Sect. \ref{otherSpurious}).

When this selection criterion was applied to the reference crossmatched sample, we kept 303 sources out of 2892 as suspected periodic short-timescale variables, including 127 with $P_{\rm Lit} \leq 0.5\,$d, 74 with $0.5 < P_{\rm Lit} \leq 1\,$d, 88 with $P_{\rm Lit} > 1\,$d, and 3 non-periodic sources. The remaining 11 objects are sources whose type is compatible with periodic variability, for which no period information is available, however.
For periodic variables with $P_{\rm Lit} \leq 1\,$d, the recovery rate was at about 32\% with respect to the known short-period variables that are flagged as short-timescale candidates from the variogram (36\% for $ P_{\rm Lit} \leq 0.5\,$d, 26\% for $0.5 < P_{\rm Lit} \leq 1\,$d), and at about 18\% when compared to the processed known short-period variables (19\% for $ P_{\rm Lit} \leq 0.5\,$d, 16\% for $0.5 < P_{\rm Lit} \leq 1\,$d). The contamination rate was about 1\% from non-periodic sources and 30\% from longer-period variables. However, the latter corresponds mostly to variables with periods between $1$ and $5\,$d and with amplitudes of a few tenths of magnitudes, whose detection at the short-timescale level is justified, as discussed previously.
Finally, when we applied our selection criteria for candidates with suspected periodic variability to the 3.9 million short-timescale candidates from the variogram analysis, we obtained  a preliminary sample of 16,703 short-timescale candidates with suspected periodic variability.

\subsection{Ground-based follow-up of some preliminary candidates}
\label{maiaFollowUp}

%\textbf{At that point of our analysis, both this monitoring campaign and inspection of specific sources, among the prelimnarily selected ones, evidenced some problematic, and probably spurious, variability behaviours, and drove to iterative filtering and refinements of our selection criteria (see Sect. \ref{envFilter} and \ref{otherSpurious}).}

At that point of our analyis, we also benefited from some supplementary ground-based observations of a few tens of candidates, chosen from the list of preliminary short-timescale variables described in Sect. \ref{stsPeriodicSelection}, so as to confirm or invalidate their variability at the expected timescale. This photometric monitoring was performed between August and November 2017 at the $1.2\,$m Mercator Flemish Telescope at the Spanish Observatory Roque de Los Muchachos (La Palma, Canary Islands, Spain) using the Mercator Advanced Imager for Asteroseismology \citep[MAIA; ][]{Raskin2013MAIA}. The data were reduced using the \texttt{ePipe} photometric reduction pipeline developed by Sergi Blanco-Cuaresma at the Geneva Observatory \citep[see][]{Roelens2016}.
We emphasize that the goal of this follow-up campaign was not to better characterise the identified candidates with short-timescale variability, for example, by refining their period. By doing so, we would have taken advantage of our privileged access to \textit{Gaia} data to do early science, which would be in total disagreement with the \textit{Gaia} DPAC rules. Our aim was first of all to verify for each of the monitored sources whether the light curve as observed from the ground exhibited some variability that might be compatible with the features observed by \textit{Gaia}, particularly in terms of amplitude, detection timescale, and period involved to ensure that the detection at short-timescale level was justified. The idea here was to evaluate the overall quality of the preliminary sample of candidates with short-timescale variability to verify whether further filtering was required prior to publication as part of \textit{Gaia} DR2.

Taking into account the MAIA instrument capacities, we chose the sources that were to be monitored at the bright side of the preliminary sample of short-timescale candidates ($G \sim 16.5 - 17\,$mag), with \textit{Gaia} amplitudes of a few tenths of magnitudes and suspected periods from a few tens of minutes to a few hours. Accordingly, and after several tests, it appeared that the observing scheme that was best suited for validation given these characteristics involves a follow-up of about $1\,$h - $1\,$h$30\,$min, continuously in one single night, with exposure times of $120\,$s. This limits the follow-up to variability that is observed at the level of tens of minutes. Unfortunately, with shorter exposure times, which would have enabled probing faster variability phenomena, the quality of the retrieved photometry was not high enough to determine the presence or absence of variations.\\
In the end, a total of 25 preliminary candidates with short-timescale variability, either chosen to probe the reliability of specific phenomena visible in \textit{Gaia} photometry or randomly chosen, were photometrically monitored from the ground.
%The outcome of this follow-up is summarised in Table \ref{tab:stspLog}. 
For one of them, the variability inferred from \textit{Gaia} data was confirmed at the expected timescale, that is to say, what was observed from ground was compatible with both $\tau_{det}$ and $P_{Gaia}$. Nine were confirmed at timescales longer than expected, that is, with a variability compatible with $\tau_{det}$ , but likely at a period longer than $P_{Gaia}$, typically a few hours instead of a few tens of minutes. Six of them could neither be confirmed nor rejected based on their ground photometry because of the relatively poor data quality resulting from bad weather. The ground-based photometry of three of the followed candidates was contaminated by bright and/or close neighbours. For six of them, the variability seen in \textit{Gaia} was proved to be spurious, as no corresponding feature was visible in MAIA measurements.
The outcome of our follow-up campaign therefore was to show the necessity for further filtering and refinement of our selection criteria, as described in Sect. \ref{envFilter} and \ref{otherSpurious}.

\subsection{Environment filtering}
\label{envFilter}

Further investigation of some of the preliminary variable candidates we selected (Sect. \ref{stsPeriodicSelection}) showed that the \textit{Gaia} short-timescale variabiliy analysis can be significantly affected by contamination of photometry that is due to the environment of the considered sources across the sky.
We therefore decided to perform some filtering of our candidate list at that stage, based on the projected vicinity of selected sources on the celestial sphere, similarly to what was done in \cite{Wevers2018}.

\begin{figure}
\centering
\includegraphics[width=0.95\linewidth]{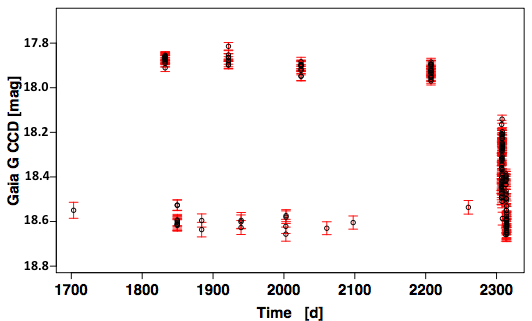}
\caption{Example of preliminary candidates for short-timescale, suspected periodic variability whose $G$ light curve is contaminated by a nearby star. The difference of magnitude between them is smaller than $1\,$mag. `Time' is expressed in BJD in $\mathrm{TBC} - 2455197.5\,$d.}
\label{fig:sts_lc_not_dominant_source}
\end{figure}

First, we removed from our preliminary sample of 16,703 candidates all the sources that were not dominant in their immediate neighbourhood, that is, objects that were not the brightest in $G$ band, by at least $1\,$mag, within a radius of $1\,$arcsecond (arcsec). Non-dominant sources can be confused with their neighbour(s) during the reconstruction of \textit{Gaia} photometry, leading to an artificial magnitude change caused by bad source identification.
Figure \ref{fig:sts_lc_not_dominant_source} shows an example candidate of a  short-timescale variablity with a suspected period as it remained after the selection of Sect. \ref{stsPeriodicSelection}, with a contaminated light curve as described. The measured magnitudes seem to alternate between two discrete levels. In this case, the considered source of mean magnitude $G = 18.42\,$mag has a neighbour with mean $G = 18.38\,$mag at a distance of $0.38\,$ arcsec. Based on the corresponding \textit{Gaia} light curves, they are likely two sources whose real $G$ magnitudes are rather around $17.9$ and $18.6\,$mag. The mistake in the mean $G$ that is effectively measured is induced by the mixing in their photometry.

\begin{figure*}
\centering
\subfloat[ ]{\includegraphics[height=0.26\linewidth]{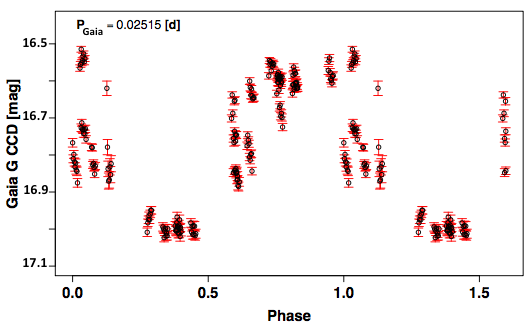}}
\subfloat[ ]{\includegraphics[height=0.25\linewidth]{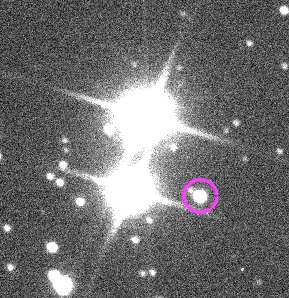}}
\subfloat[ ]{\includegraphics[height=0.27\linewidth]{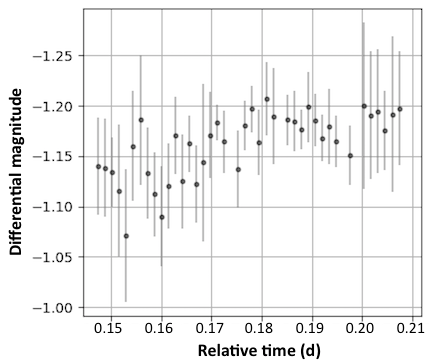}}
\caption{Example of preliminary candidates for short-timescale and suspected periodic variability whose $G$ light curve is contaminated by two nearby bright stars. (a) \textit{Gaia} $G$ per-CCD phase-folded light curve. (b) Excerpt of an image in the $R$ band from the MAIA photometer \citep{Raskin2013MAIA}; the targeted candidate is encircled in magenta. (c) MAIA differential photometry light-curve in $R$ band obtained with \texttt{ePipe}.}
\label{fig:sts_lc_bright_neighbor_contamination}
\end{figure*}

Additionally, we excluded candidates with a neighbouring object within a radius of $30\,$arcsec that had $G \leq 12\,$mag because of the a high probability that it might be contaminated by the brighter object. Here we were slightly more restrictive than \cite{Wevers2018}, since we used a limiting radius of $30\,$arcsec instead of $10\,$arcsec. This decision was motivated by some of the Mercator supplementary photometric observations performed during the \textit{Gaia} DR2 variability processing, which showed that this radius needed to be extended in the frame of our analysis.
Figure \ref{fig:sts_lc_bright_neighbor_contamination} shows an example of a probably spurious short-period candidate, for which ground-based follow-up revealed no significant variation at the expected timescale. When we inspected the corresponding MAIA images (Figure \ref{fig:sts_lc_bright_neighbor_contamination}, panel b), the considered object has two neighbours of magnitude $G \sim 10\,$mag at distances of $20$ and $26\,$arcsec, respectively. Although a slight variation of about $0.1\,$mag is visible in the MAIA $R$ differential-magnitude light curve (Figure \ref{fig:sts_lc_bright_neighbor_contamination}, panel c), it remains within the uncertainties on the measurements, and it is much smaller than the expected amplitude from the \textit{Gaia} $G$ light-curve (Figure \ref{fig:sts_lc_bright_neighbor_contamination}, panel a) at the expected timescale ($\tau_{det} = 1\,\mathrm{h}\,46\,\mathrm{min}$, $P_{\rm Gaia} = 36\,$min). This favours the hypothesis of spurious variability that is due to contamination.

%With this environment filtering, we remove 4287 short timescale suspected periodic candidates out of 22,033 in the preliminary list.

\subsection{Removing other spurious variability}
\label{otherSpurious} 

\begin{figure*}
\centering
\subfloat[ ]{\includegraphics[width=0.49\linewidth, trim = {0 0.05cm 0 0.3cm}, clip=true]{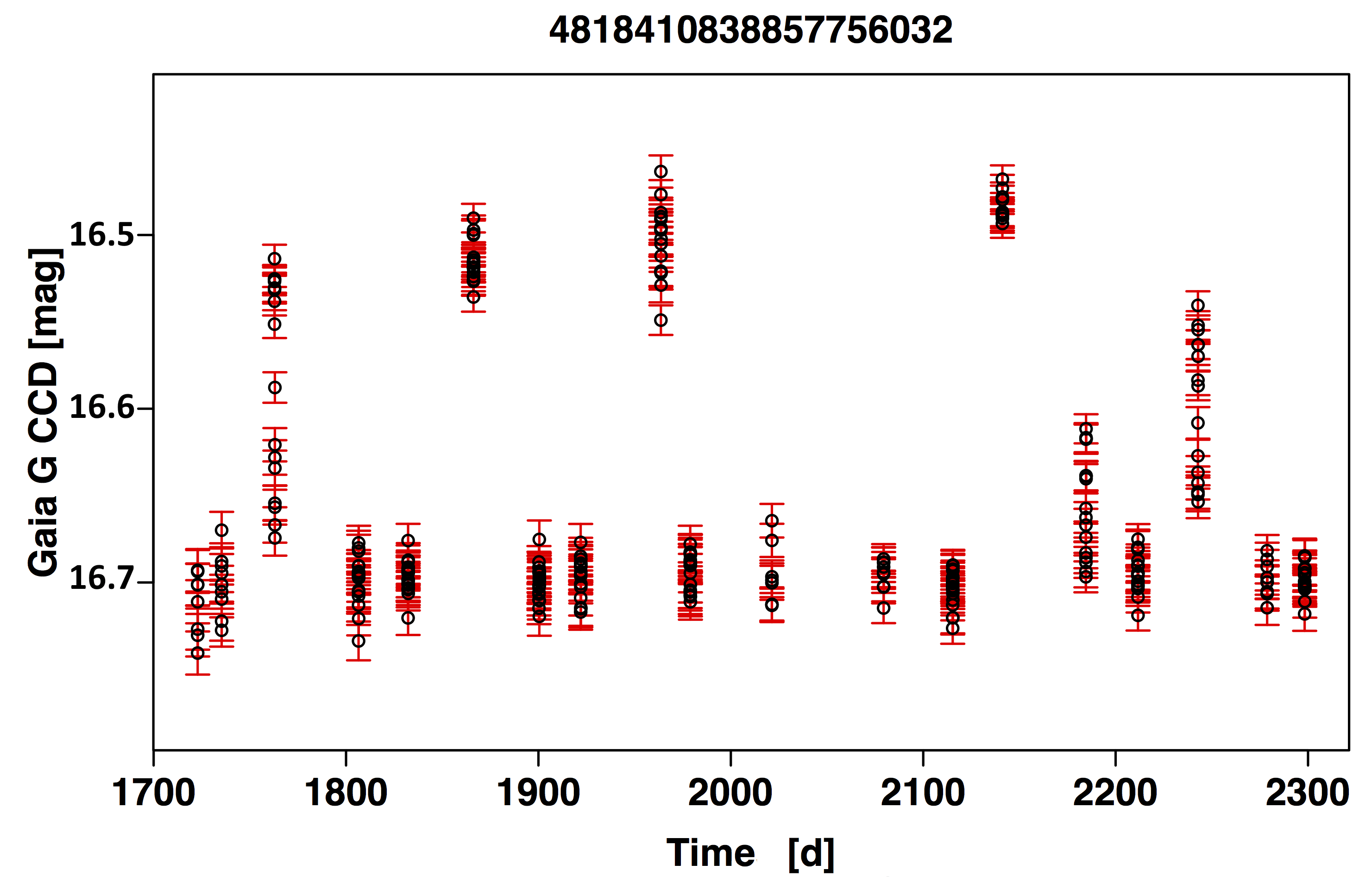}}
\subfloat[ ]{\includegraphics[width=0.49\linewidth, trim = {0 0.05cm 0 0}, clip=true]{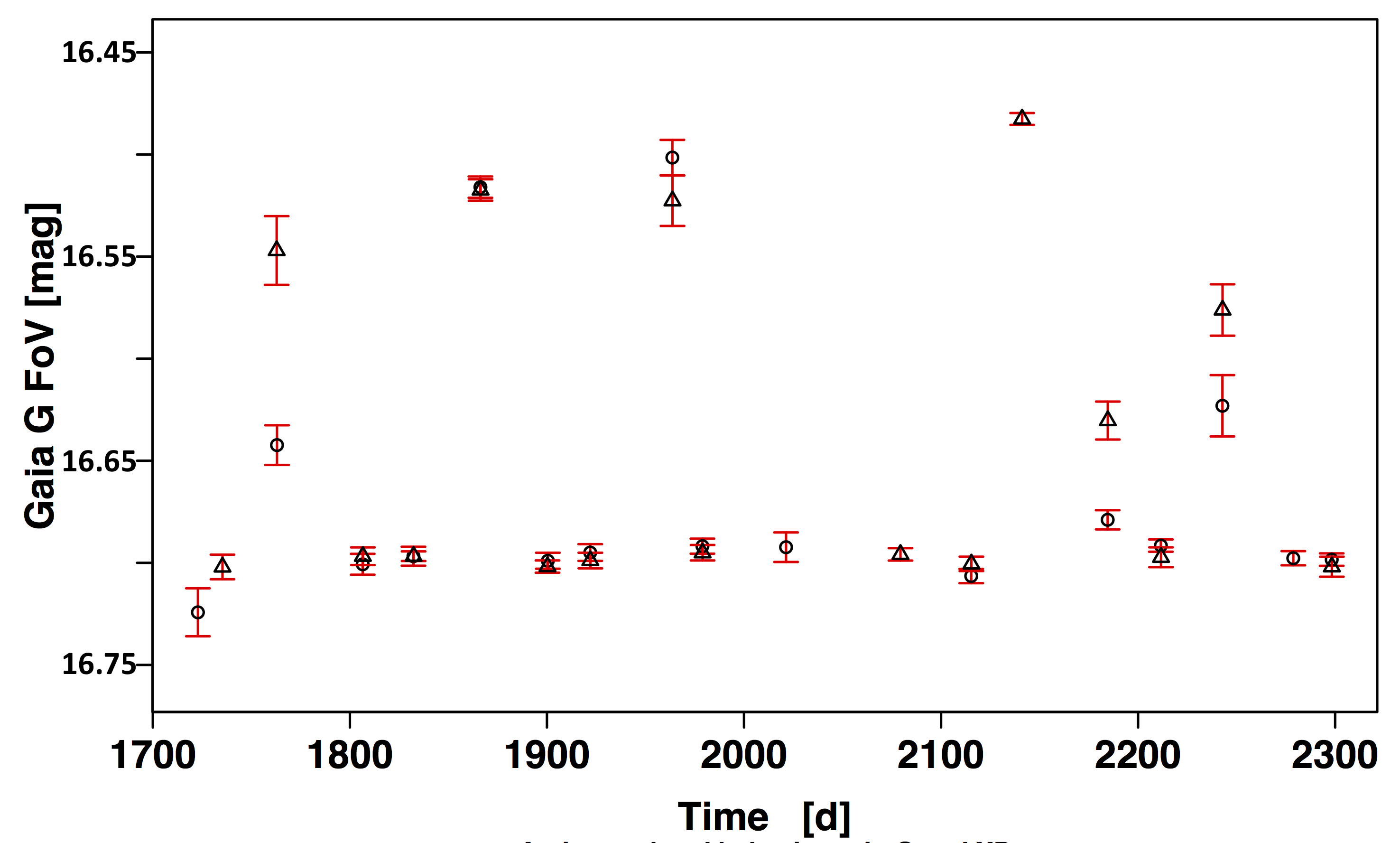}}\\
\subfloat[ ]{\includegraphics[width=0.49\linewidth, trim = {0 0.05cm 0 0}, clip=true]{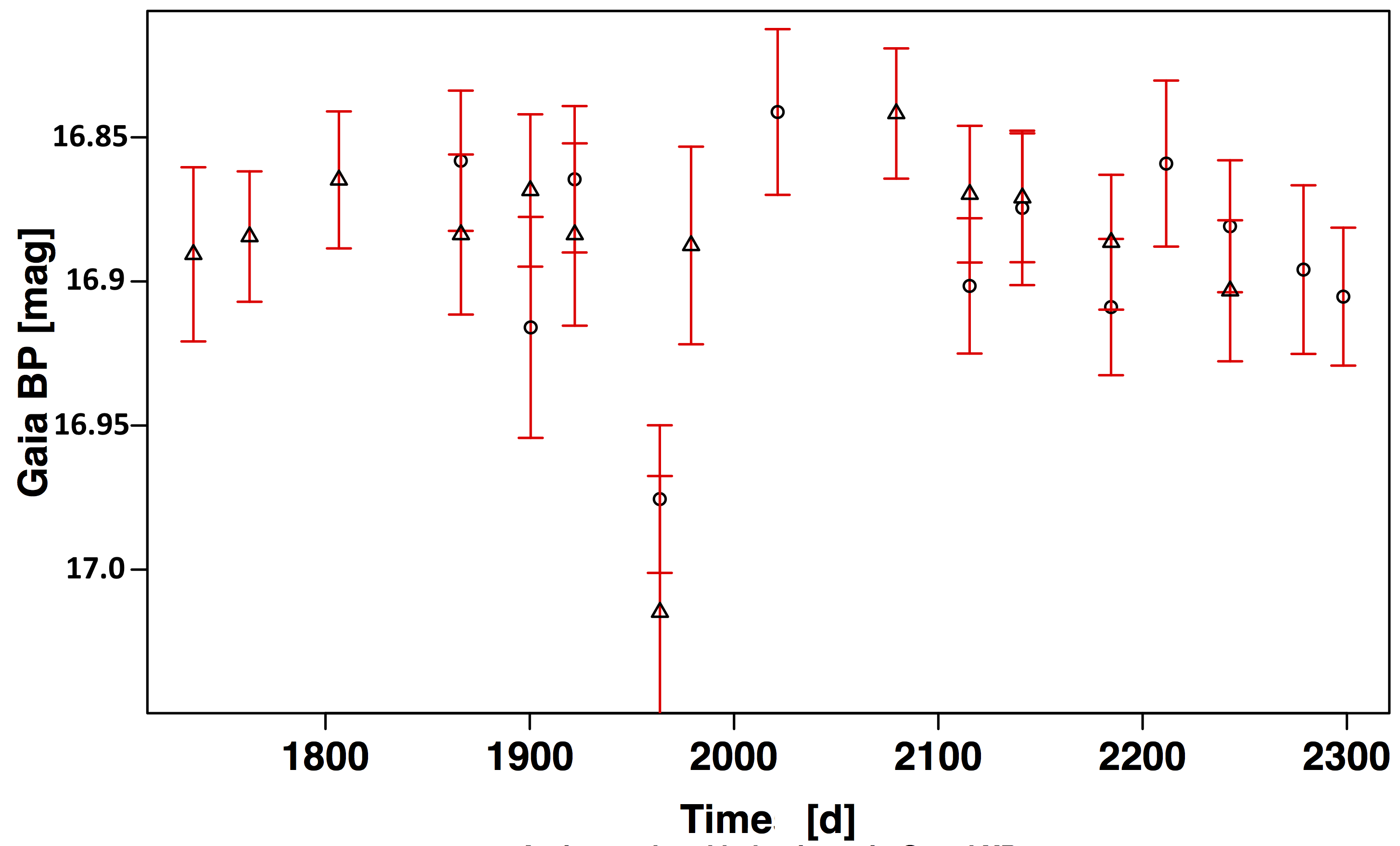}}
\subfloat[ ]{\includegraphics[width=0.49\linewidth, trim = {0 0.05cm 0 0}, clip=true]{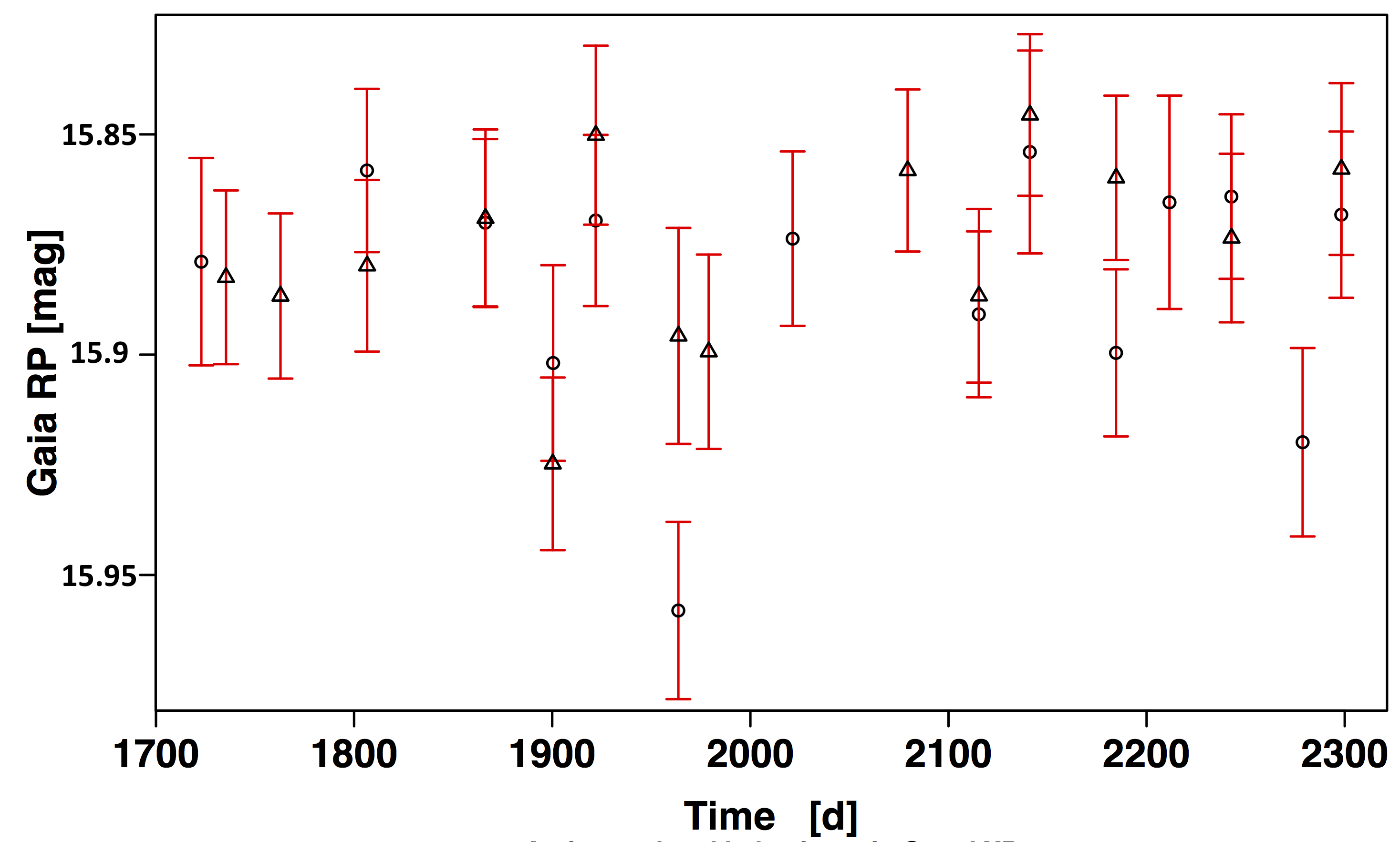}}
\caption{Example of a preliminary candidate with a short-timescale variability and a suspected period that shows anti-correlated global behaviours in $G$ and in $G_{\mathrm{BP}}$ and $G_{\mathrm{RP}}$. \textit{Gaia} light curves in $G$ CCD (a), $G$ FoV (b), $G_{\mathrm{BP}}$ (c), and $G_{\mathrm{RP}}$ (d) bands. `Time' is expressed in BJD in $\mathrm{TCB} - 2455197.5\,$d.}
\label{fig:sts_lc_anti_correlated}
\end{figure*}

By visually inspecting randomly selected light curves of the 16,703 preliminary candidates with short-timescale variability and suspected periodicity, we found several sources whose photometry globally brightens in the $G$ band, but fades in the $G_{\mathrm{BP}}$ and $G_{\mathrm{RP}}$ bands (see Figure \ref{fig:sts_lc_anti_correlated}). Such unlikely behaviour may result from photometric contamination or calibration issues. In the frame of our analysis, we removed these types of spurious candidates by mean of cuts on the skewness value $\mathcal{S}$ on light curves in the three \textit{Gaia} photometric bands. In this way, all sources with $\mathcal{S}_{G} < -1.1$ were excluded to eliminate relatively flat $G$ light curves with a few significant flares. We also removed candidates that were strongly skewed in one direction in $G$ and were strongly skewed in the other direction in both $G_{\mathrm{BP}}$ and $G_{\mathrm{RP}}$, that is, those that met the\begin{center}
removal condition
\end{center}
\begin{ceqn}
\begin{align}
(\mathcal{S}_{G} > 1~\&~\mathcal{S}_{BP} < -1~\&~\mathcal{S}_{RP} < -1)
\end{align}
\end{ceqn}
\begin{center}
or\\
$(\mathcal{S}_{G} < -1~\&~\mathcal{S}_{BP} > 1~\&~\mathcal{S}_{RP} > 1).$
\end{center}

In the meantime, this phenomenon has been investigated by the \textit{Gaia} photometry team. The sources found with these features apparently have a nearby star (at a distance of about $1$--$2\,$ arcsec) that is not necessarily bright nor of similar magnitude. This star is sometimes inside of or at the edge of the window assigned to the target for photometry integration \citep[the \textit{Gaia} windowing scheme is described e.g. in][]{Brown2016GDR1b}, and sometimes it is not. The presence or absence of the neighbouring source in the photometric window causes the brighter and fainter measurements, respectively. The dimensions of the window used for $G_{\mathrm{BP}}$/$G_{\mathrm{RP}}$ spectrophotometry are different from those of the window used for $G$ band, which explains that all three bands do not necessarily brighten or fade at the same time. More appropriate and specific treatment of such contaminated transits will be implemented in the future, enabling a transit cleaning at the time-series level rather than a rejection of the candidates, as has been done for this data release.

\begin{figure}
\centering
\includegraphics[width=0.99\linewidth]{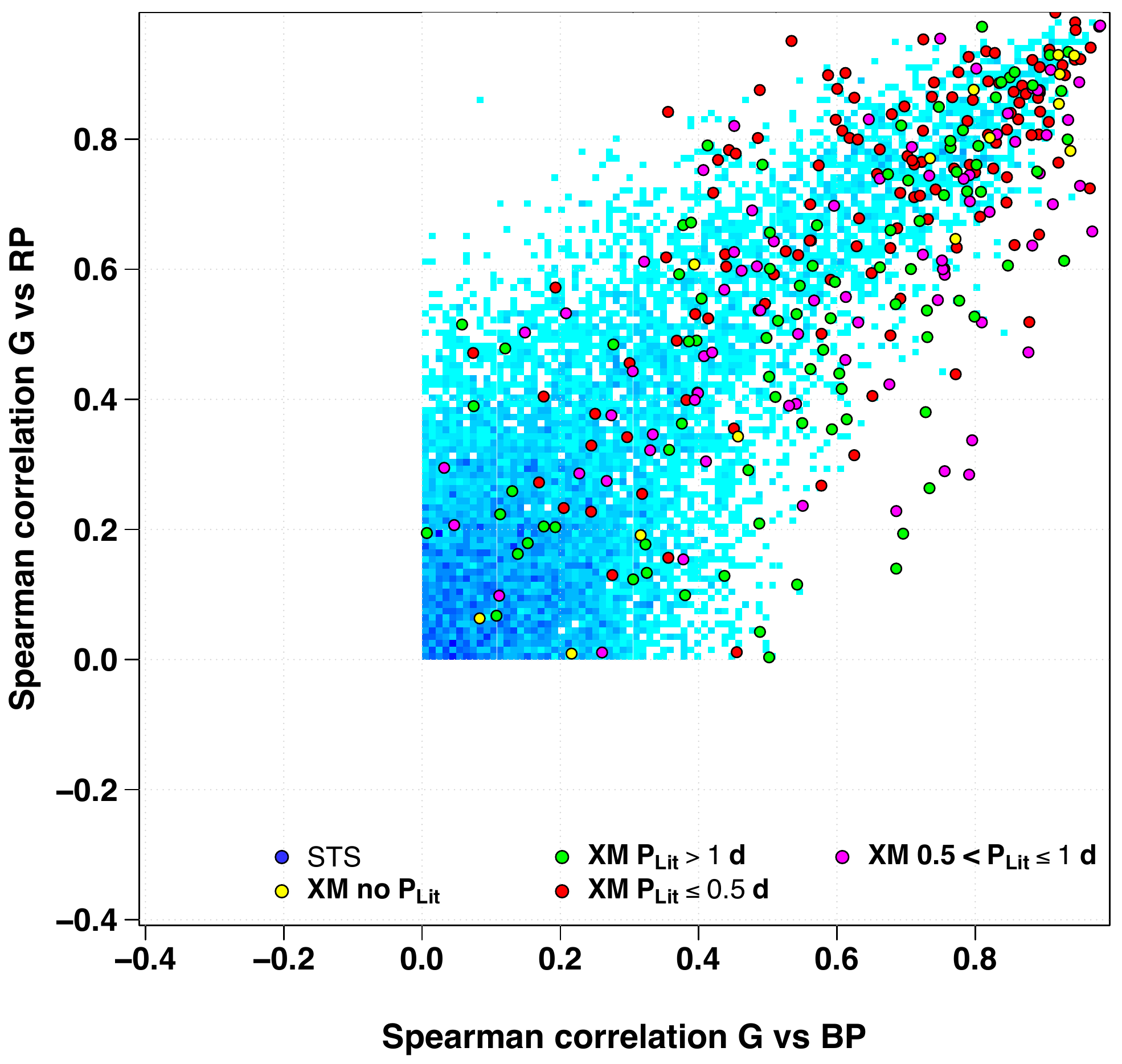}
\caption{Comparison of Spearman correlations of $G$ vs. $G_{\rm BP}$ and $G$ vs. $G_{\rm RP}$ for the sample of 16,703 preliminary candidates with short-timescale and suspected periodic variability. The filled circles represent the 303 crossmatched known variables in this sample and are colour-coded according to the period listed for them in the literature.}
\label{fig:sts_correlation_distrib}
\end{figure}

\begin{figure}
\centering
\includegraphics[width=0.99\linewidth]{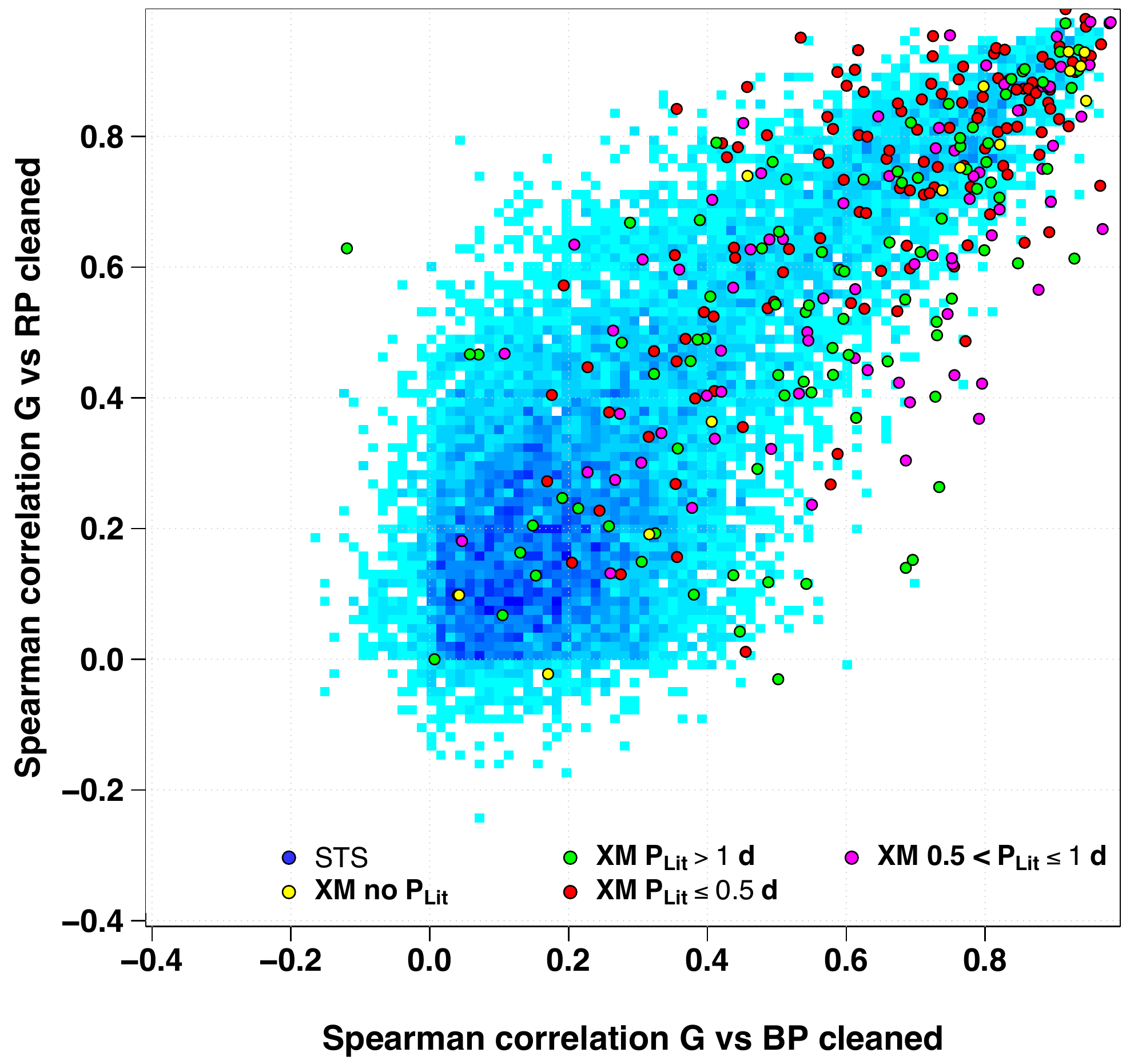}
\caption{Same as Figure \ref{fig:sts_correlation_distrib}, but with recalculated correlations with the additionally cleaned $G_{\mathrm{BP}}$ and $G_{\mathrm{RP}}$ time series.}
\label{fig:sts_new_correlation_distrib}
\end{figure}

Figure \ref{fig:sts_correlation_distrib} represents the Spearman correlation between the $G$ and $G_{\mathrm{RP}}$ time series versus the Spearman correlation between the $G$ and $G_{\mathrm{BP}}$ time series for the 16,703 preliminary candidates with short-timescale and suspected periodic variability, in particular highlighting the 303 crossmatched known variables within this sample. The preliminary cuts on the Spearman correlation values (Sect. \ref{stsPeriodicSelection}) cause the absence of candidates with negative correlations. For a significant fraction of sources in the preliminary sample, the correlations between $G$ and both $G_{\mathrm{BP}}$ and $G_{\mathrm{RP}}$ were quite low. Whereas for 80\% of the known crossmatched variables with $P_{\rm Lit} \leq 1\,$d of the set of considered candidates one of these two correlation values was higher than $0.45$ and the other was higher than $0.35$, only 27\% of the whole preliminary sample met this condition. This discrepancy caused us to question the cases of candidates with relatively low correlations: are they reliable, or should we eliminate them?
When we visually inspected the \textit{Gaia} light curves of a handful of these specific sources, we realised that although the time-series filtering is performed upstream, a few important outliers remained in the light curves. This induced spuriously low (or high) correlation values.
%\textcolor{blue}{Example of LC before and after extra time series cleaning, where the correlation value significantly changes ?}
Consequently, we applied an additional time-series cleaning operator to the $G_{XP}$ time series, based on the expected amplitude of variation (if really present) in $G$ as defined in Sect. \ref{otherParams}. This operator, specific to short-timescale variability processing, removes all the points in the $G_{XP}$ time series that lie farther away from the $G_{XP}$ median magnitude than $1.5*A_{G-CCD}*\frac{IQR_{XP}}{IQR_{G}}$, where $A_{G-CCD}*\frac{IQR_{XP}}{IQR_{G}}$ is a proxy for the expected variation amplitude in $G_{XP}$ according to what is seen in $G$ band.
Figure \ref{fig:sts_new_correlation_distrib} shows the Spearman correlation distribution for the preliminary sample, after recalculating correlation values with the additionally cleaned $XP$ time series. Since some preliminary candidates had negative correlation values after this data filtering step, it seemed clear that the cut on correlations made in Sect. \ref{stsPeriodicSelection} was not sufficient.
Again, building the appropriate correlation selection criterion relied on visual inspection of some low-correlation sources (with the new correlation calculation), ground-based photometric follow-up, and investigation of the crossmatched known variable sources in the sample.
%Moreover, additional crossmatch using the Simbad per coordinate crossmatch tool \footnote{\url{http://simbad.u-strasbg.fr/simbad/}} enables in particular to spot some extended sources within the preliminary sample (e.g. galaxies) with relatively low correlations and whose variability features visible may be suspicious. Because we aim to focus on a limited, bona fide short timescale suspected periodic variable sample, and do not target completeness for this Data Release, we decide to remove those extended objects.
In the end, we decided to keep only candidates for which one of their Spearman correlation values ($r_{G-BP}$ or $r_{G-RP}$) was higher than $0.45$ and the other value higher than $0.35$.

%\textcolor{blue}{Maybe show one example of low correlation source not confirmed by follow-up ?}

%\textcolor{blue}{Maybe position the extended sources (particularly galaxies) in the correlations plot ?}

\subsection{Removing eclipsing binaries}
\label{ebRemove}

As part of the \textit{Gaia} variability processing, a specific analysis dedicated to the identification and characterisation of eclipsing binary stars has been performed as a test case, based on the variable classification described in \cite{GDR2Rimoldini2018}. However, it has been decided to postpone the publication of the resulting candidate eclipsing binaries from all the \textit{Gaia} DR2 variability products, since reporting of new eclipsing binaries discovered from \textit{Gaia} data is planned only for Data Release 3 and onwards\footnote{For more information on the \textit{Gaia} Data Release scenario, see \url{https://www.cosmos.esa.int/web/Gaia/release}}. Hence, the candidate eclipsing binaries identified have been removed from most of the other \textit{Gaia} DR2 variability candidate lists. 

In the case of the short-timescale variability analysis, 623 candidate eclipsing binaries were found to overlap our preliminary sample of 16,703 candidates. They were therefore excluded prior to \textit{Gaia} DR2 publication.

\section{Published sample of short-timescale candidates in Gaia Data Release 2}
\label{results}

After applying all the selection criteria described in Sect.~\ref{stsSelection}, we obtained a \textbf{final list of 3,018 sources with short-timescale and suspected periodic variability that were published in Gaia DR2}. This final sample includes 138 known variables from the reference crossmatched source list presented in Sect \ref{variogram}: 71 with $P_{\rm Lit} \leq 0.5\,$d, 32 with $0.5 < P_{\rm Lit} \leq 1\,$d, 27 with $P_{\rm Lit} > 1\,$d, and 8 with no $P_{\rm Lit}$ information from the literature, whose type is compatible with short-timescale variability, however. None of the constant sources and non-periodic variables from the reference crossmatched sample remains in the final list of short-timescale candidates. The completeness of our sample is about 12\% of the 439 + 382 short-period variables in the input sample of 5.6 million sources (see Section \ref{variogram}). Based on this crossmatch, the contamination would be assessed to lie at about 19\%, but consisting of longer-period variability alone.

To proceed in the completeness and contamination analysis, we decided to focus on areas covered by both \textit{Gaia} and OGLE, thus restricted essentially to the region of the Magellanic Clouds, and to compare the variability results for these two surveys. According to the OGLE III and IV catalogues of variable stars, 45,966 sources are either in the Large Magellanic Cloud (LMC) or in the Small Magellanic Cloud (SMC), and have periods $P_{\rm Lit} \leq 1\,\mathrm{d}$. Only 24 of them are part of the published short-timescale sample, which means that the global completeness in this area is as low as 0.05\%. We recall that completeness was not the main goal of the \textit{Gaia} DR2 short-timescale analysis; it is expected to be significantly improved in further data releases.\\
In total, 48 OGLE variables in the Magellanic Clouds were identified as short-timescale candidates with suspected periods (24 with $P_{\rm OGLE} \leq 1\,$d, 24 with $P_{\rm OGLE} > 1\,$d). This results in a contamination from longer-period variables of 50\%, although these longer-period variables have periods shorter than $7\,$d (and mostly shorter than $2\,$d) and amplitudes of between $0.13$ and $0.98\,$mag, which means that they belong in the category of justified and acceptable longer-period contamination.\\
To determine whether longer-period variables were the only source of contamination remaining in the \textit{Gaia} DR2 sample of short-timescale variables, we crossmatched it with  the OGLE II photometric database, which covers parts of the LMC, and compared the $I$-band OGLE light curves to the \textit{Gaia} $G$ light curves for a few tens of sources in common. If the features seen by \textit{Gaia} are reproduced and compatible with the OGLE II photometry, then the variability is confirmed. Otherwise, the short-timescale candidate is considered as spurious. Based on this analysis, we obtained a contamination level from spurious variables of between 10\%\ and 20\% in the LMC. This region is quite dense in sources, and is therefore more likely to be affected by contamination from neighbouring stars than the Galactic halo, for example. This contamination is therefore likely an upper limit for the whole sample contamination.

\begin{figure}
\centering
\subfloat[ ]{\includegraphics[width=0.9\linewidth, page=15, trim = {0 0.5cm 0 0}, clip=true]{STS_published_known_variables_illustration.pdf}}\\
\subfloat[ ]{\includegraphics[width=0.9\linewidth, trim =  {0 0.5cm 0.5cm 2cm}, clip=true]{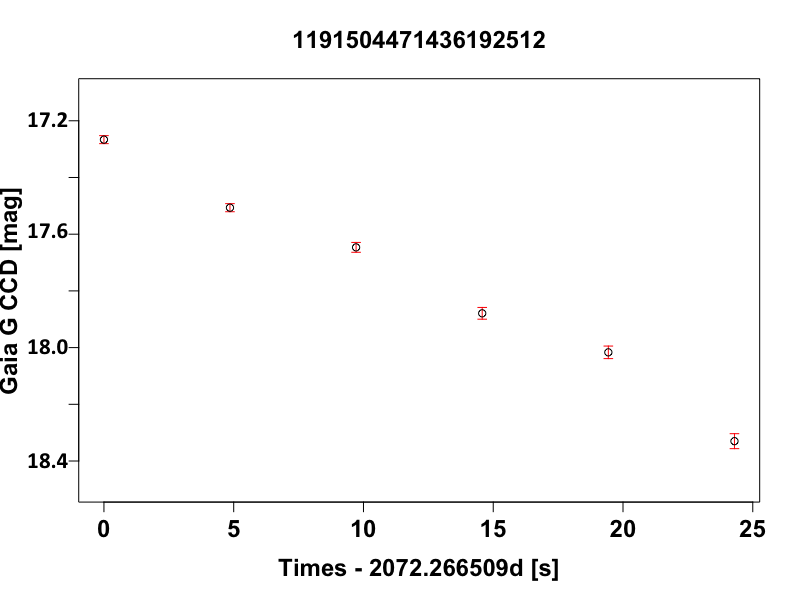}}
\caption{Phase-folded \textit{Gaia} $G$ CCD light curve of the PCEB NN Ser (a), and zoom on the fading transit that is visible at phase around 0.3 (b). `Times' are expressed in BJD in $\mathrm{TCB} - 2455197.5\,$d. The reference time used to fold the \textit{Gaia} light curve is $1738.448\,$d (in BJD in $\mathrm{TCB} - 2455197.5\,$d).}
\label{fig:sts_pub_known_nn_ser}
\end{figure}

Returning to the 138 crossmatched sources, it appears that with our short-timescale analysis, we recovered very interesting known variables, such as some post-common envelope binaries (PCEB) or cataclysmic variables (CV). The first striking example of a known PCEB within the short-timescale candidate list we wish to highlight is NN Ser, an eclipsing system whose orbital period is $3.12\,$h \citep{Haefner1989}. This system is also known to be orbited by two candidate exoplanets \citep[the presence of the exoplanets is inferred from variations in the eclipse timings]{Beuermann2010}. The \textit{Gaia} DR2 \texttt{source\_id} of NN Ser is 1191504471436192512. We investigated it because we were curious to see if we could detect evidence of the eclipse of the white dwarf by the secondary star in this well-known binary. The short period found in \textit{Gaia} short-timescale analysis exactly recovers the period from the literature. Figure \ref{fig:sts_pub_known_nn_ser} shows that NN Ser has one strongly fading FoV transit in its $G$ CCD light curve, losing more than $1\,$mag over $40\,$s. According to the ephemeris from \cite{Beuermann2010}, this transit rather corresponds to an eclipse of the binary system than to a transit of the known planets. However, we emphasise that this tremendous fading transit is removed from the cleaned $G$ FoV time series because of its relatively high $G$ magnitude uncertainty. This demonstrates the high relevance and necessity of an available $G$ per-CCD time series at the end of \textit{Gaia} mission search for short-timescale variability.

\begin{figure*}
\centering
\subfloat[ ]{\includegraphics[width=0.329\linewidth, page=14, trim = {0 0 0 1.5cm}, clip=true]{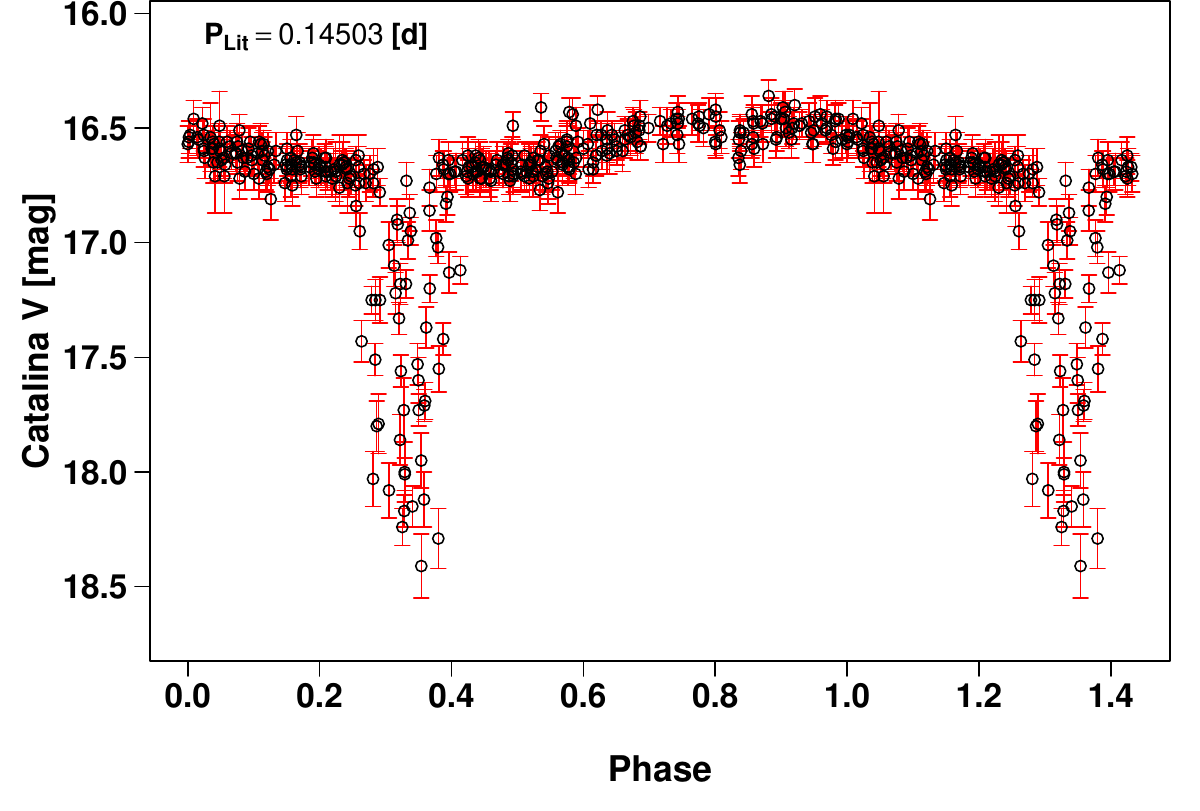}}
\subfloat[ ]{\includegraphics[width=0.329\linewidth, page=9, trim = {0 0 0 0}, clip=true]{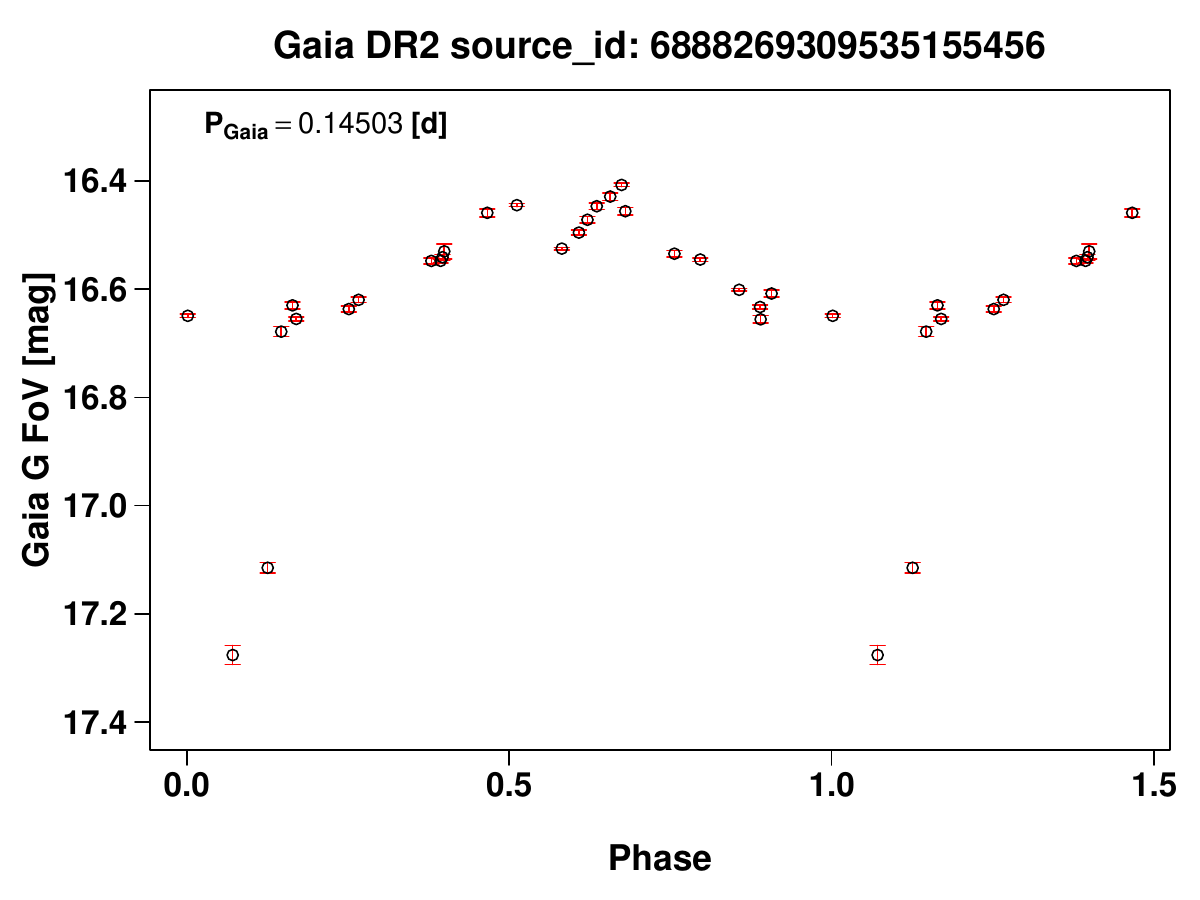}}
\subfloat[ ]{\includegraphics[width=0.329\linewidth, page=10, trim = {0 0 0 1.5cm}, clip=true]{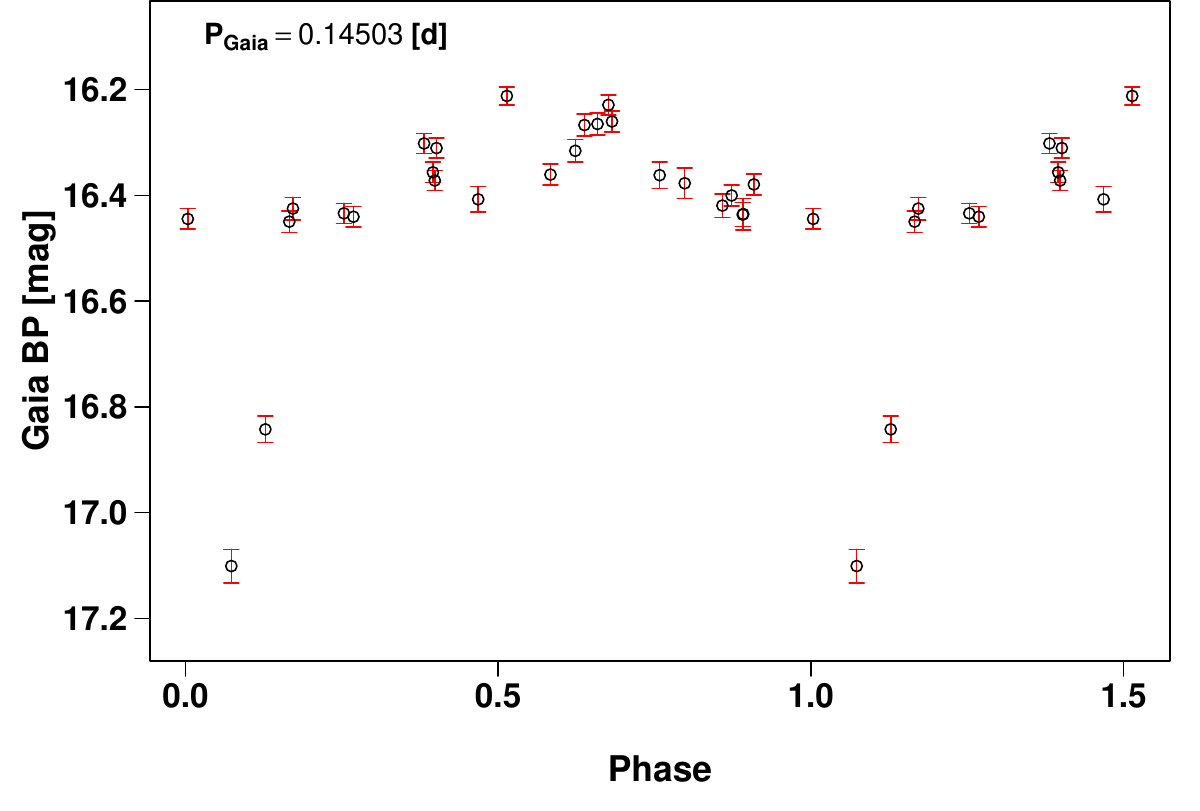}}\\
\subfloat[ ]{\includegraphics[width=0.329\linewidth, page=11, trim = {0 0 0 1.5cm}, clip=true]{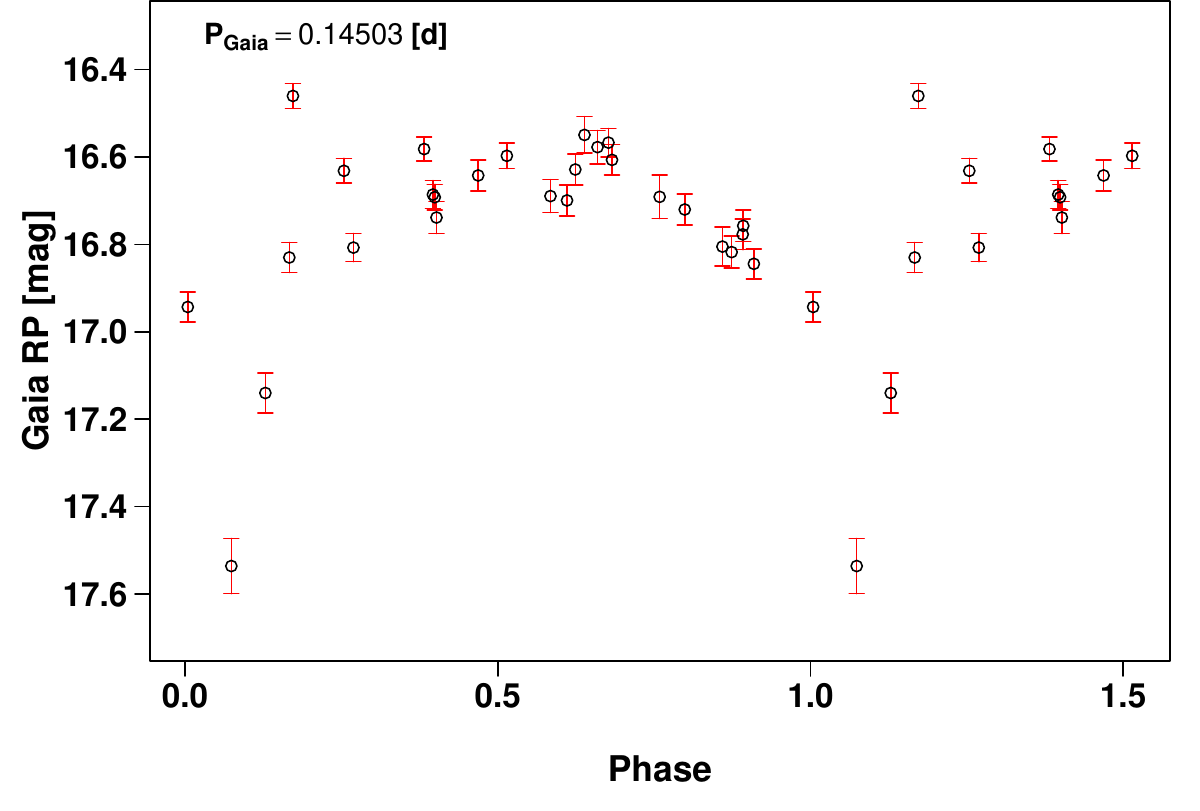}}
\subfloat[ ]{\includegraphics[width=0.329\linewidth, page=12, trim = {0 0 0 1.5cm}, clip=true]{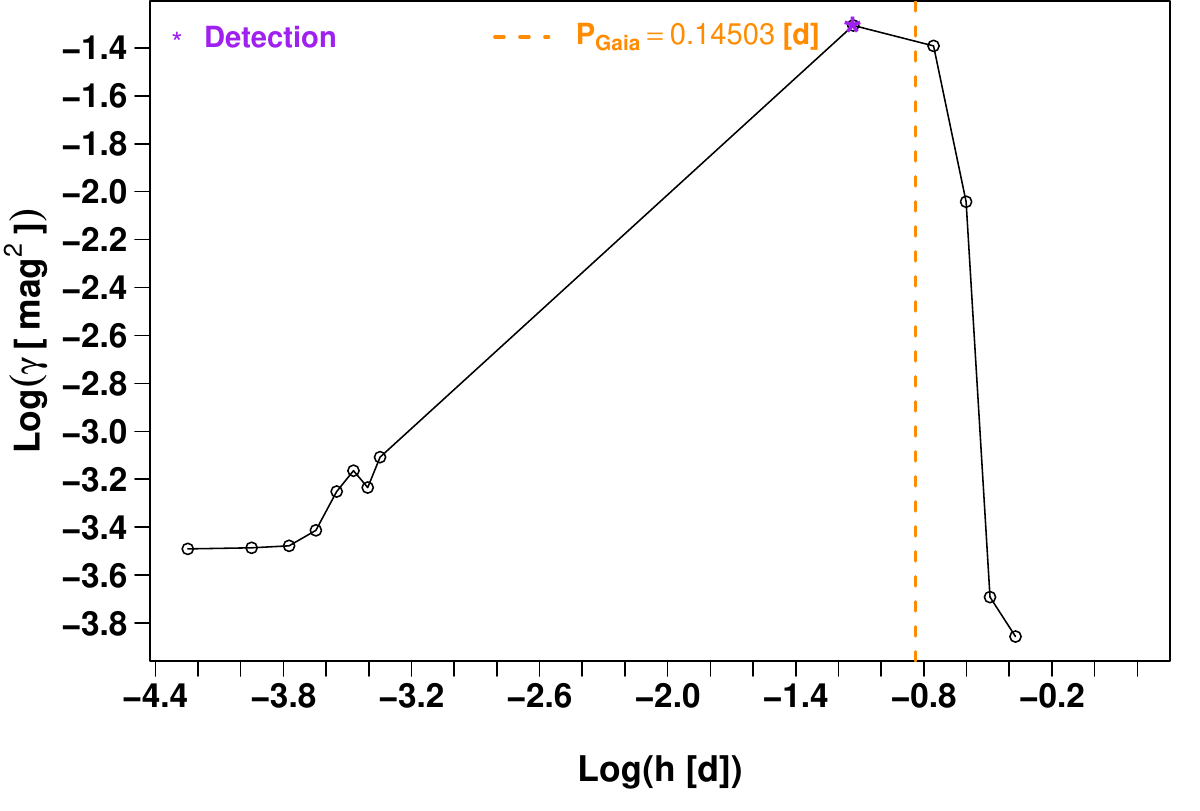}}
\caption{Phase-folded light curves and variogram of the PCEB CSS J210017. \textit{Catalina} $V$ light curve phase-folded with the period from \cite{Drake2014Catalina} (a); \textit{Gaia} $G$ FoV (b), $G_{\mathrm{BP}}$ (c), and $G_{\mathrm{RP}}$ (d) light curves phase-folded with the period from the short-timescale analysis;  and \textit{Gaia} variogram from the short-timescale analysis (e). The orange dashed line indicates the \textit{Gaia} short-timescale period, and the purple star shows the variogram point that triggered the detection. Different reference times have been used to phase-fold the \textit{Gaia} and \textit{Catalina} light curves. The reference time used to fold the \textit{Gaia} light curves is $1757.338\,$d (in BJD in $\mathrm{TCB} - 2455197.5\,$d).}
\label{fig:sts_pub_known_pceb}
\end{figure*}

\begin{figure*}
\centering
\subfloat[ ]{\includegraphics[width=0.329\linewidth, page=7, trim = {0 0 0 1.5cm}, clip=true]{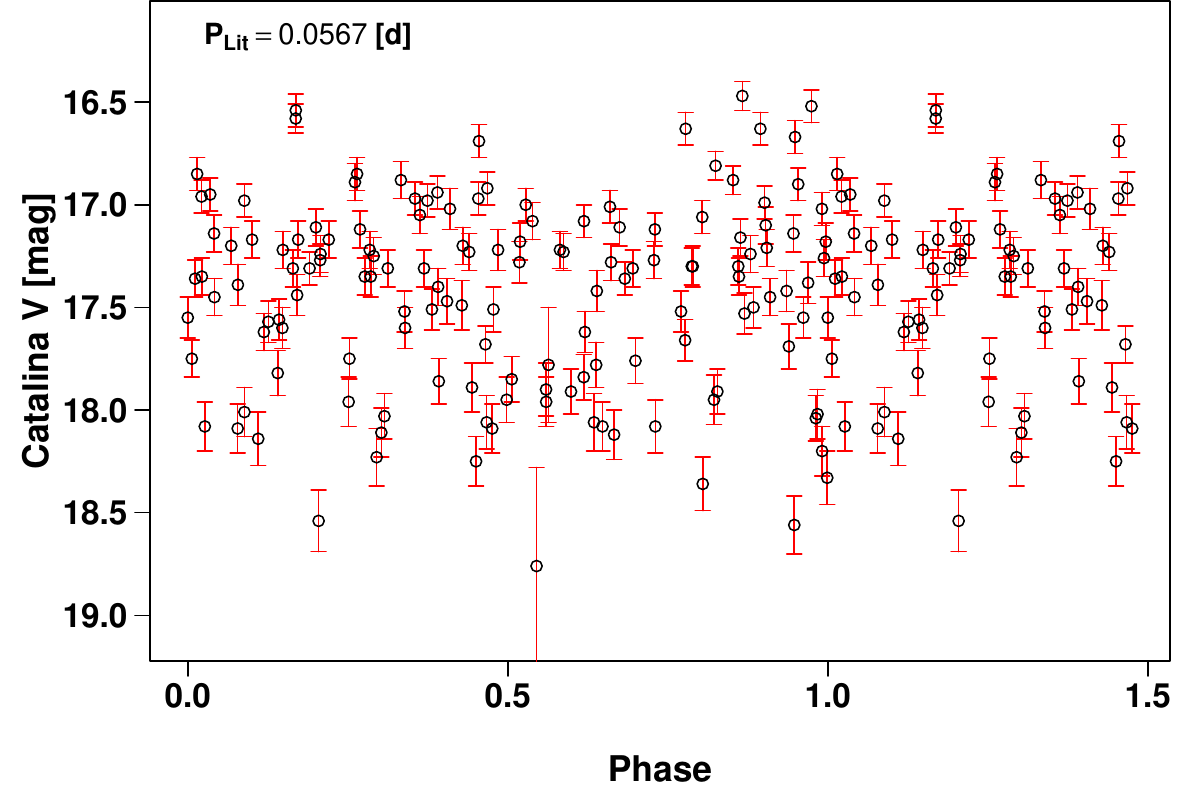}}
\subfloat[ ]{\includegraphics[width=0.329\linewidth, page=2, trim = {0 0 0 0}, clip=true]{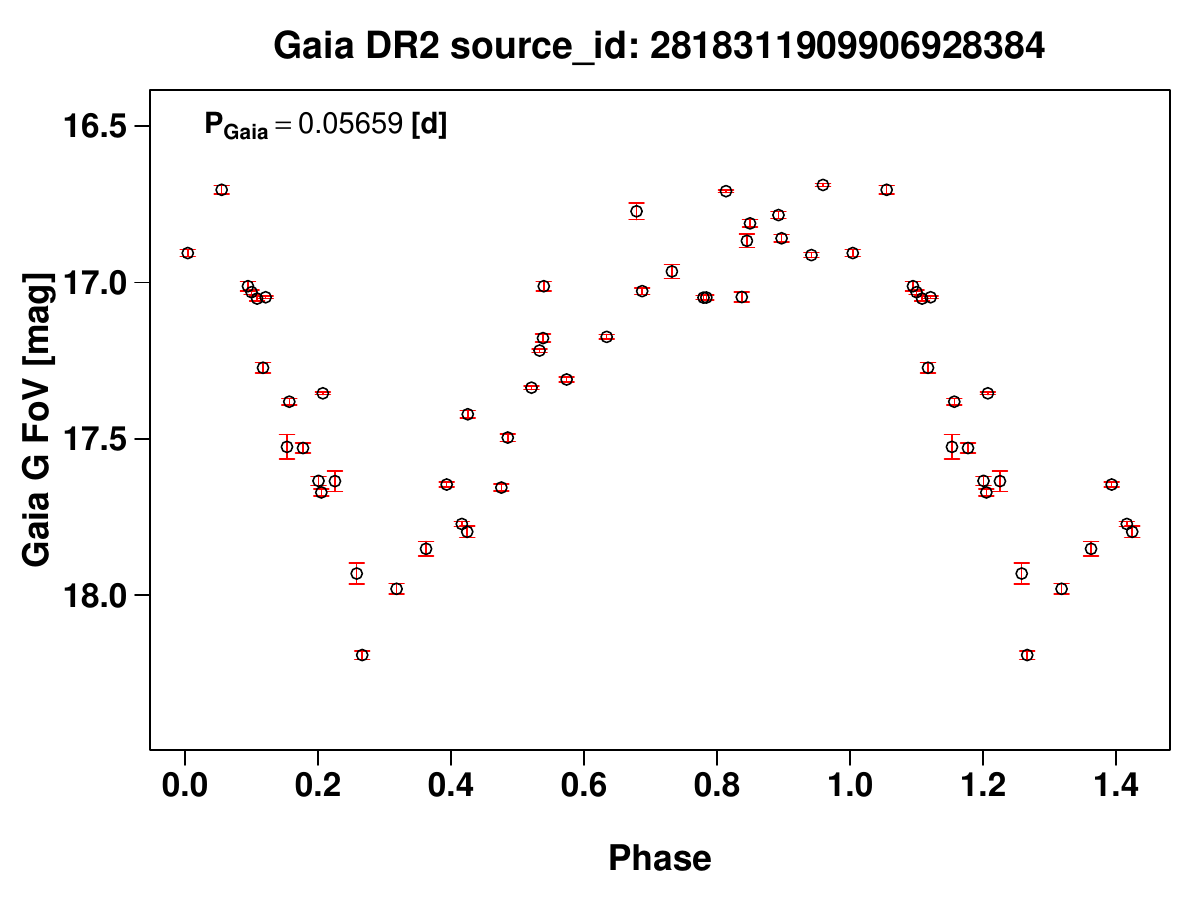}}
\subfloat[ ]{\includegraphics[width=0.329\linewidth, page=3, trim = {0 0 0 1.5cm}, clip=true]{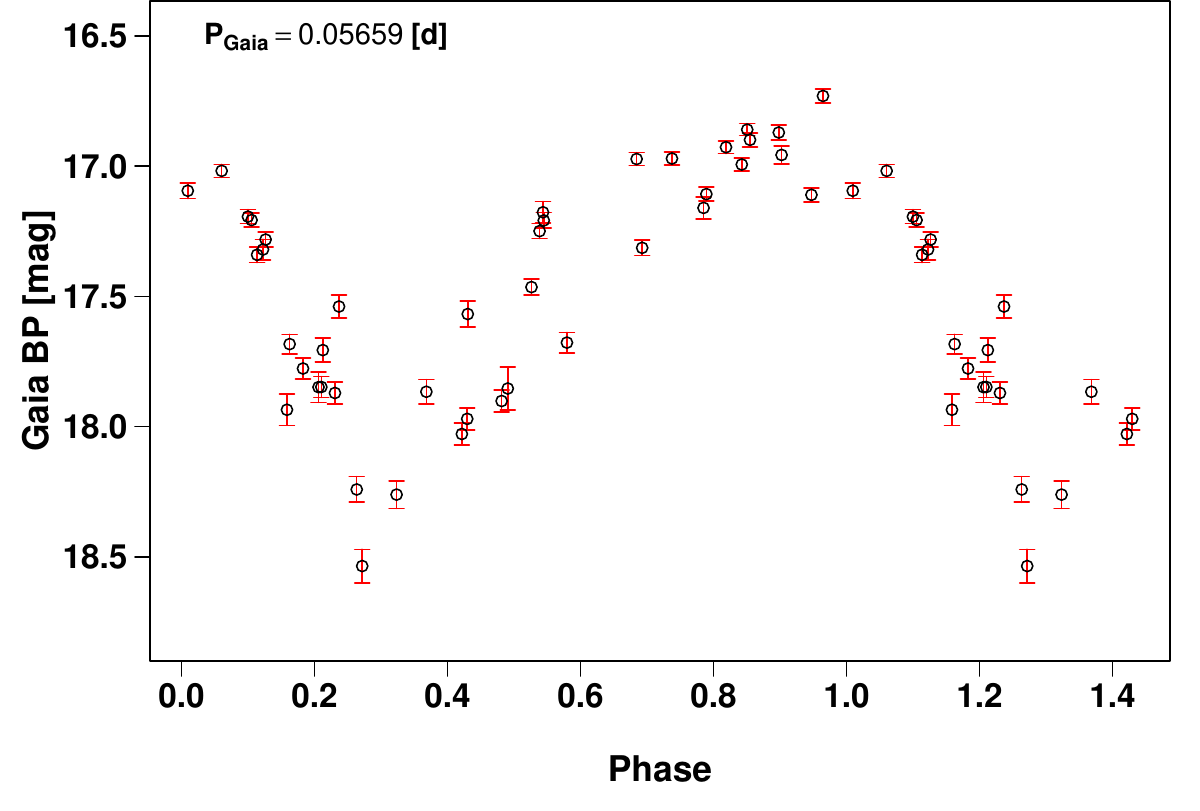}}\\
\subfloat[ ]{\includegraphics[width=0.329\linewidth, page=4, trim = {0 0 0 1.5cm}, clip=true]{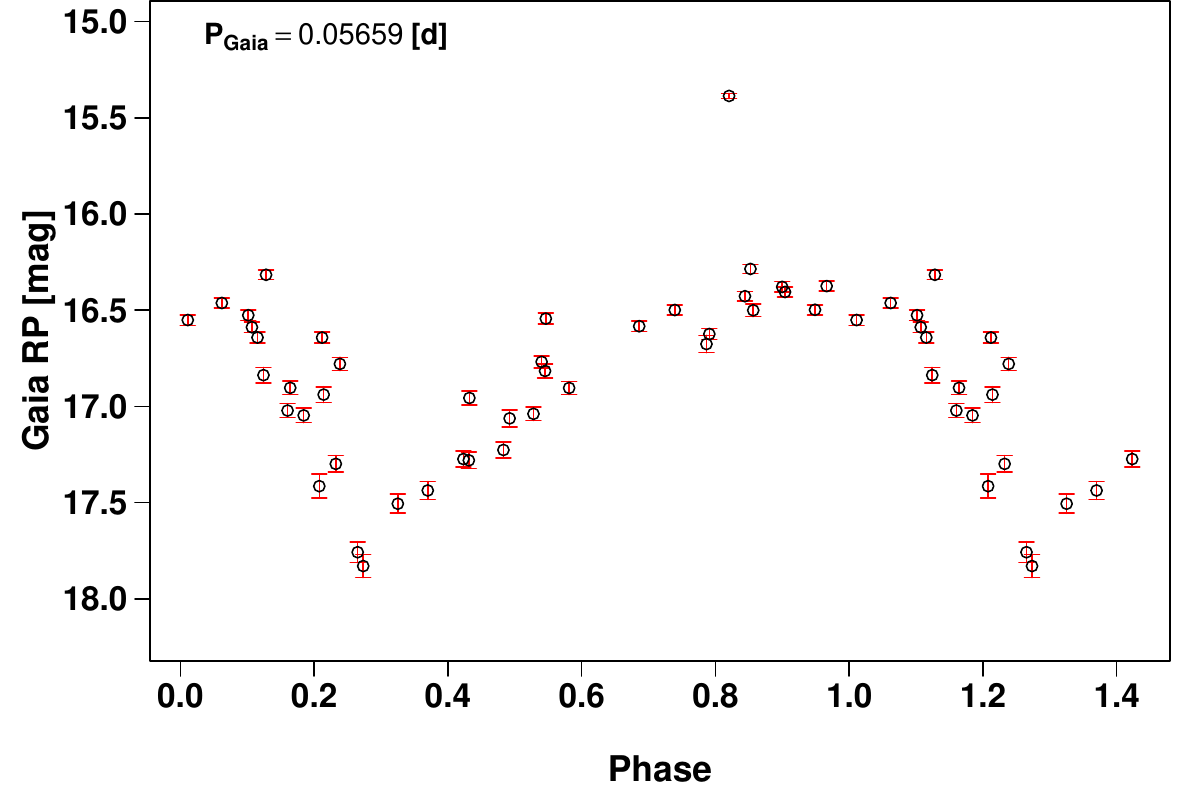}}
\subfloat[ ]{\includegraphics[width=0.329\linewidth, page=5, trim = {0 0 0 1.5cm}, clip=true]{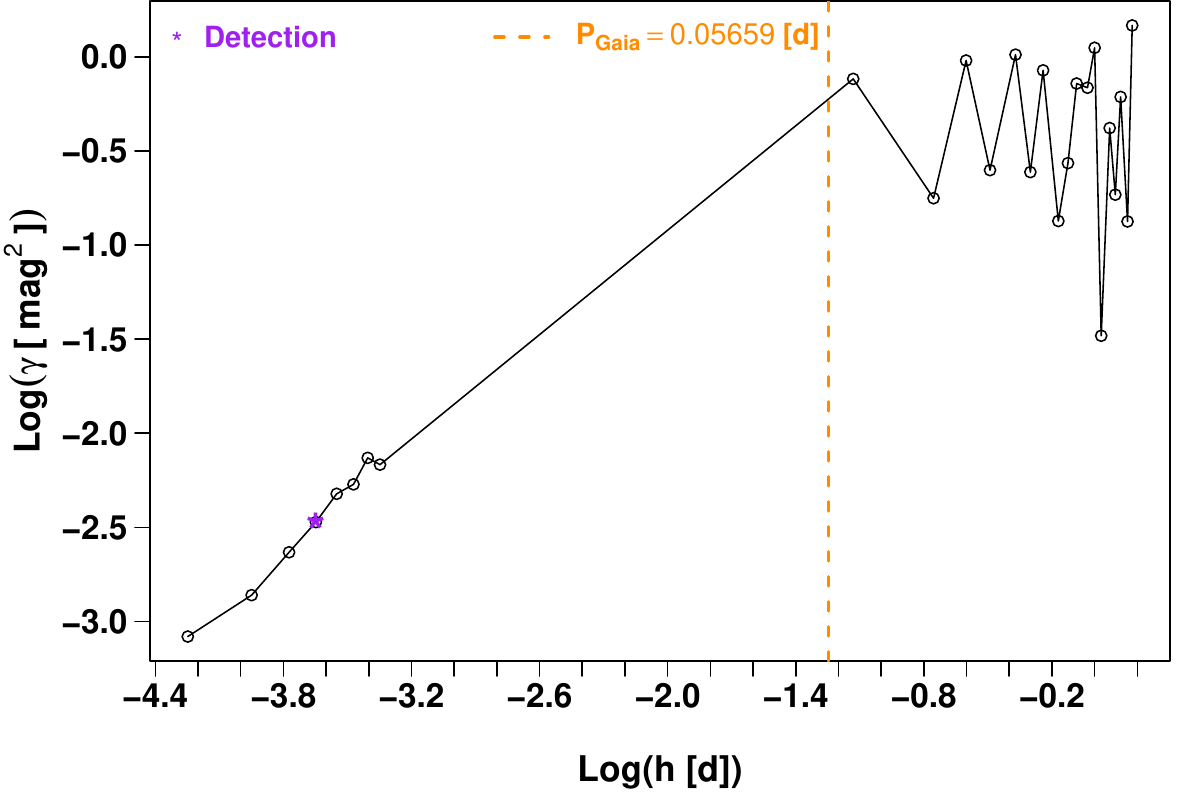}}
\caption{Same as Figure \ref{fig:sts_pub_known_pceb} for the CV of type AM Her CSS J231330, but this time with $P_{\rm Lit}$ from \cite{Margon2014}. The reference time used to fold the \textit{Gaia} light curves is $1675.660\,$d (in BJD in $\mathrm{TCB} - 2455197.5\,$d).}
\label{fig:sts_pub_known_cv}
\end{figure*}

Figures \ref{fig:sts_pub_known_pceb} and \ref{fig:sts_pub_known_cv} represent the phase-folded \textit{Gaia} $G$ FoV light curve and variogram, as well as the corresponding phase-folded \textit{Catalina} $V$ light curve from the literature\footnote{\url{http://nesssi.cacr.caltech.edu/DataRelease/}}, for a known PCEB (CSS J210017.4-141125, hereafter CSS J210017) and a known AM Her variable star (CSS J231330.8+165416, hereafter CSS J231330), respectively. The corresponding \textit{Gaia} source identifiers are 6888269309535155456 and 2818311909906928384. The \textit{Gaia} and literature periods are quite similar for both sources (for CSS J210017 $P_{Lit} = 0.14503\,$s and $P_{Gaia} = 0.14503\,$d, for CSS J231330 $P_{Lit} = 0.05670\,$d and $P_{Gaia} = 0.05659\,$d), and their phase-folded \textit{Gaia} light curves are convincing and coherent in the $G$, $G_{\mathrm{BP}}$ , and $G_{\mathrm{RP}}$ bands. The \textit{Gaia} variogram exhibits variations that are compatible with the periods we found. It is very interesting to see that in the case of CSS J210017, the eclipse is already sampled and visible despite the sparse scanning law and limited time span of the analysed data.

%Table \ref{tab:gdr2STS} shows an excerpt of the \texttt{vari\_short\_timescale} table published in the \textit{Gaia} DR2 archive\footnote{\url{https://gea.esac.esa.int/archive/}}.
The fields of the \texttt{vari\_short\_timescale} table published in the \textit{Gaia} DR2 archive\footnote{\url{https://gea.esac.esa.int/archive/}} are the following:
\begin{itemize}
\item \texttt{solution\_id}: a numeric field that unequivocally identifies the version of all subsystems and input data used to produce the table content,
\item \texttt{source\_id}: a unique numeric field identifying the source within all \textit{Gaia} products,
\item \texttt{amplitude\_estimate}: estimate of the amplitude of the variation from per-CCD time series, $A_{G-CCD}$, as defined in Sect. \ref{otherParams},
\item \texttt{number\_of\_fov\_transits}: the number of transits for the considered sources with more than 7 points in it (it can then be smaller than 20, although we work on sources with more than 20 FoVs),
\item \texttt{mean\_of\_fov\_abbe\_values}: the mean Abbe value per transit, $\mathcal{A}_{per-transit}$, as defined in Sect. \ref{otherParams},
\item \texttt{variogram\_num\_points}: the number of points constituting the variogram, that is, the number of explored lags (here it is always 26),
\item \texttt{variogram\_char\_timescales}: the variogram characteristic timescale(s) extracted for the source; by now, the only characteristic timescale retrieved is the detection timescale $\tau_{det}$ as defined in Sect. \ref{variogram},
\item \texttt{variogram\_values}: the variogram value(s) associated with the characteristic timescale(s); here it is simply the variogram value corresponding to the detection timescale,
\item \texttt{frequency}: the frequency $f_{\rm Gaia}$ resulting from a high-frequency search as defined in Sect. \ref{highFreqSearch}.
\end{itemize}
As detailed in \cite{Roelens2017}, by quantifying the averaged variation rate of the considered light curve, the variogram detection timescale and associated variogram value give clues for future ground-based follow-up of the short-timescale candidates published in \textit{Gaia} DR2. For example, the CV CSS J231330 has $\tau_{det}=19.4\,$s and $\gamma(\tau_{det}) = 0.00337\,$mag$^{2}$, which means that if the photometric instrument used for follow-up has an accuracy of about $55\,$mmag, then the observing cadence for detecting the variability should be as short as $20\,$s.

%\begin{table*}
%\begin{center}
%\begin{adjustbox}{angle=90}
%\includegraphics[width=1.3\textwidth]{gdr2_archive_sts.png}
%\end{adjustbox}
%\end{center}
%\caption{Excerpt of the \textit{vari\_short\_timescale} from the \textit{Gaia} DR2 archive}
%\label{tab:gdr2STS}
%\end{table*}

\begin{figure}
\centering
\includegraphics[width=0.99\linewidth]{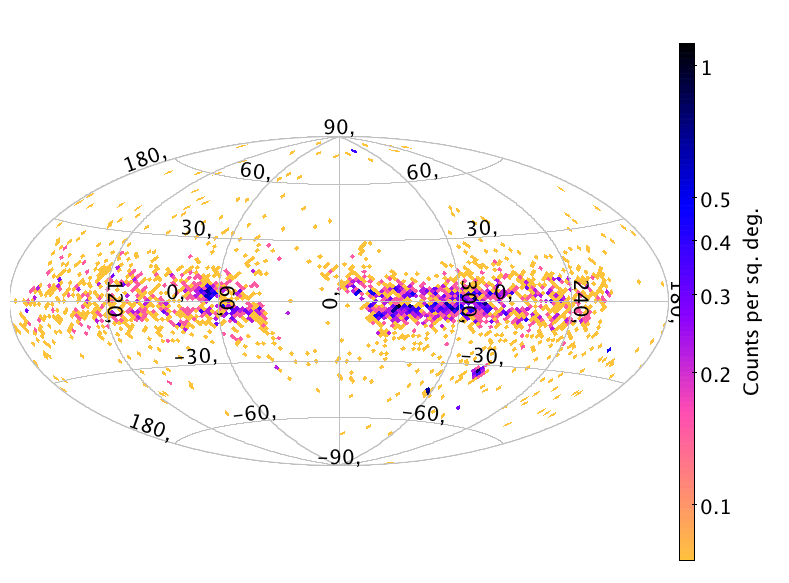}
\caption{Sky density map of the 3,018 published short-timescale candidates, in galactic coordinates. The imprint of the \textit{Gaia} scanning law and the effect of selecting sources with more than 20 FoV transits is clearly visible \citep[see][]{GDR2CU7Summary2018}}
\label{fig:sts_pub_sky_map}
\end{figure}

Figure \ref{fig:sts_pub_sky_map} represents the sky density map of the 3,018 published short-timescale candidates of \textit{Gaia} DR2. The majority of the candidates are close to the Galactic plane, with the expected lack of objects with more than 20 FoV transits around the Galactic centre. The sources found in the halo globally follow the \textit{Gaia} scanning law \citep{GDR2CU7Summary2018}. We also see slight overdensities in the southern hemisphere, corresponding to the LMC and SMC.
 
\begin{figure}
\centering
\includegraphics[width=0.99\linewidth]{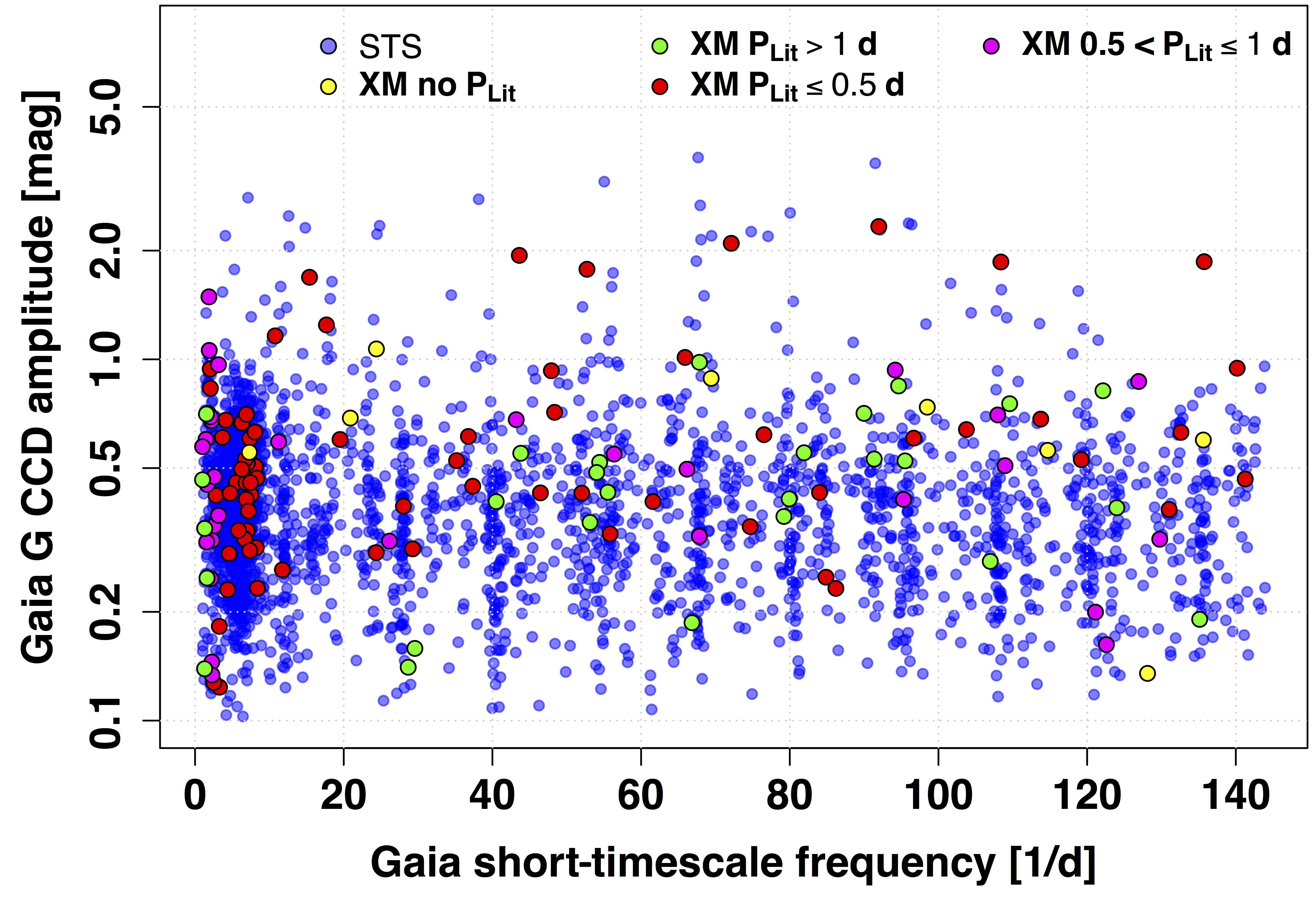}
\caption{Frequency - amplitude diagram for the 3,018 published short-timescale candidates. Known variables from the reference crossmatch catalogues are indicated by colour-coded filled circles.}
\label{fig:sts_pub_pad}
\end{figure}

\begin{figure*}
\centering
\subfloat[ ]{\includegraphics[width=0.365\linewidth, page=17, trim = {0 0 0 0}, clip=true]{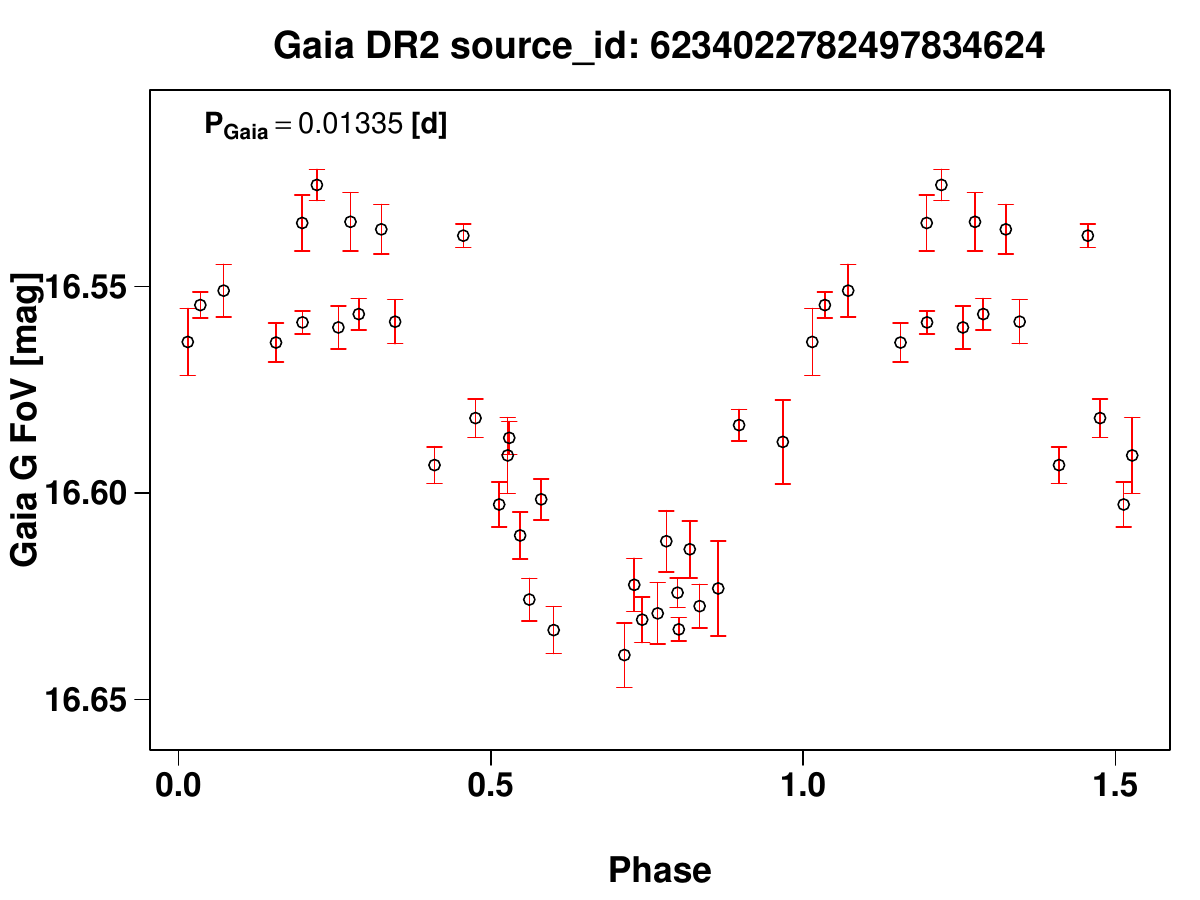}}
\subfloat[ ]{\includegraphics[width=0.365\linewidth, page=18, trim = {0 0 0 1.5cm}, clip=true]{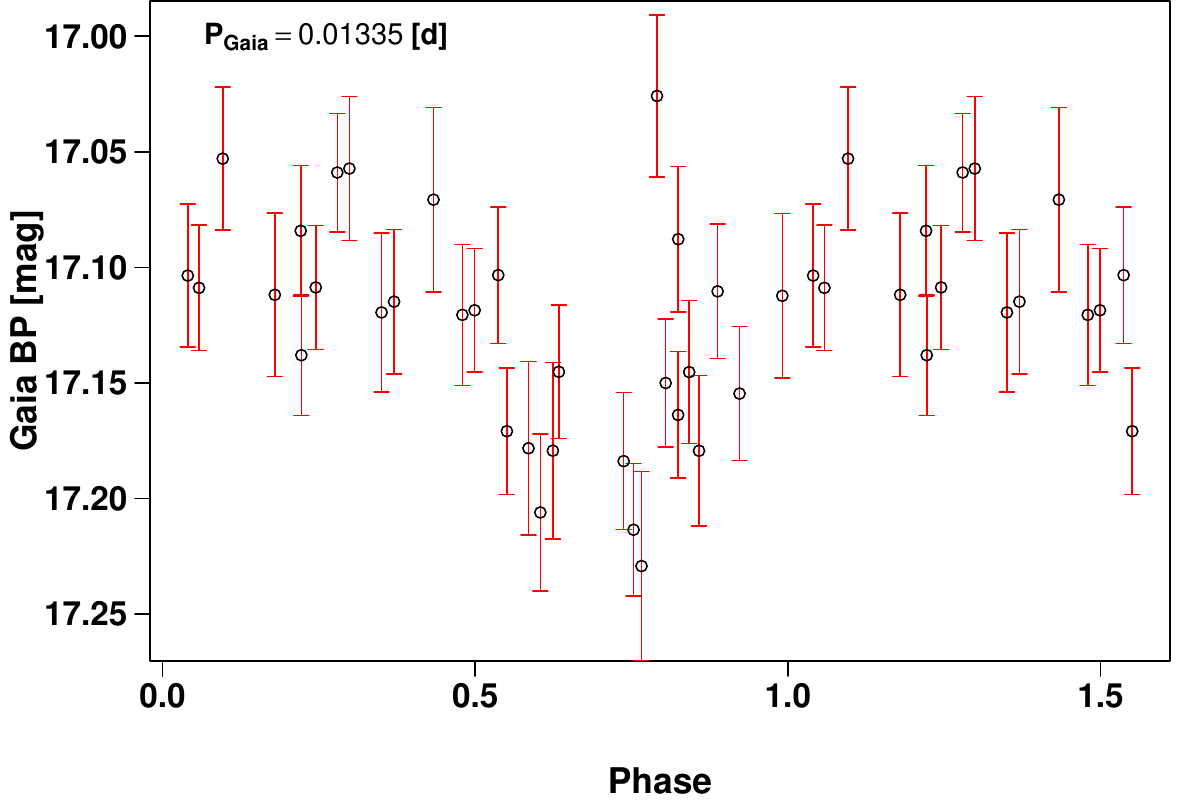}}\\
\subfloat[ ]{\includegraphics[width=0.365\linewidth, page=19, trim = {0 0 0 1.5cm}, clip=true]{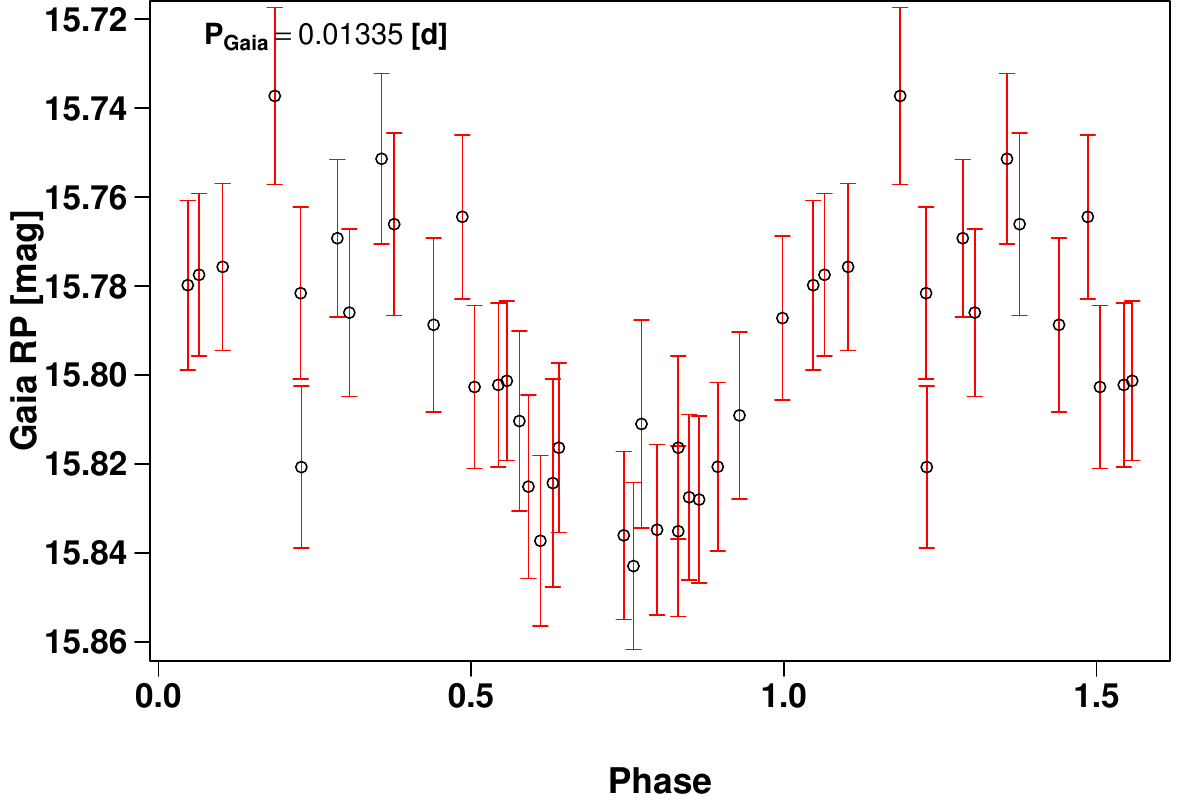}}
\subfloat[ ]{\includegraphics[width=0.365\linewidth, page=20, trim = {0 0 0 1.5cm}, clip=true]{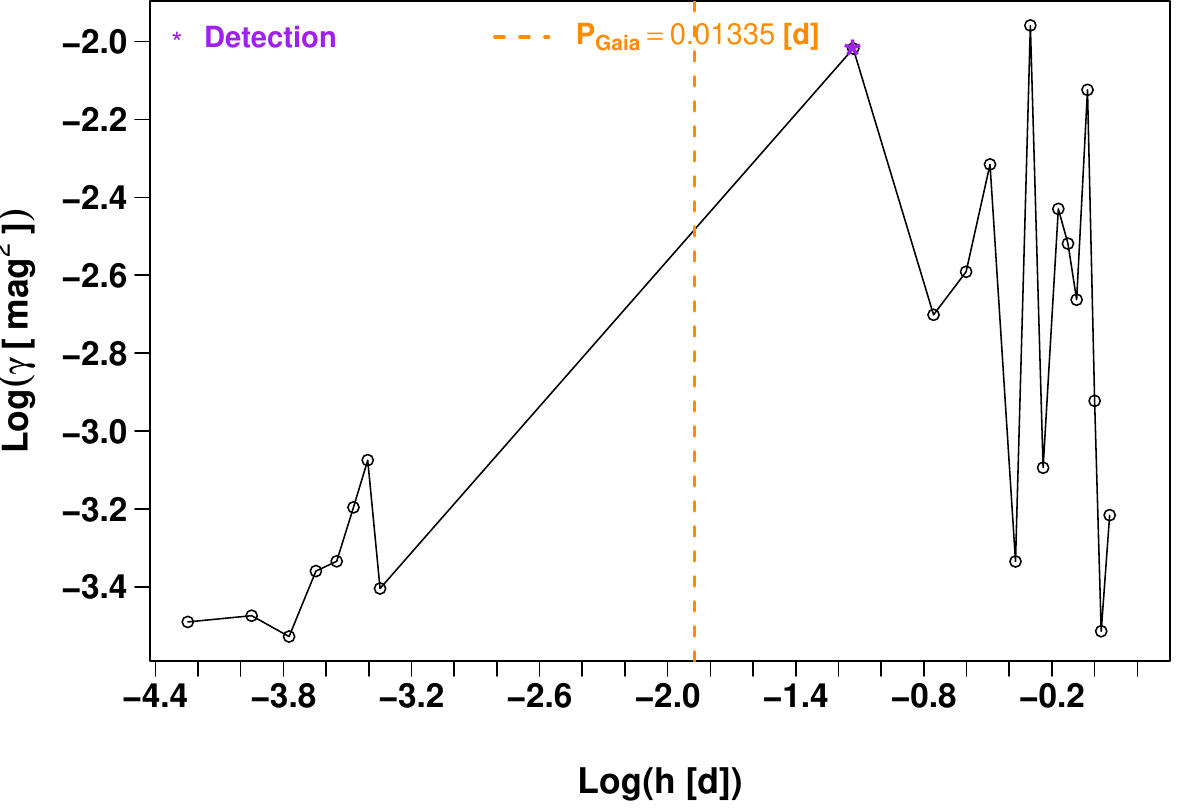}}
\caption{Phase-folded light curves in the $G$ FoV (a), $G_{\mathrm{BP}}$ (b), $G_{\mathrm{RP}}$ (c), and variogram (d) of one of the low-amplitude and short-period short-timescale candidates. Phase-folding is done using the \textit{Gaia} short-timescale period $P_{\rm Gaia}$. The reference time used to fold the \textit{Gaia} light curves is $1708.972\,$d (in BJD in $\mathrm{TCB} - 2455197.5\,$d).}
\label{fig:sts_pub_unknown_low_amp}
\end{figure*}

The frequency - amplitude diagram (Figure \ref{fig:sts_pub_pad}) shows that our final sample of candidates with short-timescale variability includes high-amplitude as well as low-amplitude variables, down to about $0.1\,$mag in G band. We note that similarly to Figure \ref{fig:sts_period_recovery}, aliasing features are clearly visible in this diagram, particularly at higher frequencies/shorter periods. However, as detailed in Sect. \ref{highFreqSearch}, we are confident that although they may not be short-period variables per se, these candidates with aliased periods are reliable periodic variable sources and have an averaged magnitude variation rate that is sufficient to justify their detection at the short-timescale level. Figure \ref{fig:sts_pub_unknown_low_amp} shows an example of a candidate with short-timescale variability from our list (\texttt{source\_id} 6234022782497834624), with a relatively low amplitude (around $0.12\,$mag) and a period of $19\,$min. Although the period may be spurious, the variogram clearly suggests a periodicity at timescales of a few hours, the variations in all three \textit{Gaia} bands are coherent, and even the phase-folded light curves look quite convincing.

\begin{figure}
\centering
\includegraphics[width=0.8\linewidth]{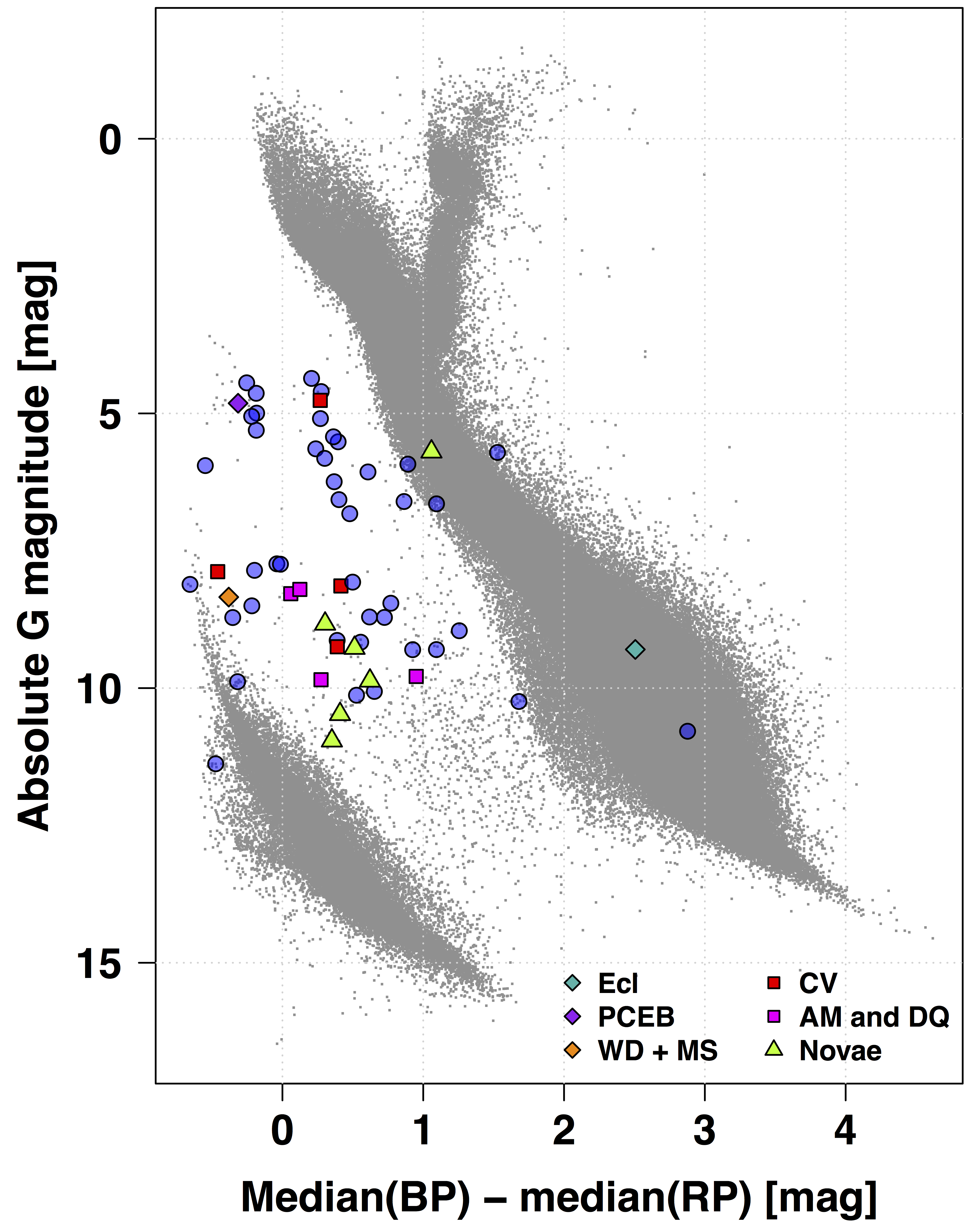}
\caption{Observational HR diagram, without correcting for extinction, of 59 of the 3,018 published short-timescale candidates, which are those with the most reliable astrometry and photometry information (blue filled circles). The filled circles of other colours represent the sources of the 59 candidates with short-timescale variability whose type is known in the literature. The candidate selection of those with a `good' parallax follows the criteria described in \cite{GDR2CU7CMD2018}, with a relative parallax precision better than 20\%. The grey background shows a subset of sources that lie closer than $200\,$pc in the HR diagram of \cite{GDR2CU7CMD2018}.}
\label{fig:sts_pub_hr}
\end{figure}

Figure \ref{fig:sts_pub_hr} shows the HR diagram of 59 of the 3,018 published candidates whose astrometry, photometry, and parallax estimates are good enough, according to the selection criterion of \cite{GDR2CU7CMD2018}, to be safely positioned in this picture. We note that no correction for extinction is applied in this plot. Some known variables among those 59 sources are also indicated. About 8 of these candidates fall on the main sequence, a few lie on the white dwarf sequence, but the majority are in between, in a region that is normally sparsely populated. However, we see that several known variables end in this area of the HR diagram, typically CV, white dwarf - main-sequence binaries, and novae. This indicates that our candidates are probably some sort of extreme binary systems, involving main-sequence stars and degenerate or semi-degenerate companions.

\begin{figure*}
\centering
\subfloat[ ]{\includegraphics[width=0.365\linewidth, page=7, trim = {0 0 0 0}, clip=true]{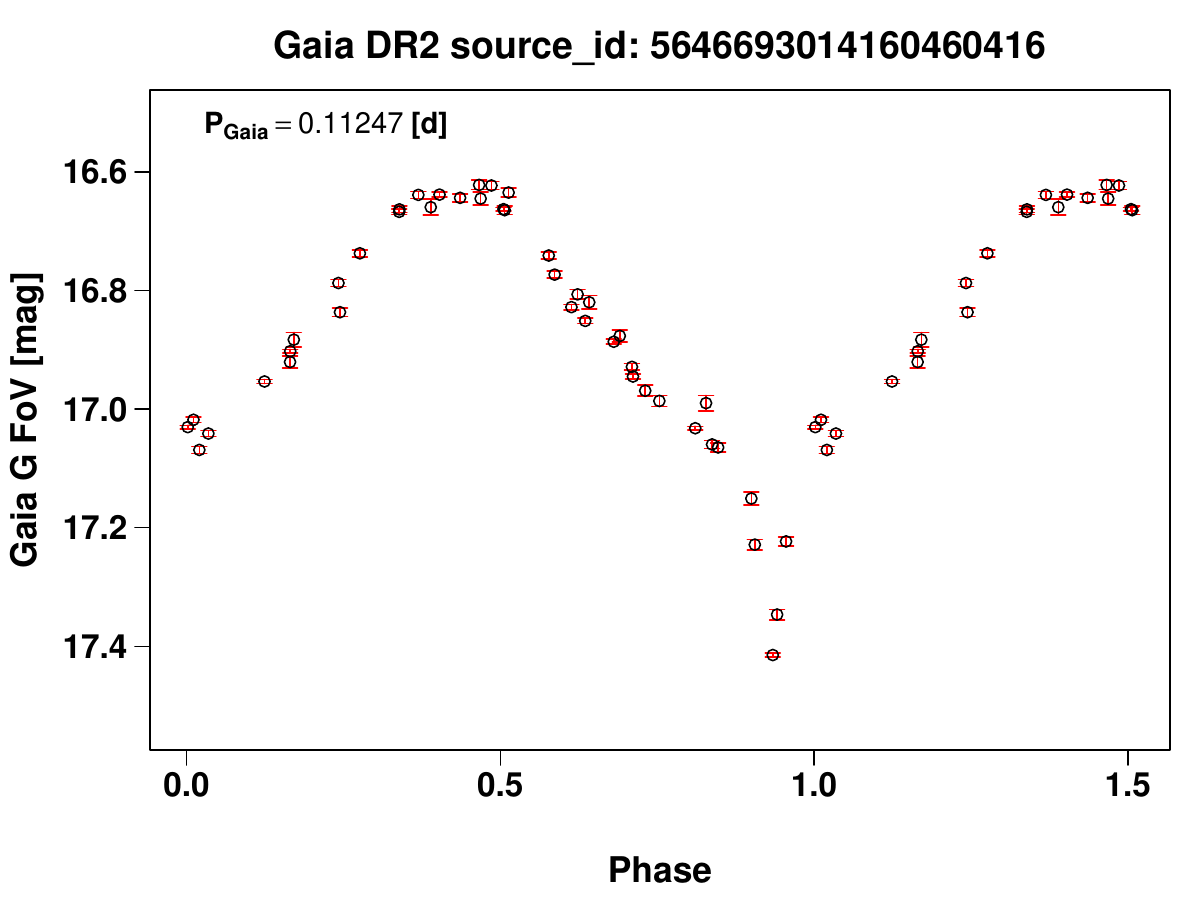}}
\subfloat[ ]{\includegraphics[width=0.365\linewidth, page=8, trim = {0 0 0 1.5cm}, clip=true]{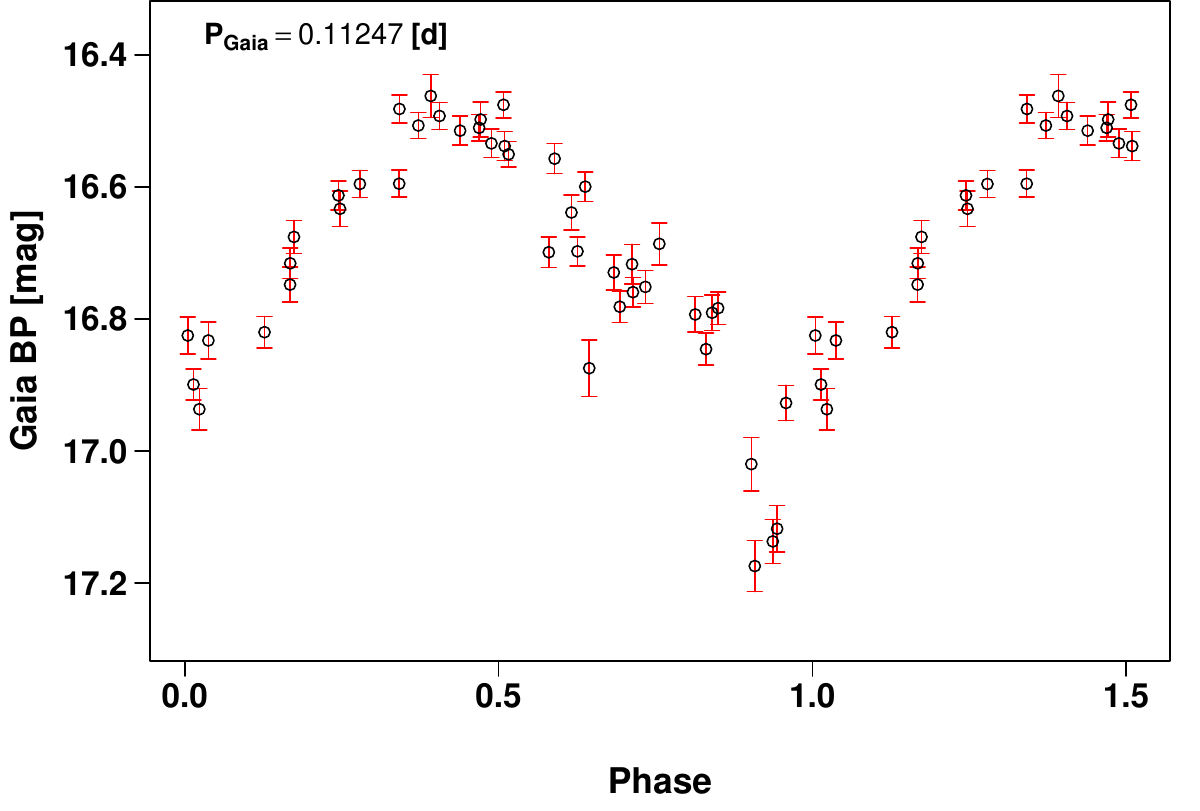}}\\
\subfloat[ ]{\includegraphics[width=0.365\linewidth, page=9, trim = {0 0 0 1.5cm}, clip=true]{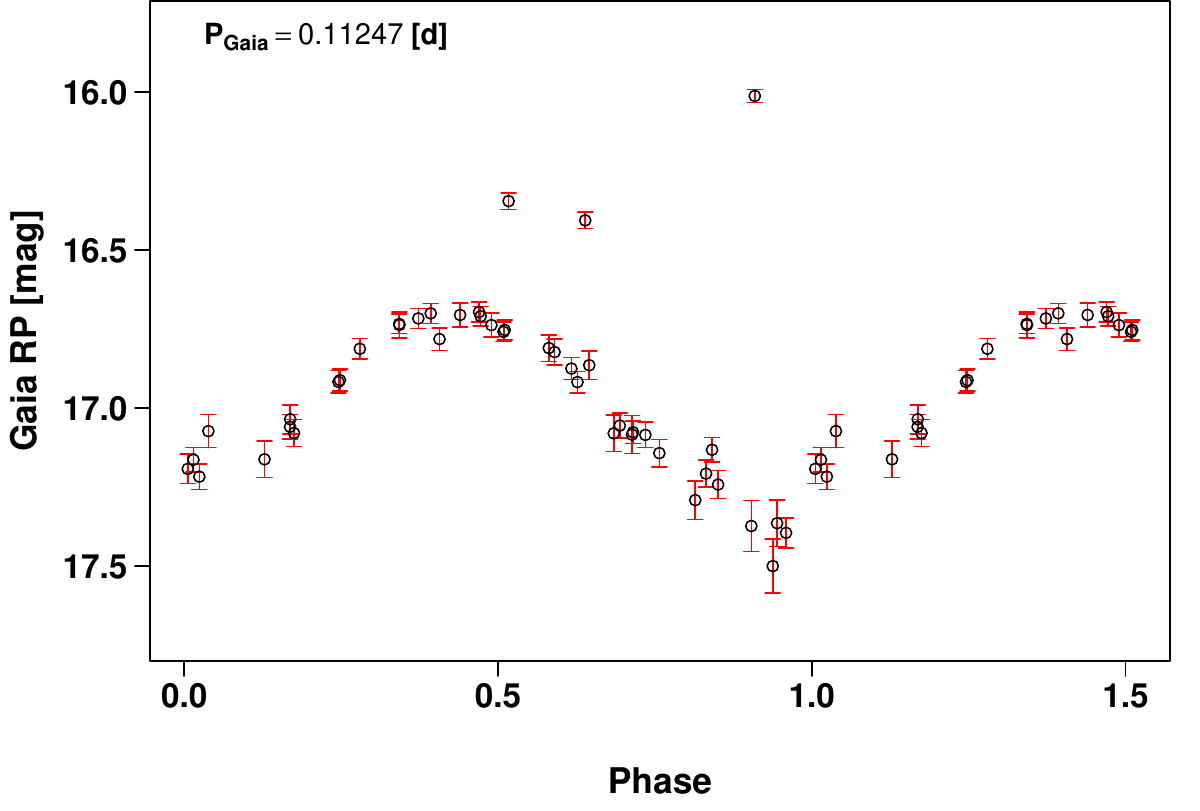}}
\subfloat[ ]{\includegraphics[width=0.365\linewidth, page=10, trim = {0 0 0 1.5cm}, clip=true]{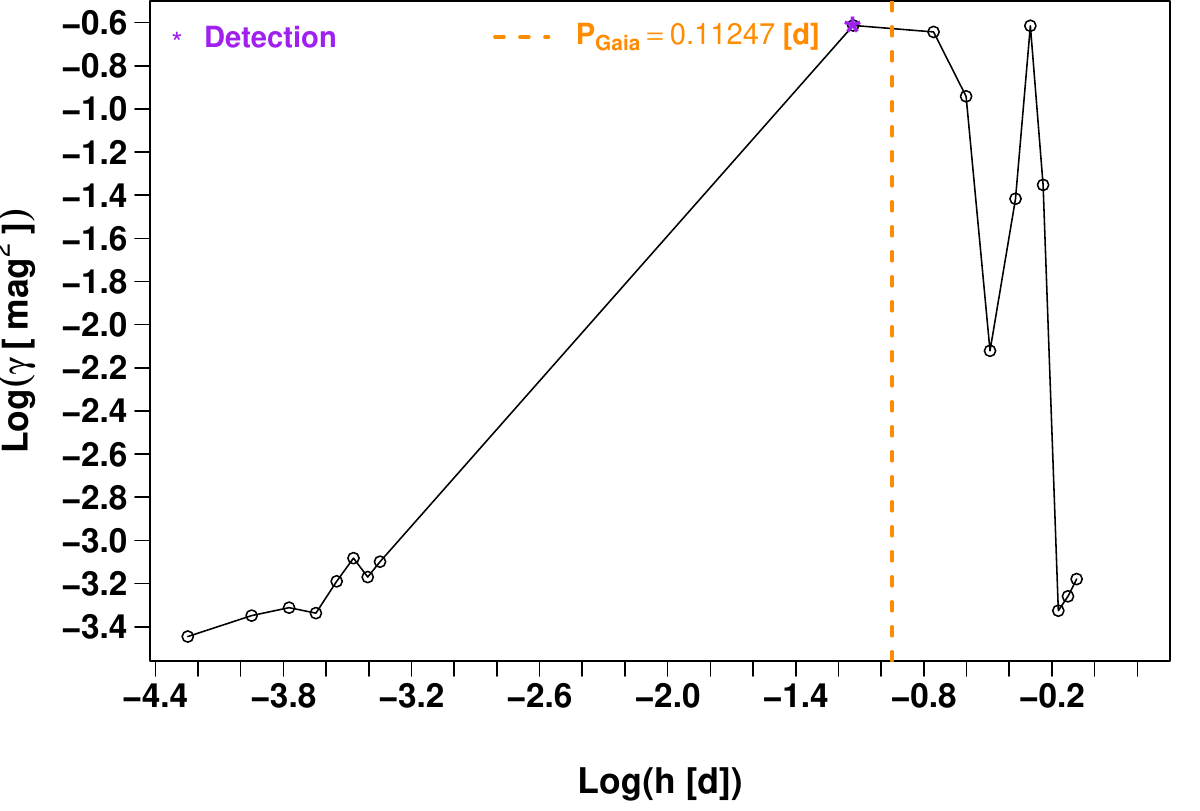}}
\caption{Same as Figure \ref{fig:sts_pub_unknown_low_amp} for a blue and bright source, a possible PCEB, of the short-timescale candidates. The reference time used to fold the \textit{Gaia} light curves is $1764.737\,$d (in BJD in $\mathrm{TCB} - 2455197.5\,$d).}
\label{fig:sts_pub_unknown_pceb}
\end{figure*}

Among the bluer and brighter short-timescale candidates in the same HR diagram region as the PCEB CSS J210017 (purple diamond in Figure \ref{fig:sts_pub_hr}), we find a very good example of a possibly unknown PCEB (\texttt{source\_id} 5646693014160460416) with a period of $2.7\,$h (Figure \ref{fig:sts_pub_unknown_pceb}).

\begin{figure*}
\centering
\subfloat[ ]{\includegraphics[width=0.365\linewidth, page=12, trim = {0 0 0 0}, clip=true]{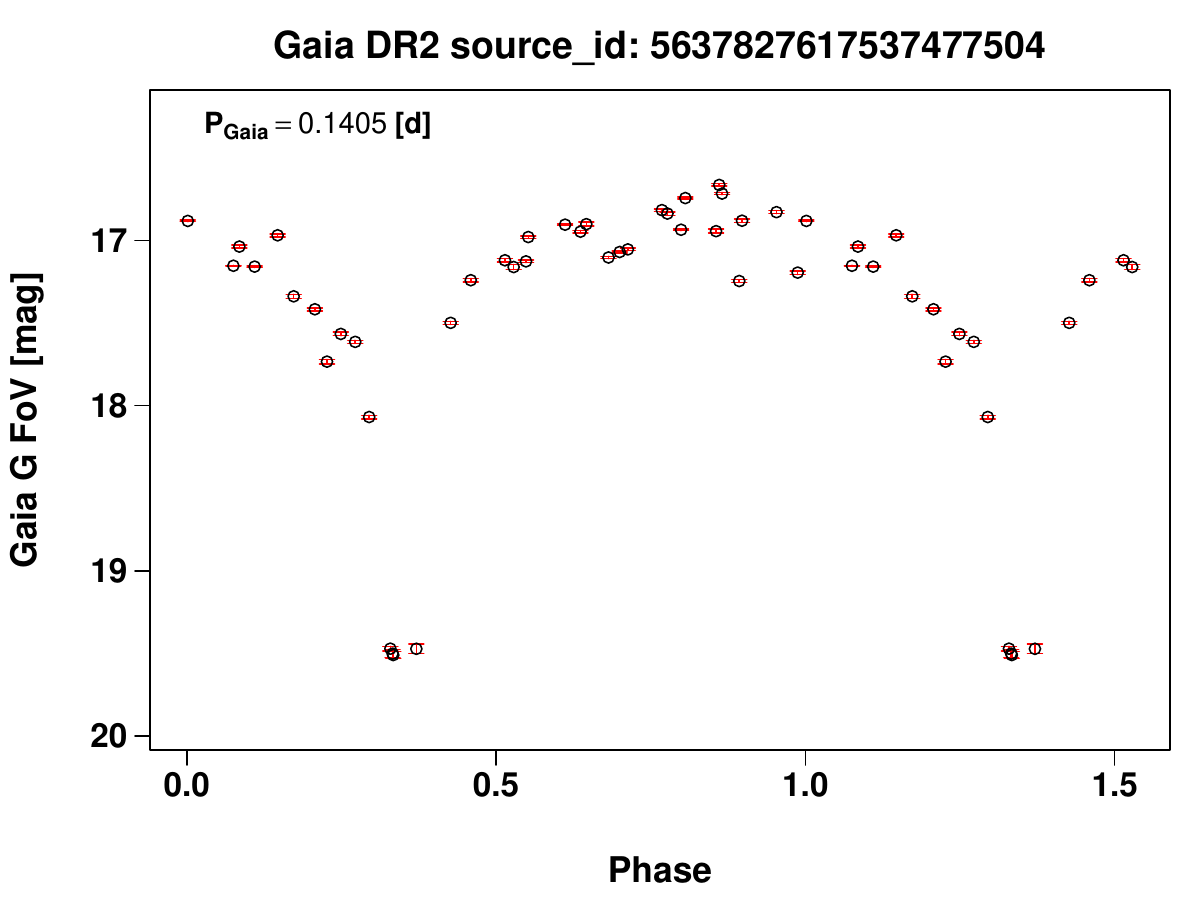}}
\subfloat[ ]{\includegraphics[width=0.365\linewidth, page=13, trim = {0 0 0 1.5cm}, clip=true]{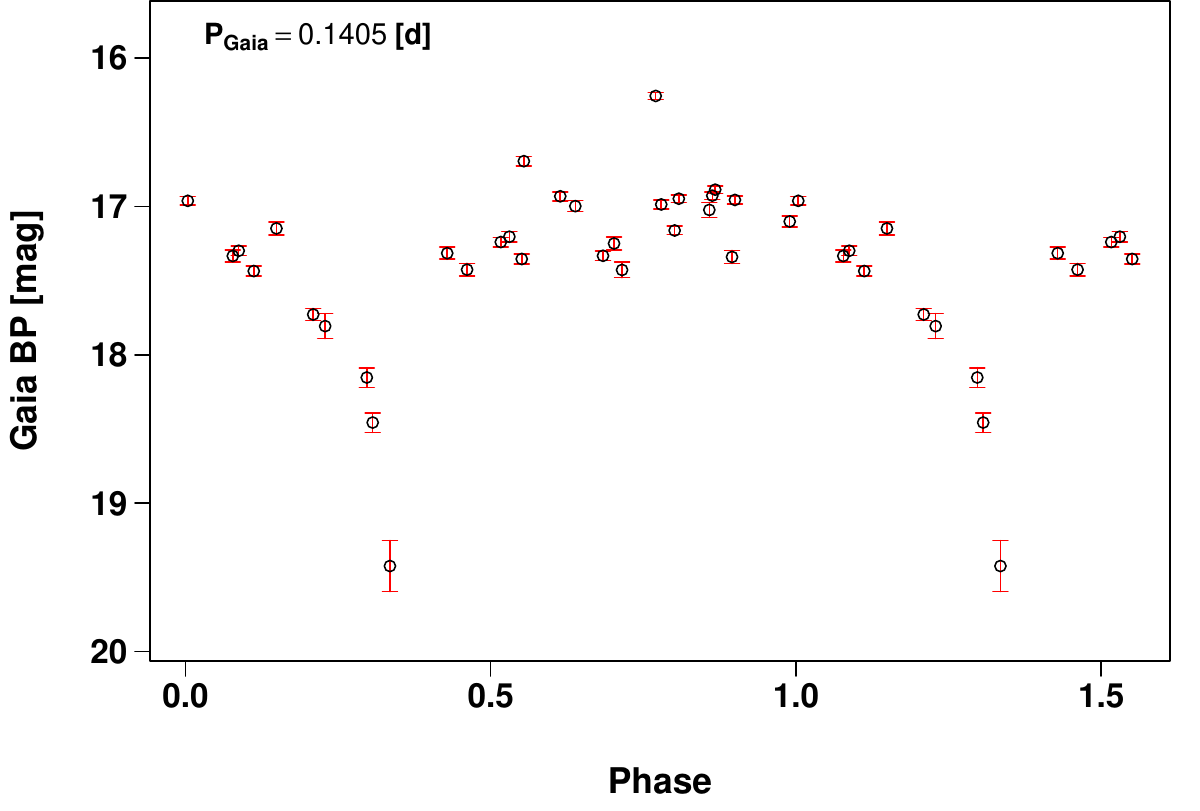}}\\
\subfloat[ ]{\includegraphics[width=0.365\linewidth, page=14, trim = {0 0 0 1.5cm}, clip=true]{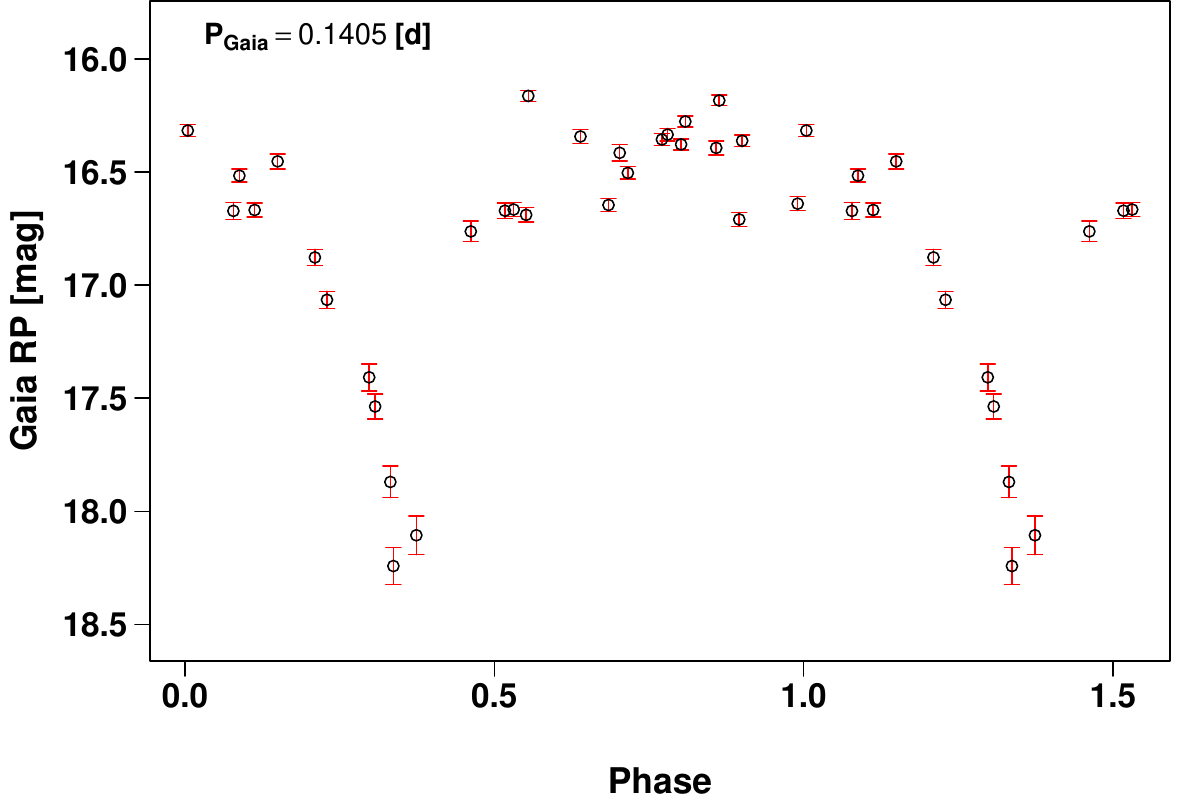}}
\subfloat[ ]{\includegraphics[width=0.365\linewidth, page=15, trim = {0 0 0 1.5cm}, clip=true]{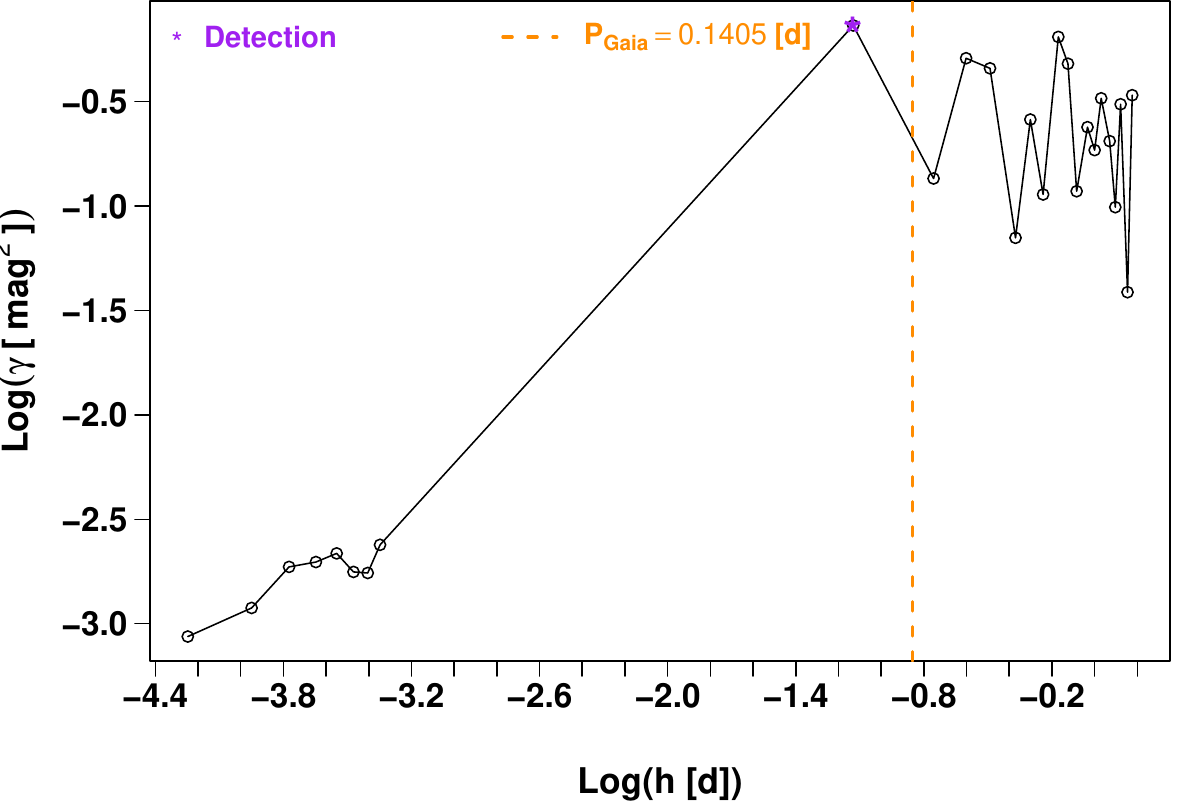}}
\caption{Same as Figure \ref{fig:sts_pub_unknown_pceb} for a curious eclipsing binary among the short-timescale candidates. The reference time used to fold the \textit{Gaia} light curves is $1765.980\,$d (in BJD in $\mathrm{TCB} - 2455197.5\,$d).}
\label{fig:sts_pub_unknown_ecl_bizarre}
\end{figure*}

By visually inspecting the light curves of all the 3,018 short-timescale candidates, we also spotted some peculiar and interesting candidates with short-timescale variability that are new discoveries, as far as we know. In particular, we wish to highlight the case of source 5637827617537477504, represented in Figure \ref{fig:sts_pub_unknown_ecl_bizarre}. It exhibits very strong eclipses of more than $1 - 1.5\,$mag in the $G$, $G_{\mathrm{BP}}$ , and $G_{\mathrm{RP}}$ bands and shows a significant out-of-eclipse variability, with an overall shape similar to what is expected for AM CVn stars, for instance. However, the period from the short-timescale analysis is $3.4\,$h, which is longer than the orbital periods of known AM CVn stars, which are between $5$ and $65\,$min \citep[see e.g.][]{Levitan2015}.
Further investigation and modelling is required to better understand and characterise this curious system.
 
\section{Conclusion}
\label{conclu}

By combining the variogram analysis, the least-squares high-frequency search, and selection criteria based on various metrics, involving \textit{Gaia} $G$ and $G_{\mathrm{BP}}$/$G_{\mathrm{RP}}$ photometry, we identified a first \textit{Gaia} set of 3,018 candidates with short-timescale and suspected periodic variability. The completeness of this sample is assessed to be about 0.05\% of all the known short-period variables scanned by \textit{Gaia} during its first 22 months of science operations (regardless of any selection based on per-CCD data or on variogram analysis), and to be about 12\% of the input sample (i.e. compared to the known short-period variables that have been processed by the short-timescale module). The contamination from false positives and non-periodic variable sources can be as high as 10-20\% in denser regions. Contamination from longer-period variables is more significant (about 20-50\%), but this can be justified as this contamination comes from variable objects with periods of a few days and amplitudes that are significant enough for a detection to be triggered at short timescales.

Owing to limited period-recovery capabilities when compared to the literature, the period information provided as part of the \textit{Gaia} DR2 analysis results of short-timescale variables must be used with caution, and is rather communicated for indicative purposes. For the upcoming \textit{Gaia} data releases, we plan to improve the high-frequency search method, so as to better handle the aliasing problem and obtain a significantly higher period recovery rate. In this perspective, implementing the estimation of typical timescale(s) from the variogram analysis, as described in \cite{Roelens2017}, would be a real asset that would add information complementary to frequency-search results.

This \textit{Gaia} DR2 short-timescale sample is one of the first lists of such variable candidates resulting from a global, comprehensive search for any fast periodic variability over a large fraction of the sky. Even by analysing only the first 22 months of intermediate \textit{Gaia} per-CCD photometry in $G$ band, we obtained promising results with the recovery and discovery of very interesting candidates with short-timescale variability, which shows the great potential of the \textit{Gaia} mission for fast-variability studies.

As explained throughout this paper, the aim of this analysis was not to reach a high level of completeness nor to describe the retrieved candidates with a very high level of detail, but more to open a new door on the rather unexplored domain of fast astronomical variability, encouraging further follow-up and characterisation of the identified sources of interest. For the next \textit{Gaia} data releases, our goal is not only to extend the list of published candidates, widening the explored magnitude range, and benefitting from the improved photometric calibration and longer time-span of the processed data (hence with more sources having a sufficient number of transits for variogram investigation), but also to proceed beyond in the analysis by classifying the detected candidates based on their magnitude, colour, astrometry, and any relevant information from \textit{Gaia} products (e.g. astrophysical parameter estimations).

\section*{Acknowledgments}
\label{merci}

This work was supported by the Swiss National Science Foundation (fns 200020\_166230/OB5276).

%It has made use of data from the European Space Agency (ESA) mission
%{\it Gaia} (\url{https://www.cosmos.esa.int/gaia}), processed by the {\it Gaia}
%Data Processing and Analysis Consortium (DPAC,
%\url{https://www.cosmos.esa.int/web/gaia/dpac/consortium}). Funding for the DPAC
%has been provided by national institutions, in particular the institutions
%participating in the {\it Gaia} Multilateral Agreement.

It has made use of data from the ESA space mission \textit{Gaia}, processed by the \textit{Gaia} Data Processing and Analysis Consortium (DPAC).
Funding for the DPAC has been provided by national institutions, some of which participate in the \textit{Gaia} Multilateral Agreement, 
which include, 
for Switzerland, the Swiss State Secretariat for Education, Research and Innovation through the ESA PRODEX program, the ``Mesures d'accompagnement'', the ``Activit\'{e}s Nationales Compl\'{e}mentaires'', the Swiss National Science Foundation, and the Early Postdoc. Mobility fellowship;
 Belgium, the BELgian federal Science Policy Office (BELSPO) through PRODEX grants;
for Italy, Istituto Nazionale di Astrofisica (INAF) and the Agenzia Spaziale Italiana (ASI) through grants I/037/08/0,  I/058/10/0,  2014-025-R.0, and 2014-025-R.1.2015 to INAF (PI M.G. Lattanzi);
for France, the Centre National d'Etudes Spatiales (CNES). 

This paper is partly based on observations made with the Mercator Telescope, operated on the island of La Palma by the Flemish Community, at the Spanish Observatorio del Roque de los Muchachos of the Instituto de Astrof\'isica de Canarias. In particular, it makes use of observations obtained with the MAIA camera, which was built by the Institute of Astronomy of KU Leuven, Belgium, thanks to funding from the European Research Council under the European Community's Seventh Framework Programme (FP7/2007-2013)/ERC grant agreement no 227224 (PROSPERITY, PI: Conny Aerts) and from the Research Foundation - Flanders (FWO) grant agreement G.0410.09. The CCDs of MAIA were developed by e2v in the framework of the ESA Eddington space mission project; they were offered by ESA on permanent loan to KU Leuven.

We gratefully acknowledge Mark Taylor for creating and implementing new requested features in the astronomy-oriented data handling and visualisation software TOPCAT \citep{Taylor2005}.

\bibliographystyle{aa}
\bibliography{sts_dr2_squelette_mybib}

\onecolumn

\begin{appendix}

\section{Catalogue retrieval}
We summarise below the ADQL queries to be used in the web interface to the Gaia DR2 archive (\url{https://gea.esac.esa.int/ archive/})  for retrieving the \textit{Gaia} DR2 candidates with short-timescale variability, their attributes, and their light curves.\\

- Retrieving the attributes of the candidates with short-timescale variability:
\begin{verbatim}
SELECT sts.*
FROM gaiadr2.vari_short_timescale AS sts
\end{verbatim}

- Retrieving the light curves of the short-timescale candidates. The following query retrieves the URL at which the photometry of an individual \textit{Gaia} source, in the example with \texttt{source\_id} 5637827617537477504, can be downloaded:
\begin{verbatim}
SELECT gaia.epoch_photometry_url
FROM gaiadr2.gaia_source AS gaia
INNER JOIN gaiadr2.vari_short_timescale AS sts
ON gaia.source_id = sts.source_id
AND sts.source_id = 5637827617537477504
\end{verbatim}

- Retrieving the statistics on the \textit{Gaia} photometric light curves (from the \texttt{vari\_time\_series\_statistics} table), for example, the mean, median, or the IQR of $G$ FoV, $G_{\mathrm{BP}}$ , and $G_{\mathrm{RP}}$ magnitudes, for all the DR2 candiates with short-timescale variability, together with their short-timescale attributes:
\begin{verbatim}
SELECT stat.mean_mag_g_fov, stat.median_mag_g_fov, stat.iqr_mag_g_fov,
stat.mean_mag_bp, stat.median_mag_bp, stat.iqr_mag_bp,
stat.mean_mag_rp, stat.median_mag_rp, stat.iqr_mag_rp,
sts.*
FROM gaiadr2.vari_time_series_statistics AS stat
INNER JOIN gaiadr2.vari_short_timescale AS sts
ON stat.source_id = sts.source_id
\end{verbatim}

- Retrieving the coordinates, parallaxes, and attributes for the candidates with short-timescale variability:
\begin{verbatim}
SELECT gaia.source_id, ra, dec, parallax, parallax_error, sts.*
FROM gaiadr2.gaia_source AS gaia
INNER JOIN gaiadr2.vari_short_timescale AS sts
ON gaia.source_id = sts.source_id
\end{verbatim}

\end{appendix}

\end{document}